\documentclass[12pt,oneside,letterpaper]{article}
\usepackage{caption}
\usepackage{slashed}
\usepackage{float}
\restylefloat{table}
\usepackage{tabu}
\usepackage{boldline,multirow}
\usepackage{multirow,array}
\usepackage{arydshln}
\usepackage{makecell}
\usepackage{longtable}
\usepackage{amssymb}
\usepackage{amsmath}
\numberwithin{equation}{section}
\usepackage[dvips]{graphicx}
\usepackage[dvipsnames]{xcolor}
\usepackage{setspace}
\usepackage{amsfonts}
\usepackage{fancyhdr}
\usepackage{xcolor}
\usepackage{graphicx}
\usepackage{rotating}
\usepackage{comment}
\usepackage{color}
\usepackage{cite}
\usepackage{braket}
\usepackage{amssymb,amsmath,amsthm,mathrsfs,latexsym,amsxtra,graphicx,appendix,epstopdf,feynmf,subfig,setspace,fix-cm,color,cite}
\usepackage[nottoc,numbib]{tocbibind}
\usepackage{pgfplots}
\usepgfplotslibrary{polar}
\usepgfplotslibrary{fillbetween}
\pgfplotsset{compat=newest}
\usepackage{mathtools}
\usepackage{tensor}
\usepackage{siunitx}
\usepackage{tikz}
\usepackage{tikzlings}
\usetikzlibrary{trees}
\usetikzlibrary{decorations.pathmorphing}
\usetikzlibrary{decorations.markings}
\usetikzlibrary{decorations.markings}
\usepackage{float}
\usepackage{rotating}
\usepackage{simplewick}
\usepackage{amssymb}
\usepackage{amsmath}
\usepackage{setspace}
\usepackage{amsfonts}
\usepackage{fancyhdr}
\usepackage{xcolor}
\usepackage{graphicx}
\usepackage{rotating}
\usepackage{comment}
\usepackage{color}
\usepackage{cite}
\usepackage{braket}
\usepackage[colorlinks=true, linkcolor  = blue,urlcolor=blue,citecolor=violet]{hyperref}
\usepackage{simpler-wick}

\definecolor{darkgreen}{rgb}{0,0.5,0}
\definecolor{darkblue}{rgb}{0,0,0.6}
\definecolor{purple}{rgb}{0.4,0.15,0.21}
\definecolor{black}{rgb}{.2,.2,.2}

\DeclareSymbolFont{myletters}{OML}{ztmcm}{m}{it}
\DeclareMathSymbol{\uplambda}{\mathord}{myletters}{"15}

\begin{document} 

\tikzset{middlearrow/.style={
        decoration={markings,
            mark= at position 0.5 with {\arrow{#1}} ,
        },
        postaction={decorate}
    }
}

\newcommand{\Vast}{\bBigg@{5}}
\newtheorem*{theorem}{Theorem}
\newtheorem{conjecture}{Conjecture}[section]
\newtheorem{corollary}{Corollary}
\newtheorem{lemma}{Lemma}
\newtheorem*{remark}{Remark}
\newtheorem{claim}{Claim}
\newtheorem{prop}{Proposition}
\theoremstyle{definition}
\newtheorem{definition}{Definition}[section]
\renewcommand\qedsymbol{$\blacksquare$}

\newcommand{\be}{\begin{equation}}
\newcommand{\ee}{\end{equation}}
\newcommand{\bea}{\begin{eqnarray}}
\newcommand{\eea}{\end{eqnarray}}
\newcommand{\ba}{\begin{align}}
\newcommand{\ea}{\end{align}}
\newcommand{\bb}{\mathbb}
\newcommand{\mrm}{\mathrm}
\newcommand{\scr}{\mathscr}
\newcommand{\p}{\partial}
\newcommand{\dd}{\mathrm{d}}
\newcommand{\tr}{\mathrm{tr}}
\newcommand{\RR}{\mathrm{R}}
\newcommand{\FF}{\mathrm{F}}
\newcommand{\DD}{\mathrm{D}}
\newcommand{\ff}{\mathrm{f}}
\newcommand{\rr}{\mathrm{r}}
\newcommand{\Tr}{\mathrm{Tr}}
\def\e{{\rm e}}
\def\bz{{\bar z}}
\def\bw{{\bar w}}
\def\p{{$|\Phi\rangle$}}
\def\pp{{$|\Phi^\prime\rangle$}}
\def\cO{{\mathcal O}}
\def\cH{\mathcal{H}}
\def\cF{\mathcal{F}}
\def\cL{\mathcal{L}}
\def\Xint#1{\mathchoice
   {\XXint\displaystyle\textstyle{#1}}%
   {\XXint\textstyle\scriptstyle{#1}}%
   {\XXint\scriptstyle\scriptscriptstyle{#1}}%
   {\XXint\scriptscriptstyle\scriptscriptstyle{#1}}%
   \!\int}
\def\XXint#1#2#3{{\setbox0=\hbox{$#1{#2#3}{\int}$}
     \vcenter{\hbox{$#2#3$}}\kern-.5\wd0}}
\def\ddashint{\Xint=}
\def\dashint{\Xint-}

\newcommand{\CC}{\mathbb{C}} 
\newcommand{\ZZ}{\mathbb{Z}} 
\newcommand{\II}{\mathrm{I}} 
\newcommand{\NN}{\mathbb{N}} 
\newcommand{\QQ}{\mathrm{Q}}
\newcommand{\xx}{\mathrm{x}}
\newcommand{\Pp}{\mathrm{p}}

\newcommand{\T}{R}

\newcommand{\lcm}{\mathrm{lcm}}

\newcommand{\ie}{{\it i.e.~}}
\def\eg{{\it e.g.~}}

\onehalfspacing

\begin{center}

~
\vskip4mm
{{\huge {
\quad Matrix integrals \& finite holography
 }
  }}
\vskip5mm

\vskip2mm

\vskip10mm


{\color{black}D}ionysios Annino{\color{black}s}$^{\,{\color{magenta}\mathcal{I}^+}}$\, \& ~Be{\color{black}atrix M}\"uhlmann$^{{\color{magenta}N^2}}$ \\ 

\end{center}
\vskip4mm
\begin{center}
{
\footnotesize
{$^{{\color{magenta}\mathcal{I}^+}}$ Department of Mathematics, King's College London, Strand, London WC2R 2LS, UK \newline\newline
$^{{\color{magenta}N^2}}$ Institute for Theoretical Physics and $\Delta$ Institute for Theoretical Physics, University of Amsterdam, \\Science Park 904, 1098 XH Amsterdam, The Netherlands\\
}}
\end{center}
\begin{center}
{\textsf{\footnotesize{
dionysios.anninos@kcl.ac.uk \& ~
b.muhlmann@uva.nl}} } 
\end{center}
\vskip5mm

\vspace{4mm}

\begin{abstract}
We explore the conjectured duality between a class of large $N$ matrix integrals, known as multicritical matrix integrals (MMI), and the series $(2m-1,2)$ of non-unitary minimal models on a fluctuating background. We match the critical exponents of the leading order planar expansion of MMI, to those of the continuum theory on an $S^2$ topology. From the MMI perspective this is done both through a multi-vertex diagrammatic expansion, thereby revealing novel combinatorial expressions, as well as through a systematic saddle point evaluation of the matrix integral as a function of its parameters. From the continuum point of view the corresponding critical exponents are obtained upon computing the partition function in the presence of a given conformal primary. Further to this, we elaborate on a Hilbert space of the continuum theory, and the putative finiteness thereof, on both an $S^2$ and a $T^2$ topology using BRST cohomology considerations. Matrix integrals support this finiteness.
 \end{abstract}
\vspace{.2in}

\pagebreak
\pagestyle{plain}
\setcounter{tocdepth}{2}
{}
\vfill

\tableofcontents


\section{Introduction}

Being $0+0$ systems, matrix integrals are of a more finite nature than large $N$ quantum field theories traditionally explored in holography.
In this work we explore, discuss and review in detail a particular class of matrix integrals, known as multicritical matrix integrals (MMI) \cite{Kazakov:1989bc, Staudacher:1989fy}. MMI are built out of a single Hermitian $N\times N$ matrix organised in an even polynomial of order $2m$ with $(m-1)$  free parameters (couplings). Despite being constructed from a single matrix, MMI admit $(m-1)$ distinct critical exponents in the leading order planar expansion, which are encoded in the non-analytic behaviour of the matrix integral as a function of its couplings. 
In the large $N$ limit and upon tuning the couplings to a set of special values, MMI are conjectured to be dual to the series ($2m-1,2$) of minimal models  \cite{Polyakov:1984yq, Friedan} --- which we denote by  $\mathcal{M}_{2m-1,2}$ --- on a fluctuating background. This can be described by coupling $\mathcal{M}_{2m-1,2}$ to two-dimensional quantum gravity, where the theory of quantum gravity at hand upon fixing the Weyl gauge is given by Liouville theory \cite{Polyakov:1981rd}. In part, our work extends to general $m$ the analysis and approach of \cite{Staudacher:1989fy} which dealt specifically with the case $m=3$.

 For $m\geq 2$, $\mathcal{M}_{2m-1,2}$ is a non-unitary two-dimensional CFT consisting of $(m-1)$ distinct Virasoro primaries, each accompanied with an infinite tower of Virasoro descendants. The conformal dimensions of the Virasoro primaries are increasingly negative, with the highest being the  vanishing conformal dimension of the identity operator. While the norm of the Virasoro primaries of $\mathcal{M}_{2m-1,2}$ is positive, the norm of the Virasoro descendants is negative, leading to the non-unitarity of the models. 
We note that the non-unitary minimal models $\mathcal{M}_{2m-1,2}$ are related to integrable lattice models. The Lee-Yang singularity \cite{LeeYang} characterising the zeroes of the partition function of the Ising model in an imaginary magnetic field in the thermodynamic limit has been identified with $\mathcal{M}_{5,2}$  \cite{Cardy:1985yy}. On the other side $\mathcal{M}_{7,2}$ has been conjectured \cite{vonGehlen:1994rp}  to correspond to the tricritical phase (the crossing point of the three lowest energy levels) of a generalisation of the Ising model with three state classical variables, known as the Blume-Capel model \cite{Blume}. 

As explored extensively throughout this work, an important piece of evidence in establishing the conjectured duality between MMI and Liouville theory coupled to $\mathcal{M}_{2m-1,2}$ is the matching of critical exponents between MMI and the continuum theory. We uncover the relation between the continuum theory on an $S^2$ topology and the explicit form of the MMI through its (finite) coupling dependence in the leading order planar expansion, as well as its perturbative multi-vertex  expansion. 

A more expansive and in some ways orthogonal direction has been pursued in \cite{Moore:1991ir, Moore:1991ag, Belavin:2014hsa, Belavin:2005af, Belavin:2008kv}. The authors compare correlation functions of integrated operators (correlation numbers) to analogous quantities in the matrix integral. In our analysis of the partition function of the continuum theory we are turning on a single operator at finite coupling of the minimal model, the calculation of correlation numbers involves turning on multiple operators with an infinitesimal coupling. 

MMI have also made an appearance, see e.g. \cite{Saad:2019lba,Johnson:2019eik}, in the context of JT gravity \cite{Jackiw:1984je}. In that context the continuum theory is studied on manifolds with boundary as compared to our analysis on $S^2$ and more generally on compact Riemann surfaces. 

One of the key motivations of our work is the existence of a semiclassical limit exhibited by Liouville theory coupled to $\mathcal{M}_{2m-1,2}$. This is the large $m$ limit,  and was first observed in \cite{Zamolodchikov:1982vx}. Specifically, upon fixing the area of the physical metric, restricting to an $S^2$ topology, and turning on only the identity operator of $\mathcal{M}_{2m-1,2}$ one finds a round two-sphere geometry as the saddle point solution. This is the geometry of Euclidean two-dimensional de Sitter space. 
Two-dimensional de Sitter space supports finiteness \cite{Erik_dS,Dong:2010pm, Banks_dS}, and its conjectured entropy is finite \cite{Gibbons:1977mu, Gibbons:1976ue}. 

\subsubsection*{Outline}
In section \ref{singlePD} and \ref{multiplePD} we study the diagrammatic expansion of MMI, providing new combinatorial expressions for Feynman diagrams whose vertices emanate an arbitrary even number of edges. As an example there are 243180821057048715513033825248570669471308484796973569520429442294243 32116879409838986729881600000000000000000 diagrams consisting of fifteen distinct vertices emanating an even number between four and thirty-two edges. 
We provide a concrete framework identifying each of the $(m-1)$ distinct planar critical exponents of the MMI in section \ref{nonanalyticsec}. Geometrically these critical exponents are living on distinct fine-tuned ``hypersurfaces'' in coupling space. We match the critical exponents of the MMI to those of the continuum theory of Liouville theory coupled to $\mathcal{M}_{2m-1,2}$ in section \ref{continuum_sec}. This matching comes with an important subtlety. Whereas for unitary minimal models the identification of critical exponents in the matrix integral with critical exponents of the emergent large $N$ continuum theory uses the KPZ relation \cite{Knizhnik:1988ak, David:1988hj, Distler:1988jt}, the minimal models at hand are non-unitary and require a generalisation of the KPZ formula \cite{Staudacher:1989fy, Brezin:1989db}. 
In section \ref{sec:Hilbert} we consider the operator content of $\mathcal{M}_{2m-1,2}$ on a fluctuating background. On a fluctuating background the number of operators of the $\mathcal{M}_{2m-1,2}$ is subject to the Virasoro constraints. Gauge fixing to the Weyl gauge further introduces the $\mathfrak{b}\mathfrak{c}$-ghost system. Initiated by work of Lian-Zuckerman (LZ) \cite{Lian:1991gk} and subsequent work \cite{Imbimbo:1991ia, Bouwknegt:1991mv, Kutasov:1991qx} it was observed that the resulting BRST cohomology admits an infinite number of operators with non-vanishing ghost number and matter and Liouville descendants. The infinite set of LZ operators is still  much smaller than the infinite tower of Virasoro descendants arising for each primary operator of $\mathcal{M}_{2m-1,2}$ on a fixed sphere. As a consequence of the Riemann-Roch theorem we do however infer that LZ operators do not lead to additional critical exponents on an $S^2$ topology. This may render the non-unitarity of $\mathcal{M}_{2m-1,2}$ on $S^2$ less severe. 
On the other hand the LZ operators contribute to the torus partition function \cite{Bershadsky:1990xb,Kutasov:1990sv}, which we match to the leading non-planar result of the MMI. We observe that the partition function on $S^2$ dominates (in absolute value) over the partition function on $T^2$ only for a sufficiently large cosmological constant $\Lambda$, whereas for small $\Lambda>0$, the partition function on $T^2$ dominates. We do not yet have a clear understanding of this phenomenon but it would be interesting to explore its consequences for the Hartle-Hawking picture \cite{Hartle:1983ai,Chen:2020tes}. Some more open questions we present in section \ref{sec:discussion}.

\section{Multicritical matrix integrals}\label{sec:definitionsmatrix}\label{sec:MMI}
In \cite{Staudacher:1989fy, Gross:1989vs, Brezin:1989db} it has been conjectured that a certain class of matrix integrals --- known as multicritical matrix integrals \cite{Kazakov:1989bc} ---  in the large $N$ limit and upon tuning certain couplings are dual to two-dimensional quantum gravity coupled to $\mathcal{M}_{2m-1,2}$. 
We  explore this conjecture by drawing explicit connections between the $(m-1)$ primaries of $\mathcal{M}_{2m-1,2}$ and properties of the multicritical matrix integral. 

We consider the matrix integral 
\begin{equation}\label{eq:partfunction}
\mathcal{M}_N^{(m)}(\boldsymbol{\alpha})= \int_{\mathbb{R}^{N^2}}[\dd M]\, e^{-N \Tr V_m(M,\boldsymbol{\alpha})}~,\quad m\geq 2~,
\end{equation}
known as the $m^{\mathrm{th}}$ multicritical matrix integral, in the planar large $N$ limit.  $M$ is a Hermitian $N\times N$ matrix and the measure factor is given by 
\begin{equation}
[\dd M]\equiv \prod_J \dd M_{JJ} \prod_{I<J}\dd \mathrm{Re}M_{IJ}\prod_{I<J}\dd \mathrm{Im}M_{IJ}~.
\end{equation}
For $V_m(M,\boldsymbol{\alpha})$ we choose the even, order $2m$ polynomial
\begin{equation}\label{eq:defVm}
V_m(M,\boldsymbol{\alpha})= \sum_{n=1}^m\frac{1}{2n}\alpha_n M^{2n}~, \quad \alpha_1\equiv 1~,
\end{equation}
with $\boldsymbol{\alpha} \equiv ( \alpha_2, \ldots , \alpha_m) \in \mathbb{R}^{m-1}$. We will denote the set of numbers $\boldsymbol{\alpha}$ as the couplings of the polynomial (\ref{eq:defVm}).  We highlight that the number of free parameters $\boldsymbol{\alpha}$ is equal to the number of primaries of $\mathcal{M}_{2m-1,2}$.

Upon diagonalisation of $M$, we can analyse the planar contribution  of (\ref{eq:partfunction}) in the large $N$ limit using a saddle point approximation. This reduces the exponent of (\ref{eq:partfunction}) to
\begin{equation}\label{eq:actioneig}
S[\rho^{(m)}_{\mathrm{ext}}(\lambda,\boldsymbol{\alpha})] =\frac{1}{2}\int_{-a}^a\dd \lambda\rho^{(m)}_{\mathrm{ext}}(\lambda,\boldsymbol{\alpha})V_m(\lambda,\boldsymbol{\alpha})- 2\int_0^a\dd \lambda\rho^{(m)}_{\mathrm{ext}}(\lambda,\boldsymbol{\alpha})\log(\lambda)~,
\end{equation}
where we assumed that the eigenvalues $\lambda \in \mathrm{spec}(M)$ are distributed in the interval $[-a,a]$ and $\rho^{(m)}_{\mathrm{ext}}(\lambda,\boldsymbol{\alpha})$ is the eigenvalue density obtained as the solution of
\begin{equation}\label{eq:extremalisation}
V'_m(\lambda,\boldsymbol{\alpha})= 2 \int_{-a}^a \dd \mu\frac{\rho^{(m)}(\mu,\boldsymbol{\alpha})}{\lambda-\mu}~.
\end{equation}
The prime indicates a derivative with respect to $\lambda$. For more details we refer to \cite{DiFrancesco:2004qj, DiFrancesco:1993cyw, DioBeatrix}.
Another important quantity is the resolvent \cite{Brezin:1977sv}
\begin{equation}\label{eq:defres}
\RR_N(z)\equiv \frac{1}{N}\text{Tr} \left( z \, \mathbb{I}_N-M \right)^{-1} = \frac{1}{N}\sum_{I=1}^N\frac{1}{z-\lambda_I}~, \quad\quad z\in \mathbb{C}/\{\lambda_I\}~.
\end{equation}
Sending $N\rightarrow \infty$ the sum can be replaced by an integral, where each eigenvalue is weighted by its average density
\begin{equation} \label{RlargeN}
\lim_{N\rightarrow \infty}\RR_N(z)\equiv \RR(z)= \int_{-a}^{a}\dd \mu \,\frac{\rho(\mu)}{z-\mu}~.
\end{equation}
For a higher order even polynomial it is convenient to express the resolvent  as \cite{Polyakov:1980ca} 
\begin{equation}\label{eq:resolvent}
\RR(z)= \int_0^a\frac{\dd x}{\pi}\frac{x V'(x)}{z^2-x^2}\frac{\sqrt{z^2-a^2}}{\sqrt{a^2-x^2}}~.
\end{equation}
From the definition  of the resolvent (\ref{RlargeN}) one obtains its large $z$ scaling $\RR(z)\sim 1/z$ reducing (\ref{eq:resolvent}) to the condition\footnote{To evaluate (\ref{eq:BCgeneral}) the following integral identity is useful:
\begin{equation}\label{eq:integralid}
\int_0^y \frac{\dd x}{\pi}\frac{x^{2n}}{z^2-x^2}\frac{\sqrt{z^2-y^2}}{\sqrt{y^2-x^2}}= \frac{1}{2 B(n,1/2)}\int_0^{y^2}\dd A \frac{A^{n-1}}{\sqrt{z^2-A}}~.
\end{equation}}
\begin{equation}\label{eq:BCgeneral}
1= \int_0^a\frac{\dd x}{\pi}\frac{x V'(x)}{\sqrt{a^2-x^2}}~.
\end{equation}
For the polynomial $V_m(M,\boldsymbol{\alpha})$ (\ref{eq:defVm}), (\ref{eq:BCgeneral}) implies
\begin{equation}\label{eq:BC}
0= \mathcal{N}_m(\boldsymbol{\alpha})\equiv 1- \sum_{n=1}^m\frac{\alpha_n u^n}{2n B(n,1/2)}~, \quad\quad u\equiv a^2~.
\end{equation}
 $B(n,1/2)$ denotes the beta function. We will call the condition $\mathcal{N}_m(\boldsymbol{\alpha})=0$ the normalisation condition for our matrix integral. For a particular choice of $\boldsymbol{\alpha} $ we can turn the normalisation condition into 
\begin{equation}\label{eq:mth_order0}
(u-4m)^m=0~,
\end{equation}
in other words $u=4m$ is an $m^{\mathrm{th}}$ order zero. The values of $\boldsymbol{\alpha} $ leading to this behaviour can be easily obtained by recursively solving the discriminant of $\mathcal{N}_m(\boldsymbol{\alpha})=0$ \cite{Ambjorn:2016lkl, Kazakov:1989bc}:
\begin{equation}\label{eq:defalphac}
\alpha_{n,c}^{(m)} \equiv (-1)^{n+1} \binom{m}{n}\frac{2n}{(4m)^{n}} B(n,1/2)~, \quad 2\leq n\leq m~.
\end{equation}
Finally we define the expectation value of the loop operator $W_\ell$ \cite{Polyakov:1980ca}, which is related to the resolvent (\ref{RlargeN}) through a Laplace transform
\begin{equation}\label{eq:laplaceRW}
\langle W_\ell\rangle\equiv \int_{-a}^a \dd \lambda \rho_{\mathrm{ext}}(\lambda)\,e^{\lambda \ell}~, \quad \RR(z)= \int_0^{\infty}\dd \ell\, \langle W_\ell\rangle\, e^{-\ell z}~.
\end{equation}
We use (\ref{eq:integralid}) to obtain the large $z$ expansion of the resolvent for the multicritical matrix integrals (\ref{eq:defVm}):
\begin{equation}\label{eq:largezResolvent}
\RR_m(z,\boldsymbol{\alpha}  )= \sum_{k\geq 0}\frac{1}{4^k}\binom{2k}{k}z^{-2k-1}\sum_{n=1}^m \frac{\alpha_n u^{n+k}}{2(n+k)B(n,1/2)}~.
\end{equation}
Using (\ref{eq:laplaceRW}), we relate the large $z$ expansion of the latter to the small $\ell$ expansion of the loop operator.
For small $\ell$ we find
\begin{equation}\label{eq:loopopexp}
\langle W_\ell^{(m)}(\boldsymbol{\alpha} )\rangle = \sum_{n\geq 0}\frac{\ell^{2n}}{(2n)!4^n}\binom{2n}{n}\sum_{k=1}^m \frac{\alpha_k u^{n+k}}{2(n+k)B(k,1/2)}= \sum_{n\geq 0}\omega^{(m)}_n(\boldsymbol{\alpha}) \ell^{2n}~,
\end{equation}
where we defined 
\begin{equation}\label{eq:omegank}
\omega_{n}^{(m)}(\boldsymbol{\alpha})\equiv \frac{1}{(n!)^24^n}\sum_{k=1}^m \frac{\alpha_k u^{n+k}}{2(n+k)B(k,1/2)}~.
\end{equation}
As a final remark we note that evaluating (\ref{RlargeN}) close to the real axis $z= x\pm i\epsilon$ we obtain the important relations 
\begin{equation}\label{resbc}
\mathrm{res}_a:~~ \rho(x)= \frac{1}{2\pi i}\left(\RR(x-i\epsilon)- \RR(x+i\epsilon)\right)~,\quad
\mathrm{res}_b:~~ V'(x)=\RR(x-i\epsilon)+ \RR(x+i\epsilon)~.
\end{equation}
Combining $\mathrm{res}_a$ with the definition of the resolvent (\ref{RlargeN}) and the integral identity  (\ref{eq:integralid}) we obtain the extremal eigenvalue density for $V_m(\lambda,\boldsymbol{\alpha} )$ 
\begin{align}\label{eq:eigdistrV}
\rho^{(m)}_{\mathrm{ext}}(z,\boldsymbol{\alpha})&= \frac{1}{\pi}\sum_{n=1}^m \frac{\alpha_n}{B(n,1/2)}z^{2n-2}\,_2 F_1\left(\frac{1}{2},1-n; \frac{3}{2}; 1- \frac{u}{z^2}\right)\sqrt{u-z^2}~\cr
&=\frac{1}{\pi}\sum_{n=1}^m \frac{\alpha_n}{B(n,1/2)}z^{2n-2}\sum_{k=0}^\infty \frac{(-1)^k}{2k+1}\binom{n-1}{k}\left(1- \frac{u}{z^2}\right)^k\,\sqrt{u-z^2}~,
\end{align}
where we used that the hypergeometric $_2 F_1(a,b;c;z)$ can be written as a polynomial as soon as either $a$ or $b$ become non-positive integers.
 At the multicritical point $\boldsymbol{\alpha}_c\equiv (\alpha_{2,c}^{(m)}, \ldots , \alpha_{m,c}^{(m)})$ (\ref{eq:defalphac}) and close to the boundary of the eigenvalue interval where $u=4m$ (\ref{eq:mth_order0}) the density scales as $\rho^{(m)}_{\mathrm{ext}}(z,\boldsymbol{\alpha}_c)\propto (4m-z^2)^{(2m-1)/2}$ generalising the well-known exponent $3/2$ \cite{Brezin:1977sv} in the quartic $m=2$ model. 
\begin{figure}[H]
\begin{center}
\includegraphics[scale=.25]{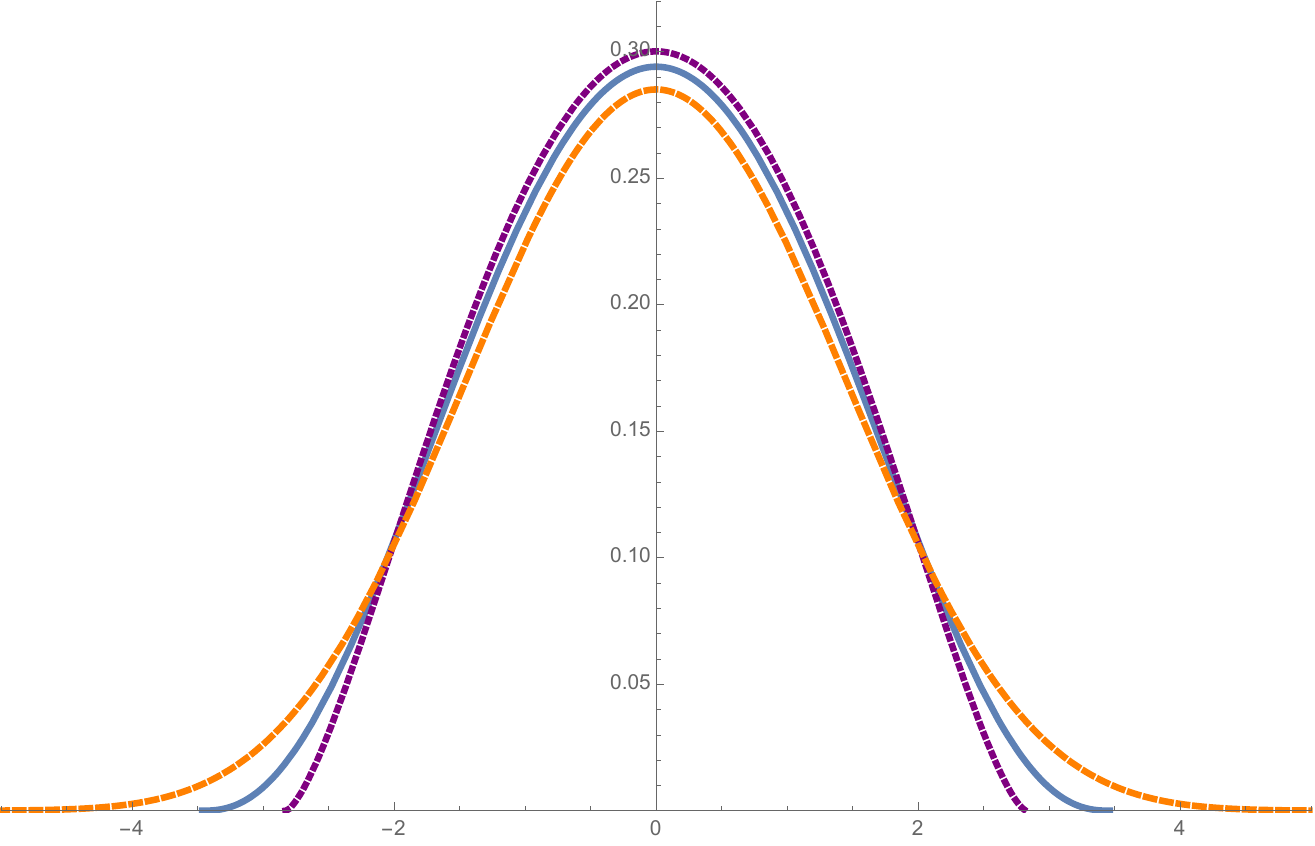}
\end{center}
\caption{Extremal eigenvalue density for $\boldsymbol{\alpha}= \boldsymbol{\alpha}_c$ and $m=2, 3 $ and $m=12$ ( purple dotted, teal, orange dashed). At the edges the eigenvalue distribution scales as $3/2, 5/2$ and $23/2$ respectively.}
\end{figure}
\noindent
Using (\ref{eq:loopopexp}) we obtain the planar on-shell action (\ref{eq:actioneig}) for arbitrary $m$:
\begin{equation}\label{eq:Srhoext}
S[\rho^{(m)}_{\mathrm{ext}}(\lambda,\boldsymbol{\alpha})] 
= \sum_{n=1}^m \frac{(2n)!}{4n}\,\alpha_n\,\omega_{n}^{(m)}+ \sum_{n=1}^m\frac{\alpha_n u^{n}}{4n^2 B(n,1/2)}-\frac{1}{2}\log u+ \log 2~,
\end{equation}
At the critical point we find 
\begin{equation}\label{eq:Scriticality}
S_c[\rho^{(m)}_{\mathrm{ext}}(\lambda,\boldsymbol{\alpha}_c)]= \frac{1}{2}H_{2m}- \frac{1}{2}\log 4m+\log 2~,
\end{equation}
where $H_n$ denotes the $n^{\text{th}}$ harmonic number. The subscript $c$ indicates that we zoom into criticality (\ref{eq:defalphac}). 
At large $m$  (\ref{eq:Scriticality}) scales as 
\begin{equation}
\lim_{m\rightarrow \infty} S_c[\rho^{(m)}_{\mathrm{ext}}(\lambda,\boldsymbol{\alpha}_c)]= \frac{1}{2}(\gamma+ \log 2)+ \frac{1}{8m}- \sum_{k=1}^\infty\frac{B_{2k}}{4k(2m)^{2k}}~,
\end{equation}
where $\gamma$ denotes the Euler-Mascheroni constant and $B_{k}$ the $k^{\text{th}}$ Bernoulli number.

\section{Planar diagrams  with a single vertex}\label{singlePD}

In this section we discuss the diagrammatic expansion of the matrix integral (\ref{eq:partfunction}).  We expand the normalisation condition (\ref{eq:BC}) and the planar on-shell action (\ref{eq:Srhoext}) for small couplings $\boldsymbol{\alpha}$. For the $m=2$ model with a single coupling this was first explored by \cite{Brezin:1977sv}. To account for the two indices of the matrix $M$ one uses the 't Hooft double line notation \cite{tHooft:1974pnl}. We will denote the resulting diagrams as ribbon diagrams. Whereas for the $m=2$ model one only encounters ribbon diagrams whose vertices emanate four edges, for the multicritical matrix integrals (\ref{eq:defVm}) we have to deal with vertices emanating an arbitrary even number of edges \cite{DiFrancesco:2004qj}.

\subsection{An $m=2$ refresher}\label{m2review}
Before delving into the multi-coupling perturbation theory we quickly review the $m=2$ case. For more details we refer to \cite{DioBeatrix}.
For $m=2$ we have the polynomial (\ref{eq:defVm})
\begin{equation}\label{eq:V2}
V_2(M,\alpha_2)= \frac{1}{2}M^2+ \frac{1}{4}\alpha_2 M^4~.
\end{equation}
Normalisation of the eigenvalue density implies the vanishing of $\mathcal{N}_2(\alpha_2)$ in (\ref{eq:BC})
with solutions $u_\pm^{(2)}$ given by
\begin{equation}\label{eq:u_form2a}
u_{\pm}^{(2)} =-\frac{2}{3\alpha_2}\pm \frac{2}{3\alpha_2}\sqrt{1+12\alpha_2} ~.
\end{equation}
Of the two solutions, only $u_+^{(2)}$ is well-behaved near $\alpha_2 = 0$. The other solution $u_-^{(2)}$ exhibits a pole at $\alpha_2 = 0$, and is ordinarily discarded. Nonetheless, it is worth noting that knowledge of the residue of the pole at $\alpha_2$ is sufficient to reconstruct $u_+^{(2)}$ from $u_-^{(2)}$. We also note that (\ref{eq:u_form2a}) exhibits a non-analytic behaviour close to $\alpha_2=-1/12$ which we recognise as the $m=2$ multicritical point (\ref{eq:defalphac}). For $\alpha_2<\alpha_{2,c}^{(2)}$ the normalisation condition has no real solution. 

We now discuss the $m=2$ model from a perturbative perspective in small $\alpha_2$. 
For $\alpha_2> \alpha_{2,c}^{(2)}$ we obtain the small $\alpha_2$ expansion of $u_+^{(2)}$
\begin{equation}\label{uplus2}
u_+^{(2)}= 4\sum_{k =0}^\infty(-1)^{k}\frac{\left(3{\alpha_2}\right)^{k}}{k+1}\binom{2k}{k}=4 \,_2F_1\left(\frac{1}{2}, 1;2;~\frac{\alpha_2}{\alpha_{2,c}^{(2)}} \right)~.
\end{equation}
Inspecting the above expression, one recognises the Catalan numbers 
\begin{equation}\label{eq:defCatalan}
\mathcal{C}_k^{(2)}\equiv \frac{1}{k+1}\binom{2 k}{k}~.
\end{equation}
Further to this, we see that the critical value $\alpha_{2,c}^{(2)}= -1/12$ controls the radius of convergence of the power series.
At large $k$, the summand in (\ref{uplus2}) goes as $\sim k^{-\boldsymbol{3/2}} (\alpha_2/\alpha_{2,c}^{(2)})^{k}$. This behaviour encodes the fact that there is a \textbf{square root} non-analyticity in the solution $u^{(2)}_+$ and as we shall soon see, it is intimately related to the growth of planar diagrams.

\begin{figure}[H]
\begin{center}
\begin{tikzpicture}[scale=.7]
\draw[middlearrow={<},line width=0.15mm] (.1,1) --(1,1)node[pos=3,scale=.6]{L};
\draw[line width=0.15mm]  (.1,.6) --(1,.6)node[pos=3,scale=.6]{K};
\draw[middlearrow={>},line width=0.15mm]  (1,.6) --(1,-.3)node[pos=1.3,scale=.6]{J};
\draw[middlearrow={<},line width=0.15mm]  (1.4,.6) --(1.4,-.3)node[pos=1.3,scale=.6]{K};
\draw[line width=0.15mm]  (1.4,.6) --(2.3,.6)node[pos=-2,scale=.6]{J};
\draw[middlearrow={<},line width=0.15mm]  (1.4,1) --(2.3,1)node[pos=-2,scale=.6]{I};
\draw[line width=0.15mm]  (1.4,1) --(1.4,1.9)node[pos=1.4,scale=.6]{L};
\draw[line width=0.15mm]  (1,1) --(1,1.9)node[pos=1.4,scale=.6]{I};
\draw[middlearrow={<},line width=0.15mm] (-3.5,1) --(-5,1)node[pos= -.2, scale=.6]{$I$}node[pos= 1.2,scale=.6]{$I$};
\node[scale=.8] at (4.2,.8)   {$\sim \frac{1}{4}\alpha_2\, N~.$};
\draw[middlearrow={>},line width=0.15mm]  (-3.5,.6) --(-5,.6)node[pos= -.2,scale=.6]{$J$}node[pos= 1.2,scale=.6]{$J$};
\node[scale=.8] at (-2, .8)   {$\sim  {N^{-1}}$ ~, };
\end{tikzpicture}
\end{center}
\caption{Propagator and quartic vertex.}
\label{fig:diagramsValpha2}
\end{figure}
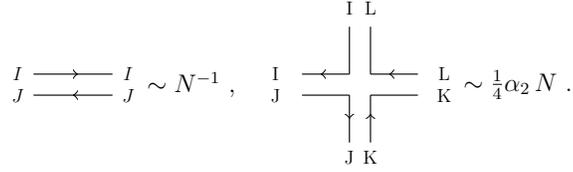

Defining $\mathcal{F}_2^{(0)}(\alpha_2)\equiv - \log \mathcal{M}_N^{(2)}(\alpha_2)/\mathcal{M}_N^{(2)}(0)$ we obtain for small $\alpha_2$ \cite{Brezin:1977sv}
\begin{equation}\label{eq:F0m2}
\mathcal{F}_2^{(0)}(\alpha_2)= -\sum_{k=1}^\infty (-1)^{k}(3\alpha_2)^{k}\frac{(2k-1)!}{k!(k+2)!}=\frac{1}{2}\alpha_2\,_3F_2\left(1,1,\frac{3}{2}; 2, 4;~\frac{\alpha_2}{\alpha^{(2)}_{2,c}} \right)~.
\end{equation} 
$\mathcal{F}_2^{(0)}(\alpha_2)$ is the generating function of the connected planar bubble diagrams generated by the matrix integral with a quartic interaction. From a small $\alpha_2$ expansion of 
\begin{equation}
\mathcal{M}_N^{(2)}(\alpha_2)= \int_{\mathbb{R}^{N^2}}[\dd M]\,e^{-\frac{N}{2}\Tr M^2}\sum_{k=0}^\infty\frac{(-1)^k}{k!}\left(\frac{\alpha_2N}{4}\right)^{k}\left(\Tr M^4\right)^{k}~,
\end{equation}
we can read off the propagator and quartic vertex (fig. \ref{fig:diagramsValpha2})
\noindent
Explicitly the propagator is given by
\begin{equation}\label{eq:Mprop}
\langle M_{IK}M_{JL}\rangle =\mathcal{M}^{-1}_N(0) \, \int_{\mathbb{R}^{N^2}} [\dd M] \, e^{-\frac{N}{2}\Tr \, M^2}M_{IJ}M_{KL} =\frac{1}{N}\delta_{IL}\delta_{KJ}~.
\end{equation}
$\mathcal{F}_2^{(0)}(4\alpha_2)$ counts  planar diagrams with four edges emanating from each vertex, with the shift $\alpha_2 \rightarrow 4 \alpha_2$ accounting for the $1/4$ weighting each vertex. We have
\begin{equation}
\mathcal{F}_2^{(0)}(4\alpha_2)= 2\alpha_2- 18\alpha_2^2+288\alpha_2^3+ \ldots~.
\end{equation}
The summand in (\ref{eq:F0m2}) scales as $\sim k^{-\bold{7/2}} (\alpha_2/\alpha_{2,c}^{(2)})^k$ at large $k$. This behaviour  encodes the growth of discrete Riemann surfaces with a fixed number of vertices $k$ \cite{Eguchi:1981kk}. 

\subsection{Binomial matrix integrals}
We discuss the polynomials \cite{DiFrancesco:2004qj,Caracciolo}
\begin{equation}\label{eq:tVk}
\tilde{V}_n(M,\alpha_n)\equiv \frac{1}{2}M^2+ \frac{1}{2n}{\alpha}_n M^{2n}~.
\end{equation}
 For $n=2$ the above polynomial is equal to the $m=2$ multicritical polynomial (\ref{eq:V2}) discussed in the last section. 
By setting all of the couplings but $\alpha_n$ to zero in (\ref{eq:eigdistrV}) we obtain the normalisation condition
\begin{equation}\label{eq:NC_tVk}
0=\tilde{\mathcal{N}}_n(\alpha_n)\equiv 1- \frac{1}{4}u- \frac{{\alpha}_n}{2nB(n,1/2)}u^n~.
\end{equation}
For $\alpha_n=\tilde{\alpha}_{n,c}$, where
\begin{equation}\label{eq:deftilde_ac}
\tilde{\alpha}_{n,c}\equiv -\frac{2n}{(4n)^n}(n-1)^{n-1}B(n,1/2)~, \quad 2\leq n\leq m~,
\end{equation}
(\ref{eq:NC_tVk}) has a second order zero at $u= 4n/(n-1)$, whereas for any other non-vanishing value of the coupling $\tilde{\alpha}_{n}$, (\ref{eq:NC_tVk}) has $n$ distinct solutions.
We further note that
\begin{equation}
|\tilde{\alpha}_{2,c}|<|\alpha_{2,c}^{(m)}|~, \quad |\tilde{\alpha}_{n,c}|>|\alpha_{n,c}^{(m)}|~,\quad m\geq 3~.
\end{equation}
For small $\alpha_n$ only one of the solutions of (\ref{eq:NC_tVk}) can be uniformly approximated by a perturbative expansion which is a power series in the coupling $\alpha_n$. 

To obtain the leading expression in the perturbative expansion (\ref{eq:NC_tVk}) we set ${\alpha}_n=0$, however 
this prevents us from obtaining the other $n-1$ solutions. Solutions which cannot be obtained in a perturbative expansion when setting the perturbation parameter to zero are discussed within the field of singular perturbation theory. 
\newline\newline
{\textbf{Singular perturbation theory.}} 
To recover the perturbative expansion of the $n-1$ solutions of  (\ref{eq:NC_tVk}) singular for ${\alpha}_n\rightarrow 0$ we start by rescaling $u\rightarrow {\alpha}_n^{-\nu}u$, $\nu\in \mathbb{R}_+$. For the case at hand  (\ref{eq:NC_tVk}) we obtain the rescaled equation
\begin{equation}\label{eq:ukstar_rescaled}
0={\alpha}_n^\nu- \frac{1}{4}u-\frac{1}{2nB(n,1/2)}{\alpha}_n^{1-\nu(n-1)}u^n~.
\end{equation}
For small ${\alpha}_n$ and $0<\nu<1/(n-1)$ or $\nu>1/(n-1)$ (\ref{eq:ukstar_rescaled}) we only obtain the trivial solution $u=0$. We are left with two special points $\nu\in \{0,1/(n-1)\}$, where we find a non-trivial solution for $u$. The perturbative solution for $\nu=0$ is the regular solution $u_\star^{(n)}$.{\footnote{Here and hereafter we introduce the subscript $\star$ to indicate the solution regular at the origin in coupling space. }}
A superscript indicates that $u$ solves (\ref{eq:NC_tVk}). 
For $\nu=1/(n-1)$ we obtain 
\begin{equation}\label{eq:rescaled_tildeu}
0=- \frac{1}{4}u-\frac{1}{2nB(n,1/2)}u^n~,
\end{equation}
solved by the $n-1$ roots of unity 
\begin{equation}
{u}^{(n)}_{1,\ldots , n-1}= e^{i\frac{\pi (2\ell+1)}{n-1}}\left(\frac{nB(n,1/2)}{2\alpha_n}\right)^{\frac{1}{n-1}}~,\quad  \quad \ell\in \{0,\ldots , n-2\}~.
\end{equation}
We then obtain the $n$ distinct perturbative expressions
\begin{align}
{u}^{(n)}_\star&= 4- 4 \binom{2n-1}{n-1}\alpha_n+ \mathcal{O}(\alpha_n^2)~, \cr
{u}^{(n)}_{1,\ldots , n-1}&= e^{i\frac{\pi (2\ell+1)}{n-1}}\left(\frac{nB(n,1/2)}{2\alpha_n}\right)^{\frac{1}{n-1}}- \frac{4}{n-1}+ \mathcal{O}\left(\alpha_n^{1/(n-1)}\right)~, \quad \ell\in \{0,\ldots , n-2\}~,
\end{align}
 approximating all $n$ solutions of (\ref{eq:NC_tVk}).

To discuss the perturbative analysis  of (\ref{eq:NC_tVk}) we take the regular solution ${u}^{(n)}_\star$. Its small $\alpha_n$ expansion reads
\begin{equation}\label{eq:u_gm_puregravity}
{u}^{(n)}_\star = 4\sum_{k=0}^\infty (-1)^{k}\mathcal{C}_k^{(n)}\,{\binom{2n-1}{n-1}^{k}}\alpha_n^{k}
=4\,_{n-1}F_{n-2}\left[\begin{matrix}\frac{1}{n}& \frac{2}{n} &  \ldots & \ldots & \frac{n-1}{n} \\&\frac{2}{n-1}& \ldots  & \frac{n-2}{n-1} & \frac{n}{n-1} \end{matrix};~\frac{\alpha_n}{\tilde{\alpha}_{n,c}} \right]~,\quad n\geq 3~.
\end{equation}
where we defined $\tilde{\alpha}_{n,c}$ in (\ref{eq:deftilde_ac}) and
\begin{equation}
\mathcal{C}_k^{(n)}\equiv \frac{1}{(n-1)k+1}\binom{nk}{k}~,
\end{equation}
are known as Pfaff-Fuss-Catalan numbers generalising the Catalan numbers (\ref{eq:defCatalan}) for $n=2$. 
For large $k$ the summand in (\ref{eq:u_gm_puregravity}) scales as 
\begin{equation}\label{eq:u_binomial}
\sim k^{-\bold{3/2}}e^{-k(n-1)\log(n-1)}\tilde{\alpha}_{n,c}^{-k}~.
\end{equation}
Note that the exponent $\boldsymbol{3/2}$ is universal for all binomial matrix integrals. 
Introducing $\mathcal{\tilde{F}}^{(0)}_n(\alpha_n)$  analogously to $\mathcal{F}^{(0)}_2(\alpha_2)$ and using (\ref{eq:u_gm_puregravity}) we obtain the perturbative $\alpha_n$ expansion 
\newline
\begin{multline}\label{eq:Fk_single_a}
\mathcal{\tilde{F}}^{(0)}_n(\alpha_n)= 
- \sum_{k=1}^{\infty}(-1)^{k}\frac{\left(\binom{2n-1}{n-1}\alpha_n\right)^{k}}{k!}\frac{(nk-1)!}{((n-1)k+2)!}\cr
= \frac{1}{2n}\,\mathcal{C}_n^{(2)}\,\alpha_n \,_{n+1}F_{n}\left[\begin{matrix} 1& 1 &\frac{n+1}{n}  &\ldots  & \ldots& \frac{2n-2}{n}& \frac{2n-1}{n} \\&2& \frac{n+2}{n-1} &\ldots & \ldots& \frac{2n-1}{n-1} & \frac{2n}{n-1} \end{matrix};~\frac{\alpha_n}{\tilde{\alpha}_{n,c}} \right]~.
\end{multline}
Since $\tilde{\alpha}_{2,c}= \alpha_{2,c}^{(2)}$ and using $\mathcal{C}_2^{(2)}= 2$, (\ref{eq:Fk_single_a}) reduces  to (\ref{eq:F0m2}) for $n=2$. 
For large $k$ the summand of this expression scales universally as
\begin{equation}\label{eq:7/2_binomial}
\sim k^{-\bold{7/2}}e^{-k(n-1)\log(n-1)}\tilde{\alpha}_{n,c}^{-k}~.
\end{equation}


\section{Planar diagrams with multiple vertices}\label{multiplePD}

In this section we consider the diagrammatic multi-vertex expansion of the multicritical matrix integrals (\ref{eq:defVm}). We discuss the $m=3$ and $m=4$ (\ref{eq:defVm}) cases in some detail since the normalisation condition (\ref{eq:BC}) for these matrix integrals is a cubic and a quartic polynomial whose roots admit explicit expressions. For $m\geq 5$ the normalisation condition is a quintic or  higher polynomial, and in general not solvable by radicals. The general $m$ case can be dealt with from a perturbative perspective by employing the Lagrange inversion theorem \cite{DiFrancesco:2004qj}. 

\subsection{$m=3$ analysis}
\begin{figure}[H]
\begin{center}
\begin{tikzpicture}[scale=1.2]
\draw (-9,-.08) to[out=-80,in=-100] (-7.6,-.08);
\draw (-9,-.08) to[out=80,in=100] (-7.6,-.08);
\draw(-9.095,-.07) ellipse (.1cm and .12cm);
\draw(-9.2,-.07) ellipse (.2cm and .22cm);
\draw(-7.4,-.07) ellipse (.2cm and .2cm);
\draw (-6,-.08) to[out=-80,in=-100] (-4.6,-.08);
\draw (-6,-.08) to[out=80,in=100] (-4.6,-.08);
\draw(-6.2,-.07) ellipse (.2cm and .12cm);
\draw(-5.795,-.07) ellipse (.2cm and .12cm);
\draw(-4.4,-.07) ellipse (.2cm and .2cm);
\draw (-3.05,-.08) to[out=-80,in=-100] (-1.7,-.08);
\draw (-3.05,-.08) to[out=-30,in=-150] (-1.7,-.08);
\draw (-3.05,-.08) to[out=30,in=150] (-1.7,-.08);
\draw (-3.05,-.08) to[out=80,in=100] (-1.7,-.08);
\draw (-3.25,-.08) ellipse (.2cm and .15cm);
\draw (-.5,-.08) to[out=-80,in=-100] (.6,-.08);
\draw (-.5,-.08) to[out=80,in=100] (.6,-.08);
\draw (.8,-.08) ellipse (.2cm and .15cm);
\draw[rotate=-39.8] (-.539,-.375) ellipse (.2cm and .08cm);
\draw[rotate=40] (-.648,.25) ellipse (.2cm and .08cm);
\node[scale=.7] at (1.3, -.2) {.};
\node[scale=.7] at (-1.4, -.2) {,};
\node[scale=.7] at (-3.9, -.2) {,};
\node[scale=.7] at (-6.9, -.2) {,};
\node[scale=.7] at (-8.2, -.8) {$48\alpha_2\alpha_3$};
\node[scale=.7] at (-5.2, -.8) {$24\alpha_2\alpha_3$};
\node[scale=.7] at (-2.4, -.8) {$24\alpha_2\alpha_3$};
\node[scale=.7] at (0.3, -.8) {$48\alpha_2\alpha_3$};
\end{tikzpicture}
\end{center}
\caption{Diagrams combining vertices with four and six edges emanating. Each line represents a thick double line. The number below each diagrams counts the number of diagrams including the symmetry factor.}
\label{fig:mixing_a2a3}
\end{figure}
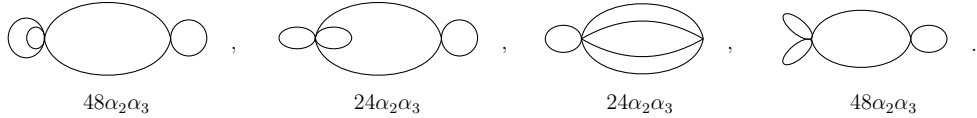

The normalisation condition $\mathcal{N}_3(\alpha_2,\alpha_3)=0$ is the cubic equation (\ref{eq:BC})
\begin{equation}\label{eq:N3a}
1-\frac{1}{4}u- \frac{3}{16}\alpha_2u^2- \frac{5}{32}\alpha_3u^3=0~,
\end{equation}
whose general solutions can be expressed as
\begin{equation}\label{u3solns}
u^{(3)}_\ell= \frac{32}{15 \alpha_3} \left(-\frac{3}{16} \alpha_2+{\Delta_0}\, { \zeta^{-1}_\ell \,\Gamma^{-1/3}}+{\zeta_\ell \, \Gamma^{1/3}} \right)~, \quad\quad \ell = 1,2,3~.
\end{equation}
Here, $\zeta_\ell= e^{2\pi i (\ell-1)/3}$ is a third root of unity and we have defined
\begin{eqnarray}
\Delta_0 &\equiv& \frac{3}{256} \left(3 \alpha_2^2-10\alpha_3\right)~, \\
\Delta_1 &\equiv& -\frac{27}{2048} \left(\alpha_2^3-5 \alpha_2 \alpha_3-50\alpha_3^2\right)~, \\
\Gamma &\equiv& \frac{1}{2}\left( \Delta_1+ \sqrt{\Delta_1^2 - 4 \Delta_0^3} \right)~.
\end{eqnarray}
We further define
\begin{equation}\label{eq:discriminant3}
D_3 \,\equiv \,\Delta_1^2 - 4 \Delta_0^3 = \frac{675}{4194304} \alpha_3^2 \left(-9 (12 \alpha_2+1) \alpha_2^2+20 (27 \alpha_2+2)\alpha_3+2700 \alpha_3^2\right)~.
\end{equation}
Expanding $D_3$ at small $\alpha_3$, we identify $\alpha_2 = -1/12$ as a special value, corresponding to $\alpha_{2,c}^{(2)}$. Expanding $D_3$ at small $\alpha_2$ reveals $\alpha_3 =- 2/135$ as a special value, corresponding to $\tilde{\alpha}_{3,c}$ (\ref{eq:deftilde_ac}). Near both $(\alpha_2,\alpha_3 ) = (-1/12,0)$ and $(\alpha_2,\alpha_3 ) = (0,-2/135)$, where $D_3=0$,  $\Delta_1$ remains non-vanishing such that the non-analytic behaviour of the solutions $u_\ell^{(3)}$ near these points is that of a \textbf{square root}. 
On the other hand, expanding $D_3$ near $\alpha_2=-1/9$ reveals  $\alpha_3 = 1/270$ as a special value, the multicritical point (\ref{eq:defalphac}). At $\alpha_2 = -1/9$ we have that $\Delta_1 = (135 \alpha_3-1) (270 \alpha_3-1)/55296$, which vanishes for $\alpha_{3,c}^{(3)}$. This implies that the non-analytic behaviour of $u_\ell^{(3)}$ near the multicritical point is that of a \textbf{cubic root}.
\newline\newline
{\textbf{Single-variable perturbation theory.}} We start by discussing the normalisation condition (\ref{eq:N3a}) along the path  $\gamma_\star^{(3)}$ in coupling space \cite{Ambjorn:2016lkl}
\begin{equation}\label{eq:gamma_star}
\gamma_\star^{(3)}: ~[0,1]\rightarrow \mathbb{R}^2~, \quad\quad ~t \mapsto \begin{pmatrix}  \alpha_{2,c}^{(3)}\, t\\  \alpha_{3,c}^{(3)}\, t^2\end{pmatrix}~,
\end{equation} 
leading to the rescaled normalisation condition 
\begin{equation}\label{eq:n3t}
0= 1- \frac{1}{4}u -\frac{3}{16}\alpha_{2,c}^{(3)}tu^2- \frac{5}{32}\alpha_{3,c}^{(3)}t^2u^3~.
\end{equation}
Of all paths, the path $\gamma_\star^{(3)}$ is special in that upon rescaling $u\rightarrow u/t$, $t^{-1}$ takes the role of an overall pre-factor for $V_3(M, \boldsymbol{\alpha})$ (\ref{eq:defVm}). 
The solution of (\ref{eq:n3t}) regular for $t\in [0,1]$ reads
\begin{equation}\label{eq:u3_t}
u_\star^{(3)}= \frac{12}{t}\left(1-(1-t)^{1/3}\right)=12\sum_{k=0}^\infty(-1)^k\binom{1/3}{k+1}t^{k}~.
\end{equation}
At large $k$ the summand scales as $\sim k^{-\boldsymbol{4/3}}$,  different from the $\sim k^{-\boldsymbol{3/2}}$ for $m=2$ (\ref{uplus2}). 
Note that (\ref{eq:u3_t}) converges for $|t|\leq 1$. 
$\,$\newline\newline
{\textbf{Two-variable perturbation theory.}} The solution of $\mathcal{N}_3(\alpha_2,\alpha_3)=0$ regular near the origin in coupling space is 
\begin{equation}\label{eq:u3}
u^{(3)}_\star=  4\sum_{k_1,k_2=0}^\infty \frac{(-1)^{k_1+k_2}}{(1+k_1+2k_2)}\frac{\prod_{s=1}^{k_2}(k_1+s)}{k_2!}\binom{3k_2+2k_1}{2k_2+k_1}(10\alpha_3)^{k_2}(3\alpha_2)^{k_1}~.
\end{equation}
Performing the substitution  $k= k_1+2k_2$, $n= k_1+k_2$ we find the single sum expression
\begin{equation}\label{eq:u3_a2a3}
u^{(3)}_\star=4\sqrt{\pi} \sum_{k=0}^\infty\left(\frac{20\alpha_3}{3\alpha_2}\right)^{k}\frac{1}{\Gamma(2+k)}\,_3\tilde{F}_2\left(1,-k,1+k;\frac{1}{2}-\frac{k}{2}, 1-\frac{k}{2}; \frac{9\alpha_2^2}{40\alpha_3}\right)~.
\end{equation}
Along the path $\gamma_\star^{(3)}$ 
we recover (\ref{eq:u3_t}),
as shown in appendix \ref{app:bestiary}. 
%
Equation (\ref{eq:u3_a2a3}) provides a perturbative expansion regular for small couplings $\alpha_2$ and $\alpha_3$. Depending on the range of the couplings (\ref{eq:u3_a2a3}) arises from a different solution (\ref{u3solns}) of the normalisation condition $\mathcal{N}_3(\alpha_2,\alpha_3)=0$. Switching for simplicity to polar coordinates $(\alpha_2,\alpha_3) =(r\cos\phi, r\sin\phi)$, 
we observe that the function
\begin{equation}\label{eq:B3_patchwork}
B_3(r,\phi)= u_1(r,\phi)\Theta(\pi-\phi)+ u_3(r,\phi)\Theta(\phi-\pi)~,
\end{equation}
is well behaved and real near the origin.
\newline\newline
{\textbf{On-shell action for $m=3$.}}
We define 
\begin{equation}\label{eq:def_F3}
\mathcal{F}^{(0)}_3(\alpha_2,\alpha_3)\equiv -\log \frac{\mathcal{M}_N^{(3)}(\alpha_2,\alpha_3)}{\mathcal{M}_N^{(3)}(0,0)}~.
\end{equation}
Using (\ref{eq:Srhoext}) for $m=3$  and the regular solution (\ref{eq:u3}) we obtain the small $\alpha_2, \alpha_3$ expansion 
 \begin{multline}\label{eq:F3_pert}
\mathcal{F}^{(0)}_3(\alpha_2,\alpha_3)=- \sum_{k_2=1}^{\infty}(-1)^{k_2}\frac{(10\alpha_3)^{k_2}}{k_2!}\frac{(3k_2-1)!}{(2k_2+2)!}\cr
- \sum_{k_2=0}^\infty\sum_{k_1=1}^{\infty}(-1)^{k_1+k_2} \frac{(10\alpha_3)^{k_2}}{k_2!} \frac{(3\alpha_2)^{k_1}}{k_1!}\frac{(3k_2+2k_1-1)!}{(2k_2+k_1+2)!}~.
 \end{multline}
It is convenient to perform again the substitution $k= k_1+2k_2$, $n= k_1+k_2$ leading to
  \begin{equation}\label{eq:F3_pert2}
\mathcal{F}^{(0)}_3(\alpha_2,\alpha_3)=
-\sqrt{\pi}\sum_{k=1}^\infty\left(\frac{20\alpha_3}{3\alpha_2}\right)^{k}\frac{1}{k\Gamma(3+k)}\,_3\tilde{F}_2\left(1,-k,k;\frac{1}{2}-\frac{k}{2}, 1-\frac{k}{2}; \frac{9\alpha_2^2}{40\alpha_3}\right)~.
 \end{equation} 
Evaluating (\ref{eq:F3_pert2}) at the multicritical point $\alpha_{2,c}^{(3)}= -1/9$, $\alpha_{3,c}^{(3)}= 1/270$ (\ref{eq:defalphac}) we recover the value of the on-shell action at criticality (\ref{eq:Scriticality}). Note that in the definition of $\mathcal{F}^{(0)}_3(\alpha_2,\alpha_3)$ in (\ref{eq:def_F3}) we subtract the Gaussian term which evaluates to $3/4$. In appendix \ref{app:bestiary} we show that the summand scales as $\sim k^{-\boldsymbol{10/3}}$, which differs from the $\sim k^{-\boldsymbol{7/2}}$ encountered in (\ref{eq:F0m2}). 
$\mathcal{F}^{(0)}_3(4\alpha_2,6\alpha_3)$ counts planar diagrams with four or six edges emanating from each vertex
\begin{equation}\label{eq:Fm3}
\mathcal{F}^{(0)}_3(4\alpha_2,6\alpha_3)=2\alpha_2+ 5\alpha_3-18\alpha_2^2- 300\alpha_3^2- 144 \alpha_2\alpha_3+\ldots~.
\end{equation}
To see some of these coefficients explicitly we expand the exponential in (\ref{eq:partfunction})
\begin{equation}\label{eq:pertM3}
\mathcal{M}_N^{(3)}(\alpha_2,\alpha_3)= \int_{\mathbb{R}^{N^2}}[\dd M]\,e^{-\frac{N}{2}\Tr M^2}\sum_{k_1,k_2=0}^\infty\frac{(-1)^{k_1+k_2}}{k_1!k_2!}\left(\frac{\alpha_2N}{4}\right)^{k_1}\left(\frac{\alpha_3N}{6}\right)^{k_2}\left(\Tr M^4\right)^{k_1}\left(\Tr M^6\right)^{k_2}~,
\end{equation}
leading to the graphical representation of the propagator, the quartic and the sextic vertex
\begin{figure}[H]
\begin{center}
\begin{tikzpicture}[scale=.6]
\draw[middlearrow={<},line width=0.15mm] (.1,1) --(1,1)node[pos=3,scale=.6]{L};
\draw[line width=0.15mm]  (.1,.6) --(1,.6)node[pos=3,scale=.6]{K};
\draw[middlearrow={>},line width=0.15mm]  (1,.6) --(1,-.3)node[pos=1.3,scale=.6]{J};
\draw[line width=0.15mm]  (1.4,.6) --(1.4,-.3)node[pos=1.3,scale=.6]{K};
\draw[middlearrow={>},line width=0.15mm]  (1.4,.6) --(2.3,.6)node[pos=-2,scale=.6]{J};
\draw[line width=0.15mm]  (1.4,1) --(2.3,1)node[pos=-2,scale=.6]{I};
\draw[middlearrow={>},line width=0.15mm]  (1.4,1) --(1.4,1.9)node[pos=1.4,scale=.6]{L};
\draw[line width=0.15mm]  (1,1) --(1,1.9)node[pos=1.4,scale=.6]{I};
\draw[middlearrow={<},line width=0.15mm] (-3.5,1) --(-5,1)node[pos= -.2, scale=.6]{$I$}node[pos= 1.2,scale=.6]{$I$};
\node[scale=.8] at (4.4,.8)   {$\sim \frac{1}{4}\alpha_2\, N~,$};
\draw[middlearrow={>},line width=0.15mm]  (-3.5,.6) --(-5,.6)node[pos= -.2,scale=.7]{$J$}node[pos= 1.2,scale=.7]{$J$};
\node[scale=.8] at (-2, .8)   {$\sim  {N^{-1}}$ ~, };
\draw[line width=0.15mm]  (7.4,1.4) --(7.4,2.2)node[pos=1.4,scale=.6]{I};
\draw[line width=0.15mm]  (7.8,1.4) --(7.8,2.2)node[pos=1.4,scale=.6]{J};
\draw[middlearrow={<},line width=0.15mm]  (7.8,.4) --(7.8,-.6)node[pos=1.4,scale=.6]{S};
\draw[line width=0.15mm]  (7.4,.4) --(7.4,-.6)node[pos=1.4,scale=.6]{K};
\draw[middlearrow={<},line width=0.15mm]  (7.8,1.4) --(8.8,2.2)node[pos=1.4,scale=.6]{J};
\draw[line width=0.15mm]  (7.8,.9) --(8.8,1.7)node[pos=1.4,scale=.6]{L};
\draw[middlearrow={<},line width=0.15mm]  (7.8,.9) --(8.8,.1)node[pos=1.4,scale=.6]{L};
\draw[line width=0.15mm]  (7.8,.4) --(8.8,-.4)node[pos=1.4,scale=.6]{S};
\draw[middlearrow={<},line width=0.15mm]  (7.4,.4) --(6.4,-.4)node[pos=1.4,scale=.6]{K};
\draw[middlearrow={>},line width=0.15mm]  (7.4,1.4) --(6.4,2.2)node[pos=1.4,scale=.6]{I};
\draw[middlearrow={<},line width=0.15mm]  (7.4,.9) --(6.4,1.7)node[pos=1.4,scale=.6]{R};
\draw[line width=0.15mm]  (7.4,.9) --(6.4,.1)node[pos=1.4,scale=.6]{R};
\node[scale=.8] at (10.8,.8)   {$\sim \frac{1}{6}\alpha_3\, N~.$};
\end{tikzpicture}
\end{center}
\caption{Propagator, quartic  and sextic vertex. }
\label{fig:diagramsValpha}
\end{figure}
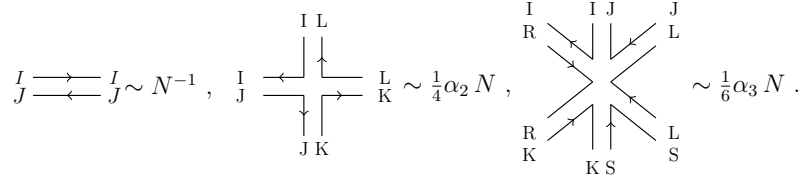
\noindent
In fig. \ref{fig:mixing_a2a3} we show the diagrams contributing to $\mathcal{O}(\alpha_2\alpha_3)$ in (\ref{eq:Fm3})
\subsection{$m=4$ analysis}
\begin{figure}[H]
\begin{center}
\begin{tikzpicture}[scale=.6]
\draw[line width=0.15mm]  (7.4,1.4) --(7.4,2.4)node[pos=1.4,scale=.6]{I};
\draw[middlearrow={>},line width=0.15mm]  (7.8,1.4) --(7.8,2.4)node[pos=1.4,scale=.6]{J};
\draw[line width=0.15mm]  (7.8,.4) --(7.8,-.8)node[pos=1.4,scale=.6]{S};
\draw[middlearrow={>},line width=0.15mm]  (7.4,.4) --(7.4,-.8)node[pos=1.4,scale=.6]{K};
\draw[line width=0.15mm]  (7.8,1.4) --(8.8,2.2)node[pos=1.4,scale=.6]{J};
\draw[middlearrow={>},line width=0.15mm]  (8,1.1) --(8.8,1.7)node[pos=1.4,scale=.6]{L};
\draw[line width=0.15mm]  (8,1.1) --(9,1.1)node[pos=1.4,scale=.6]{L};
\draw[middlearrow={>},line width=0.15mm]  (8,.7) --(9,.7)node[pos=1.4,scale=.6]{P};
\draw[line width=0.15mm]  (8,.7) --(8.8,.1)node[pos=1.4,scale=.6]{P};
\draw[middlearrow={>},line width=0.15mm]  (7.8,.4) --(8.8,-.4)node[pos=1.4,scale=.6]{S};
\draw[line width=0.15mm]  (7.4,.4) --(6.4,-.4)node[pos=1.4,scale=.6]{K};
\draw[middlearrow={>},line width=0.15mm]  (7.4,1.4) --(6.4,2.2)node[pos=1.4,scale=.6]{I};
\draw[line width=0.15mm]  (7.2,1.1) --(6.4,1.7)node[pos=1.4,scale=.6]{Q};
\draw[middlearrow={>},line width=0.15mm]  (7.2,.7) --(6.4,.1)node[pos=1.4,scale=.6]{R};
\draw[line width=0.15mm]  (7.2,.7) --(6.2,.7)node[pos=1.4,scale=.6]{R};
\draw[middlearrow={>},line width=0.15mm]  (7.2,1.1) --(6.2,1.1)node[pos=1.4,scale=.6]{Q};
\node[scale=.8] at (11.4,.8)   {$\sim \frac{1}{8}\alpha_4\, N~.$};
\end{tikzpicture}
\end{center}
\caption{Eight-order vertex.}
\label{fig:a4}
\end{figure}
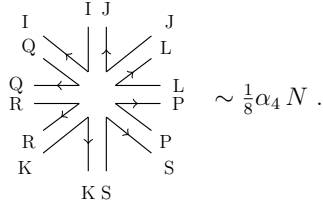

The normalisation condition $\mathcal{N}_4(\boldsymbol{\alpha})=0$ is the quartic equation (\ref{eq:BC})
\begin{equation}\label{eq:NCm4}
0=1- \frac{1}{4}u -\frac{3}{16}\alpha_2 u^2- \frac{5}{32}\alpha_3 u^3- \frac{35}{256}\alpha_4 u^4~,
\end{equation}
admitting four solutions $u^{(4)}_{\ell,\pm}$, $\ell = 1,2$ whose properties we discuss in appendix \ref{app:m4}.
\newline\newline
{\textbf{Single-variable perturbation theory.}}
Along the path $\gamma_\star^{(4)}$ 
\begin{equation}\label{eq:gamma_star4}
\gamma_\star^{(4)}: ~[0,1]\rightarrow \mathbb{R}^3~, \quad\quad ~t \mapsto \begin{pmatrix}  \alpha_{2,c}^{(4)}\, t\\  \alpha_{3,c}^{(4)}\, t^2 \\ \alpha_{4,c}^{(4)}t^3\end{pmatrix}~,
\end{equation} 
the regular solution of (\ref{eq:NCm4}) reads
\begin{equation}\label{eq:u4_t}
u_\star^{(4)}= \frac{16}{t}\left(1-(1-t)^{1/4}\right)=16\sum_{k=0}^\infty(-1)^k\binom{1/4}{k+1}t^{k}~,
\end{equation}
converging for $|t|\leq 1$.
At large $k$ the summand scales as $\sim k^{-\boldsymbol{5/4}}$ as compared to $\sim k^{-\boldsymbol{4/3}}$ for $m=3$ (\ref{eq:u3_t}) and $\sim k^{-\boldsymbol{3/2}}$ for $m=2$ (\ref{uplus2}) and the binomial matrix integrals (\ref{eq:u_binomial}). 
\newline\newline
{\textbf{Three-variable perturbation theory.}}
Solving (\ref{eq:NCm4}) and expanding the regular solution for $\alpha_2, \alpha_3$ and $\alpha_4$ close to zero we obtain 
\begin{multline}\label{eq:u4_star}
u^{(4)}_\star=4\sum_{k_1,k_2,k_3=0}^\infty \frac{(-1)^{k_1+k_2+k_3}}{(1+k_1+2k_2+3k_3)}(35\alpha_4)^{k_3}(10\alpha_3)^{k_2}(3\alpha_2)^{k_1}\cr
\times \frac{\prod_{s=1}^{k_2+k_3}(k_1+s)}{(k_2+k_3)!}\frac{\prod_{s=1}^{k_3}(k_2+s)}{k_3!}\binom{4k_3+3k_2+2k_1}{3k_3+2k_2+k_1}~.
\end{multline}
Performing the substitution $k=k_1+ 2k_2+3k_3$, $n=k_1+k_2+k_3$, $l=k_2+k_3$
\allowdisplaybreaks
\begin{align}\label{eq:u4star}
u^{(4)}_\star&=4\sqrt{\pi}\sum_{k=0}^\infty\sum_{n=0}^{k}(-1)^{n}\left(\frac{7\alpha_4}{\alpha_3}\right)^{k}\left(\frac{3\alpha_2\alpha_3}{7\alpha_4}\right)^{n}\frac{\Gamma(1+k+n)}{\Gamma(2+k)\Gamma(1+n)\Gamma(1+k-n)}\cr
&\times\, _3\tilde{F}_2\left(1,-n,-k+n; \frac{1}{2}+\frac{1}{2}(-k+n), 1+ \frac{1}{2}(-k+n); \frac{5\alpha_3^2}{21\alpha_2\alpha_4}\right)~.
\end{align}
We conjecture
\begin{align}\label{eq:u4_star_t}
&\sum_{n=0}^{k}\frac{(-3)^{n}}{8^{k}}\frac{\Gamma(1+k+n)}{\Gamma(2+k)\Gamma(1+n)\Gamma(1+k-n)}\cr
&\times\, _3\tilde{F}_2\left(1,-n,-k+n; \frac{1}{2}+\frac{1}{2}(-k+n), 1+ \frac{1}{2}(-k+n); \frac{4}{6}\right)=\frac{4}{\sqrt{\pi}}\binom{1/4}{k+1}~.
\end{align}
%
According to this conjecture (\ref{eq:u4star}) reduces to (\ref{eq:u4_t}) along $\gamma_\star^{(4)}$.
Equation (\ref{eq:u4star}) provides a perturbative expansion regular for small couplings $\boldsymbol{\alpha}$. Depending on the range of the couplings it arises from a different solution (\ref{eq:solutions_m4}) of the normalisation condition $\mathcal{N}_4(\boldsymbol{\alpha})=0$, as we elucidate further in appendix \ref{app:m4}. 
\newline\newline
\textbf{On-shell action for $m=4$.}
We define 
\begin{equation}\label{eq:def_F4}
\mathcal{F}^{(0)}_4(\alpha_2,\alpha_3,\alpha_4)\equiv -\log \frac{\mathcal{M}_N^{(4)}(\alpha_2,\alpha_3,\alpha_4)}{\mathcal{M}_N^{(4)}(0,0,0)}~.
\end{equation}
Using (\ref{eq:Srhoext}) for $m=4$ and (\ref{eq:u4star}) we obtain the small $\alpha_2, \alpha_3,\alpha_4$ expansion 
 \begin{multline}\label{eq:F4}
\mathcal{F}^{(0)}_4(\alpha_2,\alpha_3,\alpha_4)=- \sum_{k_3=1}^{\infty}(-1)^{k_3}\frac{(35\alpha_4)^{k_3}}{k_3!}\frac{(4k_3-1)!}{(3k_3+2)!}\cr
- \sum_{k_3=0}^\infty\sum_{k_2=1}^{\infty}(-1)^{k_2+k_3} \frac{(35\alpha_4)^{k_3}}{k_3!}\frac{(10\alpha_3)^{k_2}}{k_2!} \frac{(4k_3+3k_2-1)!}{(3k_3+2k_2+2)!}~\cr
- \sum_{k_3,k_2=0}^\infty\sum_{k_1=1}^{\infty}(-1)^{k_1+k_2+k_3} \frac{(35\alpha_4)^{k_3}}{k_3!}\frac{(10\alpha_3)^{k_2}}{k_2!}\frac{(3\alpha_2)^{k_1}}{k_1!} \frac{(4k_3+3k_2+2k_1-1)!}{(3k_3+2k_2+k_1+2)!}~.
 \end{multline}
 Along $\gamma^{(4)}_\star$ and using (\ref{eq:u4_t}) we obtain 
 \begin{equation}
\lim_{t\rightarrow 1-\epsilon} \mathcal{F}_{4, \mathrm{n.a.}}^{(0)}(\boldsymbol{\alpha})|_{\gamma^{(4)}_\star} \sim \epsilon^{\boldsymbol{9/4}}~.
 \end{equation}
 $\mathcal{F}^{(0)}_4(4\alpha_2,6\alpha_3,8\alpha_4)$ counts planar diagrams with four, six or eight edges emanating from each vertex
\begin{multline}
\mathcal{F}^{(0)}_4(4\alpha_2,6\alpha_3, 8\alpha_4)=2\alpha_2+ 5\alpha_3+14\alpha_4-18\alpha_2^2- 300\alpha_3^2-4900\alpha_4^2\cr
- 144 \alpha_2\alpha_3-560 \alpha_2\alpha_4
-2400\alpha_3\alpha_4+ 201600\alpha_2\alpha_3\alpha_4+\ldots~.
\end{multline}
We obtain a graphical representation of the eight-order vertex (fig.\ref{fig:a4}) following the steps in (\ref{eq:pertM3}).

\subsection{$m\geq 5$ analysis}
For general $m\geq 5$ we cannot solve the normalisation condition (\ref{eq:BC}) by radicals. Instead we conjecture generalisations of (\ref{eq:u3}) and (\ref{eq:u4_star}) for the normalisation condition and the expressions (\ref{eq:F3_pert}) and (\ref{eq:F4}) for the on-shell action relying on numerical results. \newline\newline
{\textbf{Single-variable perturbation theory.}}
We start by parametrising a path connecting the origin in coupling space to the multicritical point 
\begin{equation}\label{eq:gamma_starm}
\gamma_\star^{(m)}: ~[0,1]\rightarrow \mathbb{R}^{m-1}~, \quad t \mapsto \begin{pmatrix}  \alpha_{2,c}^{(m)}\, t\\ \vdots \\  \alpha_{m,c}^{(m)}\, t^{m-1}\end{pmatrix}~.
\end{equation} 
This leads to the regular solution of $\mathcal{N}_m(\boldsymbol{\alpha})=0$
\begin{equation}\label{eq:u_gamma_sm}
u_\star^{(m)}= \frac{4m}{t}\left(1-(1-t)^{1/m}\right)=4m \sum_{k=0}^\infty (-1)^k \binom{1/m}{k+1}t^{k}~,
\end{equation}
convergent for $|t|\leq 1$.
For large $k$ the summand scales as $\sim\Gamma(1+1/m)k^{-\boldsymbol{1-\frac{1}{m}}}$.\newline\newline
{\textbf{Multi-variable perturbation theory.}} For general small couplings $\boldsymbol{\alpha}$ the perturbative expansion of the regular solution reads
\begin{multline}\label{eq:small_alpha_gu}
u^{(m)}_\star= 4 \sum_{k_1,\ldots, k_{m-1}=0}^\infty  \frac{(-1)^{k_1+\ldots+ k_{m-1}}}{(1+k_1+ \ldots + (m-1)k_{m-1})}\prod_{\ell=2}^m\left[\binom{2\ell-1}{\ell-1}\alpha_\ell\right]^{k_{\ell-1}}\binom{2k_1+ \ldots + mk_{m-1}}{k_1+ \ldots + (m-1)k_{m-1}}\cr
\times \frac{\prod_{s=1}^{k_2+\ldots + k_{m-1}}(k_1+s)}{(k_2+\ldots + k_{m-1})!}\frac{\prod_{s=1}^{k_3+\ldots + k_{m-1}}(k_2+s)}{(k_3+\ldots + k_{m-1})!} \cdots \frac{\prod_{s=1}^{k_{m-1}}(k_{m-2}+s)}{(k_{m-1})!}~.
\end{multline}
The above expression accounts for mixing of the couplings. Note that if we set all but one of the couplings (e.g. $\alpha_\ell$) to zero (\ref{eq:small_alpha_gu}) reduces to (\ref{eq:u_gm_puregravity}). We conjecture that this reduces to (\ref{eq:u_gamma_sm}) along the path $\gamma_\star^{(m)}$. Similarly to the case $m=3$ and $m=4$ we believe that also for $m\geq 5$ there exists a smooth function which leads to (\ref{eq:small_alpha_gu}) when approached from different directions in coupling space. 

\begin{figure}[H]
\begin{center}
 \begin{tikzpicture}[scale=.35]
   \begin{polaraxis}[grid=none, axis lines=none, line width=1.2 ]
     \addplot[mark=none,domain=0:360,samples=300,brown] {  abs(cos(2*x/2))};
   \end{polaraxis}
\node at (3.45, 1) {$2\alpha_2$};
 \end{tikzpicture}
 \qquad
 \begin{tikzpicture}[scale=.35]
   \begin{polaraxis}[grid=none, axis lines=none, line width=1.2 ]
     \addplot[mark=none,domain=0:360,samples=300,purple] { abs(cos(3*x/2))};
   \end{polaraxis}
   \node at (5.7, 1.25) {$5\alpha_3$};
 \end{tikzpicture}
 \qquad
 \begin{tikzpicture}[scale=.35]
   \begin{polaraxis}[grid=none, axis lines=none, line width=1.2 ]
     \addplot[mark=none,domain=0:360,samples=300,orange] { abs(cos(4*x/2))};
   \end{polaraxis}
 \node at (6., 1.25) {$14\alpha_4$};
 \end{tikzpicture}
 \qquad
 \begin{tikzpicture}[scale=.35]
   \begin{polaraxis}[grid=none, axis lines=none, line width=1.2 ]
     \addplot[mark=none,domain=0:360,samples=300, magenta] { abs(cos(5*x/2))};
   \end{polaraxis}
 \node at (6.5, 1.4) {$42\alpha_5$};
 \end{tikzpicture}
 \qquad
  \begin{tikzpicture}[scale=.35]
   \begin{polaraxis}[grid=none, axis lines=none, line width=1.2 ]
     \addplot[mark=none,domain=0:360,samples=300,teal] { abs(cos(6*x/2))};
   \end{polaraxis}
   \node at (7.2, 1.65) {$132\alpha_6$}; 
 \end{tikzpicture}
 \end{center}
 \caption{Vertices with four to twelve edges emanating. Each line represents a thick line. The number below each diagram enumerates the number of such diagrams as counted by $\mathcal{F}_m^{(0)}(4\alpha_2, \ldots , 2m\alpha_m)$ for $m \geq 6$.}
 \label{fig:Catalan_diagrams}
 \end{figure}
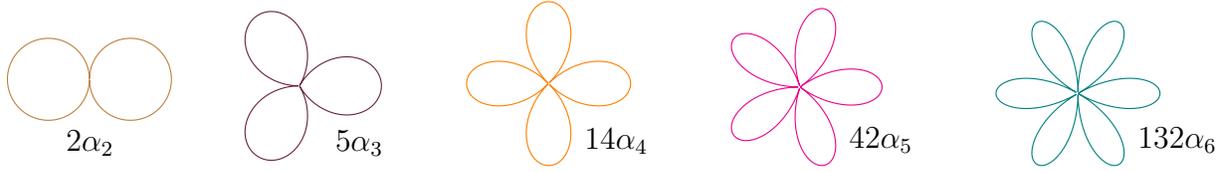

We note that we can also express (\ref{eq:small_alpha_gu}) in terms of incomplete exponential Bell polynomials \cite{Bell}. 
Using the Lagrange inversion theorem (for an introduction see e.g. \cite{enumerative_combinatorics}) to solve the normalisation condition perturbatively (\ref{eq:BC}) we obtain 
\begin{equation}\label{eq:u_intermsof_BP}
u^{(m)}_\star=\sum_{k=1}^\infty\frac{4^k}{k!}\sum_{\ell=0}^{k-1}(-1)^\ell k^{(\ell)}\mathcal{Y}_{k-1,\ell}\left(\hat{f}_1,\ldots , \hat{f}_{k-\ell}\right)~,\quad \hat{f}_k\equiv 4 k! \frac{\alpha_{k+1}}{2(k+1)B(k+1,1/2)}~,
\end{equation}
where $\hat{f}_k=0$ for $k\geq m$;  $ k^{(\ell)}\equiv k(k+1)\cdots (k+\ell-1)$ denotes the rising factorial and $\mathcal{Y}_{n,k}(x_1,\ldots ,x_{n-k+1})$ are the incomplete exponential Bell polynomials, defined recursively through
\begin{equation}\label{eq:def_BP}
\mathcal{Y}_{n,k}(x_1,\ldots ,x_{n-k+1})\equiv \sum\frac{n!}{j_1!j_2!\ldots j_{n-k+1}!}\left(\frac{x_1}{1!}\right)^{j_1}\left(\frac{x_2}{2!}\right)^{j_2}\cdots \left(\frac{x_{n-k+1}}{(n-k+1)!}\right)^{j_{n-k+1}}~.
\end{equation}
The summation is over all sequences $j_1, j_2, \ldots j_{n-k+1}$ of non-negative integers subject to the conditions
\begin{equation}
j_1+j_2+ \ldots +j_{n-k+1}=k~, \quad j_1+2j_2+ 3j_3+\ldots +(n-k+1)j_{n-k+1}= n~.
\end{equation}
The Bell polynomial encodes information on the partitions of a set. $\mathcal{Y}_{n,k}(x_1,\ldots ,x_{n-k+1})$ tells us how many partitions with block size between $1$ and $(n-k+1)$ a set with $n$ elements can have when divided into $k$ blocks. As an example 
\begin{equation}
\mathcal{Y}_{4,2}(x_1,x_2,x_3)= 3x_2^2+4 x_1 x_3~,
\end{equation}
reflects that the set $\boldsymbol{y}\equiv \{y_1,y_2,y_3,y_4\}$ can be divided into blocks of size 2 in two different ways. We can have 3 mutually, non-overlapping subsets, each consisting of a block of size two. Additionally we have 4 different ways to break $\boldsymbol{y}$ into a size 1 block and a size 3 block. 
\newline\newline
\textbf{On-shell action for general $m$.}
We define
\begin{equation}\label{eq:Fm0_general}
\mathcal{F}_m^{(0)}(\boldsymbol{\alpha})\equiv- \log\frac{\mathcal{M}_N^{(m)}(\boldsymbol{\alpha})}{\mathcal{M}_N^{(m)}(\boldsymbol{0})}~.
\end{equation}
Using (\ref{eq:Srhoext}) and (\ref{eq:small_alpha_gu}) we obtain the perturbative expansion
\begin{align}\label{eq:generalFm}
\mathcal{F}^{(0)}_m(\boldsymbol{\alpha})=
&- \sum_{k_{m-1}=1}^{\infty}(-1)^{k_{m-1}}\frac{\left(\binom{2m-1}{m-1}\alpha_m\right)^{k_{m-1}}}{k_{m-1}!}\frac{(mk_{m-1}-1)!}{((m-1)k_{m-1}+2)!}+\cr
&- \sum_{k_{m-1}=0}^{\infty}\sum_{k_{m-2}=1}^\infty(-1)^{k_{m-1}+k_{m-2}}\prod_{j=m-1}^m\frac{\left(\binom{2j-1}{j-1}\alpha_j\right)^{k_{j-1}}}{k_{j-1}!}\frac{(\sum_{j=m-2}^{m-1}(j+1)k_j-1)!}{(\sum_{j=m-2}^{m-1}jk_j+2)!}+\ldots \cr
&- \sum_{k_2,\ldots, k_{m-1}=0}^{\infty}\sum_{k_1=1}^\infty(-1)^{\sum_{j=1}^{m-1} k_{j}}\prod_{j=2}^m\frac{\left(\binom{2j-1}{j-1}\alpha_j\right)^{k_{j-1}}}{k_{j-1}!}\frac{(\sum_{j=1}^{m-1}(j+1)k_j-1)!}{(\sum_{j=1}^{m-1}jk_j+2)!}~.
\end{align}
Assuming that all but one of the couplings is very small the above reduces to (\ref{eq:Fk_single_a}).  Along $\gamma^{(m)}_\star$ (\ref{eq:gamma_starm}) and using (\ref{eq:u_gamma_sm}) we obtain 
 \begin{equation}\label{eq:naF_diagrammatics}
\lim_{t\rightarrow 1-\epsilon} \mathcal{F}_{m, \mathrm{n.a.}}^{(0)}(\boldsymbol{\alpha})|_{\gamma^{(m)}_\star} \sim \epsilon^{\boldsymbol{2+\frac{1}{m}}}~.
 \end{equation}
 Upon shifting $\alpha_\ell\rightarrow 2\ell \alpha_\ell$, $\ell = 2,\ldots , m$,  $\mathcal{F}^{(0)}_m(\boldsymbol{\alpha})$ (\ref{eq:Fm0_general}) counts planar diagrams with vertices emanating $4,\ldots, 2m$ edges. As an example we infer that $\mathcal{F}^{(0)}_{10}(\boldsymbol{\alpha})$ contains the term
 \begin{equation}
46549055536250157437879915371089100800000000\alpha_2\alpha_3\alpha_4\alpha_5\alpha_6\alpha_7\alpha_8\alpha_9\alpha_{10}~,
 \end{equation}
which counts diagrams with nine distinct vertices emanating an even number between four and twenty edges. After this shift the terms linear in the couplings have coefficients given by the Catalan numbers $\mathcal{C}_k^{(2)}$ with $k=2,\ldots ,m$.
In fig. \ref{fig:Catalan_diagrams} we illustrate diagrams with a single vertex emanating four to twelve edges.

Finally we remark that combining (\ref{eq:u_intermsof_BP}) with the on-shell action (\ref{eq:Srhoext}) one can obtain an expression of $\mathcal{F}^{(0)}_m(\boldsymbol{\alpha})$ in terms of the incomplete exponential Bell polynomials (\ref{eq:def_BP}). Our expression (\ref{eq:generalFm}) unpacks this result.

\section{Non-analytic behaviour of multicritical matrix integrals}\label{nonanalyticsec}

In this section, we uncover the non-analytic behaviour of the planar on-shell action (\ref{eq:Srhoext}) as a function of its couplings near the multicritical point (\ref{eq:defalphac}).
\newline\newline
\textbf{An $m=2$ refresher.}
For $m=2$ the polynomial  (\ref{eq:defVm}) reduces to the quartic polynomial
\begin{equation}\label{eq:V2M}
V_2(M,\alpha_2)= \frac{1}{2}M^2+ \frac{1}{4}\alpha_2 M^4~.
\end{equation}
From (\ref{eq:resolvent}) it is straightforward to obtain the resolvent and using res$_a$ (\ref{resbc}) we find the eigenvalue distribution (\ref{eq:extremalisation}): 
\begin{align}\label{eq:R_rho_2}
\RR_2(z,\alpha_2)&= \frac{1}{2}V_2'(z,\alpha_2)- \frac{1}{4}(2+ \alpha_2 u + 2\alpha_2 z^2)\sqrt{z^2-u}~, \cr
\rho_{\mathrm{ext}}^{(2)}(z,\alpha_2)&= \frac{1}{4\pi}(2+ \alpha_2 u + 2\alpha_2 z^2)\sqrt{u-z^2}~,\quad u=a^2~.
\end{align}
Combining the above leads to the planar on-shell action (\ref{eq:Srhoext})
\begin{equation}
S[\rho_{\mathrm{ext}}^{(2)}(\lambda,\alpha_2)]=  \sum_{n=1}^2 \frac{(2n)!}{4n}\,\alpha_n\,\omega_{n}^{(2)}+ \sum_{n=1}^2\frac{\alpha_n u^{n}}{4n^2 B(n,1/2)}-\frac{1}{2}\log u+ \log 2~,
\end{equation}
with $\omega_n^{(2)}$ defined in (\ref{eq:omegank}).  Using $u_+^{(2)}$ (\ref{eq:u_form2a}) this implies 
\begin{equation}\label{eq:m2_5/2}
\mathcal{F}^{(0)}_{2,\mathrm{n.a.}}(\alpha_2^\epsilon) = \frac{7}{24}- \frac{1}{2}\log 2+\epsilon-18\epsilon^2+  \frac{384}{5}\sqrt{3}\,\epsilon^{\boldsymbol{5/2}}+ \mathcal{O}(\epsilon^3)~,
\end{equation}
where the subscript n.a. indicates the leading non-analyticity.
The leading non-analytic behaviour, encoded in the critical exponent $\textbf{5/2}$, characterises the particular universality class associated to the $m=2$ (\ref{eq:V2M}) and binomial matrix integrals (\ref{eq:tVk}) and is intimately related to the exponent $\boldsymbol{7/2}$ we observed in (\ref{eq:F0m2}) at large $k$. 

%

\subsection{Critical exponents for $m=3$}
For $m=3$ we have 
\begin{equation}\label{eq:V3M}
V_3(M,\alpha_2,\alpha_3)= \frac{1}{2}M^2+\frac{1}{4}\alpha_2 M^4+ \frac{1}{6}\alpha_3M^6~.
\end{equation}
The resolvent and the eigenvalue density are given by
\begin{multline}
\RR_3(z,\alpha_2,\alpha_3)=\frac{1}{2}V_3'(z,\alpha_2,\alpha_3)-\frac{1}{16} \left(8 + 4 u\alpha_2+3 u^2 \alpha_3 + z^2(8\alpha_2+ 4 u\alpha_3)+ 8\alpha_3 z^4\right)\sqrt{z^2-u}~, \cr
\rho_{\mathrm{ext}}^{(3)}(z,\alpha_2,\alpha_3)=\frac{1}{16\pi} \left(8+ 4 u\alpha_2 + 3 u^2 \alpha_3 + z^2(8\alpha_2+4 u\alpha_3)+ 8\alpha_3 z^4\right)\sqrt{u-z^2}~. 
\end{multline}
For the $m=3$ multicritical matrix integral, in contradistinction to the $m=2$ model (\ref{eq:V2M}), we obtain two different non-analyticities --
one along a fine-tuned path, another one along a generic path in coupling space. \newline\newline
{\textbf{Fine-tuned path.}} 
Normalising the above eigenvalue density we obtain the normalisation condition $\mathcal{N}_3(\alpha_2,\alpha_3)=0$ (\ref{eq:N3a}) whose solutions we discussed in (\ref{u3solns}). Recall that the discriminant (\ref{eq:discriminant3})
\begin{equation}
D_3=\frac{675}{4194304} \alpha_3^2 \left(-9 (12 \alpha_2+1) \alpha_2^2+20 (27 \alpha_2+2)\alpha_3+2700 \alpha_3^2\right)~,
\end{equation}
vanishes at the multicritical point. More generally $D_3$ vanishes for
\begin{equation}\label{eq:alpha3expansion}
\alpha_{3,\pm} = \alpha_{3,c}^{(3)}- \frac{1}{10}\epsilon\pm \frac{1}{5}\epsilon^{3/2}~,\quad \alpha_2^{\epsilon}\equiv \alpha_{2,c}^{(3)}+ \epsilon~.
\end{equation}
and we restrict $\epsilon>0$.
Solving $\mathcal{N}_3(\alpha_{2}^{\epsilon}, \alpha_{3,\pm})=0$ we find $u=12+\tilde{x}\,\epsilon^{1/2}+ \mathcal{O}(\epsilon^2)$, 
with the three solutions $\tilde{x}_{1,2}=\pm 36$ and $\tilde{x}_3=72$. WLOG we focus on $\tilde{x}=\pm 36$.
Expanding the on-shell action leads to a leading non-analyticity of the form
\begin{equation}\label{eq:S3exact3/2}
\mathcal{F}^{(0)}_{3,\mathrm{n.a.}}(\alpha_{2}^\epsilon,\alpha_{3,\pm}) = \frac{19}{40}-\frac{1}{2}\log3+ \frac{9}{10}\epsilon\pm \frac{9}{10}\epsilon^{\boldsymbol{3/2}}-\frac{243}{40}\epsilon^2+ \mathcal{O}(\epsilon^{5/2})~.
\end{equation}
The constant term is the action at criticality (\ref{eq:Scriticality}) for $m=3$ with the Gaussian piece, which is equal to $3/4$, subtracted. The critical exponent is given by $\textbf{3/2}$, which differs from the $\textbf{5/2}$ critical exponent of the $m=2$ model (\ref{eq:m2_5/2}). Note that had we not kept the solution $\alpha_{3,\pm}$ to order $\mathcal{O}(\epsilon^{3/2})$ we would have not obtained the correct leading non-analytic behaviour in (\ref{eq:S3exact3/2}). \newline\newline
{\textbf{Generic path.}}
Further to this, one can uncover another critical exponent by zooming into criticality while adding a linear deformation to one of the couplings. This leads to the ansatz
\begin{equation}\label{eq:expansionStrS}
(\alpha_{2},\alpha_3)= (\alpha_{2,c}^{(3)}, \alpha_3^\epsilon)~, \quad u=12-36\cdot 10^{1/3} \epsilon^{1/3}+108\cdot 10^{2/3} \epsilon^{2/3}-3240 \epsilon+ \mathcal{O}(\epsilon^2)~.
\end{equation}
The solution $u$ is expanded up to order $\mathcal{O}(\epsilon)$ to avoid the appearance of spurious non-analyticities.
Expanding the action around (\ref{eq:expansionStrS}) we find
\begin{equation}\label{eq:S3exact7/3}
\mathcal{F}^{(0)}_{3,\mathrm{n.a.}}(\alpha_{2,c}^{(3)},\alpha_{3}^\epsilon)=  \frac{19}{40}-\frac{1}{2}\log 3+\frac{9}{2}\epsilon-206550\epsilon^2+\frac{11208375}{7}10^{1/3}\epsilon^{\boldsymbol{7/3}}+\mathcal{O}(\epsilon^2)~.
\end{equation}
The critical exponent is given by $\textbf{7/3}$, which again differs from the $\textbf{5/2}$ critical exponent of the $m=2$ model. 

We believe that there are no other critical exponents near the multicritical point for $m=3$ in the leading order planar expansion. 




\subsection{Critical exponents for general $m$}
Since for $m\geq 4$ the normalisation condition is a higher order polynomial, we now outline a perturbative approach for the fine-tuned path. \newline\newline
{\textbf{Fine-tuned path.}} 
We would like to deform the couplings near the multicritical point $\boldsymbol{\alpha}_c $ (\ref{eq:defalphac}) in the following manner
\begin{equation}\label{eq:ansatz_CE}
\boldsymbol{\alpha}^\epsilon  =\boldsymbol{\alpha} _c+ \boldsymbol{s}\,\epsilon~, \quad u= 4m+x~,
\end{equation}
where $x$ and $\epsilon$ are small parameters, and $\boldsymbol{s} \equiv (s_2,\ldots,s_m) \in \mathbb{R}^{m-1}$.
Expanding the normalisation condition (\ref{eq:BC}) we find
\begin{align}\label{eq:boundaryc}
\mathcal{N}_m(\boldsymbol{\alpha})
&=   \left( {-}\frac{x}{4m}\right)^m -\epsilon\sum_{n=2}^m\frac{(4m)^ns_{n}}{2nB(n,1/2)}\sum_{\ell=0}^n\binom{n}{\ell} \left(\frac{x}{4m} \right)^\ell+ \mathcal{O}(\epsilon^2)~.
\end{align}
For those $\boldsymbol{s}$ satisfying
\begin{equation}\label{eq:hypersurfacek}
\bigcup_{j=1}^{r'} \mathcal{H}_{m}^{(j)}=0~,\quad r'= 1,\ldots , m-2~,
\end{equation}
where $\mathcal{H}_{m}^{(j)}$ (hypersurfaces) are defined as
\begin{align}\label{eq:defhypersurfaces}
\mathcal{H}_{m}^{(j)}&
\equiv m^{-j-1}\sum_{n=j+1}^m \frac{(4m)^n}{2n B(n,1/2)}\binom{n-2}{j-1} s_n=0~,
\end{align}
the coefficients of $x^{\ell}$ in (\ref{eq:boundaryc}) with $0 \leq \ell\leq r'-1$ vanish. 
The proof of this can be found in appendix \ref{app:na}. Consequently on (\ref{eq:hypersurfacek}), the solutions of $\mathcal{N}_m(\boldsymbol{\alpha})=0$ in $x$ will scale as $\epsilon^{1/(m-r')}$. To indicate that the deformations $\boldsymbol{s}$ are living on (\ref{eq:hypersurfacek})
we introduce an additionial superscript to the normalisation condition $\mathcal{N}_m^{(r')}(\boldsymbol{\alpha})$ and $\mathcal{F}^{(r')}_{m,\mathrm{n.a.}}(\boldsymbol{\alpha})$, where $r'\in \{1,\ldots , m-2\}$.
\begin{figure}[H]
\begin{center}
\begin{tikzpicture}[scale=.9]
\draw[dashed,->] (-1,0) -- (6.5,0);
\draw[dashed,->] (0,-1) -- (0,5);
\draw[dashed,->] (.8,.8) -- (-1.6,-1.6);
\node[scale=.8] at (2.72,3.16)   {{\tikz\penguin[umbrella=magenta, scale=0.28,rotate=310];}};
\node[scale=.8] at (-.3,1.56)   {{\tikz\penguin[ scale=0.28];}};
\node[scale=.8] at (4.2,2.73)   {{\tikz\penguin[hat=teal, scale=0.28,rotate=20];}};
\node[scale=.8] at (6.85,0)   {$\boldsymbol{\alpha} $~};
\node[scale=.8] at (-1.8,-1.8)   {$\boldsymbol{\alpha} $};
\node[scale=.8] at (0,5.2)   {$\boldsymbol{\alpha} $};
\node[scale=.8] at (3.2,2.2)   {$\boldsymbol{\alpha} _c$};
\draw (-3,1) to[out=50,in=180] (3,4);
\draw (1,-1) to[out=50,in=180] (6,1);
\draw (-3,1) to[out=50,in=180] (1,-1);
\draw (3,4) to[out=-5,in=100] (6,1);
\draw[magenta] (1,3.77) to[out=-5,in=100] (3.35,.65);
\draw[orange] (-2,2.05) to[out=-50,in=115] (5.25,.97);
\draw[teal] (.4,-.86) to[out=70,in=205] (5.15,2.9);
\draw [fill] (3.05,1.92) circle [radius=0.05];
\end{tikzpicture}
\end{center}
\caption{We zoom into the critical point $\boldsymbol{\alpha} _c$ and then start moving into special directions to recover the different non-analyticities. }
\end{figure}
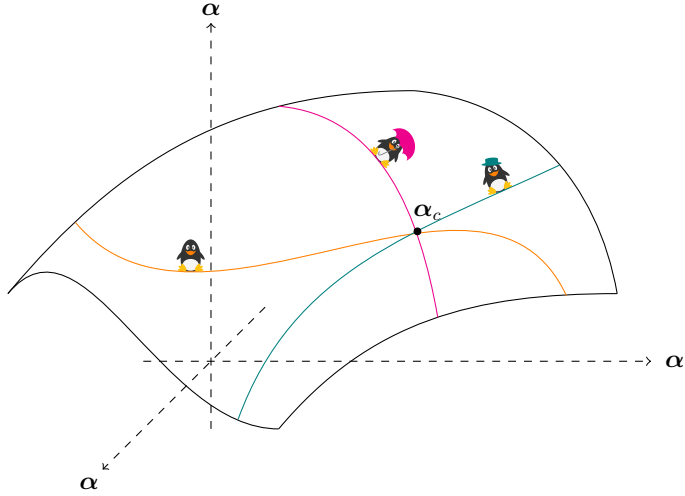
\noindent
We are thus led to the following ansatz
\begin{equation}\label{eq:movingaway}
\boldsymbol{\alpha}^\epsilon \equiv \boldsymbol{\alpha} _c+ \boldsymbol{s}\,\epsilon~, \quad u = 4m+ \tilde{x}\, \epsilon^{\frac{1}{m-r'}}~,\quad \tilde{x}\in\mathbb{R}~,
\end{equation}
giving rise to the normalisation condition
\begin{align}\label{eq:bc}
\mathcal{N}_m^{(r')}(\boldsymbol{\alpha}^\epsilon)
   =&\left[(-1)^m\frac{\tilde{x}^m}{(4m)^m}- \frac{\tilde{x}^{r'}}{4^{r'}}m^2\,\mathcal{H}_{m}^{(r'+1)}\right]\,\epsilon^{\frac{m}{m-r'}}+ \mathcal{O}\left(\epsilon^{\frac{m+1}{m-r'}}\right)~.
\end{align}
We refer to appendix \ref{app:na} for a proof of (\ref{eq:bc}). Recalling the discussion near (\ref{eq:S3exact3/2}) for $m=3$, to ensure that we obtain the correct non-analytic behaviour for $\mathcal{F}^{(0)}_{m,\mathrm{n.a.}}(\boldsymbol{\alpha})$, we must expand one of the couplings to subleading order. For $p\in \{2,\ldots ,m\}$ arbitrary we take
\begin{equation}\label{eq:finetuningBC}
\alpha^\epsilon_{n\neq p}= \alpha_{c,n\neq p}^{(m)}+  s_{n\neq p}\,\epsilon~, \quad \alpha_p^\beta = \alpha_{c,p}^{(m)}+ s_p\, \epsilon+\tilde{s}\,\epsilon^\beta~,\quad u = 4m+ \tilde{x}\,\epsilon^{\frac{1}{m-r'}}~.
\end{equation}
For $\beta< m/(m-r')$ the last term gives additional contributions spoiling the $\epsilon^{m/(m-r')}$ non-analyticity. For $\beta > m/(m-r')$ the leading non-analyticity is unaffected.  For $\beta= m/(m-r')$ we are led to 
 \begin{align}\label{eq:finalBC}
 \mathcal{N}_m^{(r')}(\alpha_{n\neq p}^\epsilon, \alpha_p^\beta)
  =&\left[(-1)^m\frac{\tilde{x}^m}{(4m)^m}- \frac{\tilde{x}^{r'}}{4^{r'}}m^2\,\mathcal{H}_{m}^{(r'+1)}- \frac{(4m)^p}{2p B(p,1/2)}\tilde{s}\right]\,\epsilon^{{{\frac{m}{m-r'}}}}+ \mathcal{O}\left(\epsilon^{\frac{m+1}{m-r'}}\right)~.
 \end{align}
It can be checked that adding subleading corrections to more than one of the $\alpha_n$ does not lead to additional non-analyticities.

We now determine the non-analytic behaviour of $\mathcal{F}^{(0)}_{m,\mathrm{n.a.}}(\boldsymbol{\alpha})$ once we move away from the critical point (\ref{eq:movingaway}). 
Using $\beta=m/(m-r')$ in (\ref{eq:finetuningBC})
we are led to
\begin{multline}\label{eq:Skm1}
S^{(r')}_m[\rho^{(m)}_{\mathrm{ext}}(\lambda,\boldsymbol{\alpha}^\epsilon)] = \mathcal{S}_m^{(r')}[\boldsymbol{s},\epsilon,\epsilon^2]
 -\frac{1}{2}H_m\left[(-1)^m\frac{\tilde{x}^m}{(4m)^m}- \frac{\tilde{x}^{r'}}{4^{r'}}m^2\,\mathcal{H}_{m}^{(r'+1)}-\frac{(4m)^p}{2p B(p,1/2)}\tilde{s}\right]\,\epsilon^{\frac{m}{m-r'}}  \\ +\frac{(4m)^p}{2p^2 B(p,1/2)}\frac{m!p!}{(m+p)!}\,\tilde{s}\,\epsilon^{\frac{m}{m-r'}}+ \mathcal{O}\left(\epsilon^{\frac{m+1}{m-r'}}\right)~,
\end{multline}
where $\mathcal{S}_m^{(r')}[\boldsymbol{s},\epsilon,\epsilon^2]$ is an $\tilde{x}$-independent expression whose explicit form we present in appendix \ref{app:action}.
Taking $\tilde{x}= \tilde{x}_*$ to be a solution of the normalisation condition $\mathcal{N}_m^{(r')}(\alpha_{n\neq p}^\epsilon, \alpha_p^\beta)=0$ at order $\epsilon^{m/(m-r')}$ we finally obtain for $r'=1,2,\ldots , m-2$
\begin{equation}\label{eq:Skm}
\mathcal{F}^{(0)}_{m,\mathrm{n.a.}}(\alpha_{n\neq p}^\epsilon, \alpha_p^\beta)-\mathcal{S}_m^{(r')}[\boldsymbol{s},\epsilon,\epsilon^2]+\frac{3}{4}
=\frac{(4m)^p}{2p^2 B(p,1/2)}\frac{m!p!}{(m+p)!}\,\tilde{s}\,\epsilon^{{\boldsymbol{\frac{m}{m-r'}}}}+ \mathcal{O}\left(\epsilon^{\frac{m+1}{m-r'}}\right)~, \quad 2\leq p \leq m ~.
\end{equation}
In summary for the $m^{\mathrm{th}}$ multicritical matrix integral (\ref{eq:defVm}) we have so far obtained $(m-2)$ distinct critical exponents given by $\boldsymbol{m/(m-r')}$, $r'=1, 2, \ldots ,m-2$.
\newline\newline
{\textbf{Example $m=3$.}} Let us take 
\begin{equation}\label{genericm3}
(\alpha_2,\alpha_3)= (\alpha_2^\epsilon,\alpha_3^\epsilon)~, \quad u=12 +x~, \quad\quad 0<\epsilon\ll 1~.
\end{equation}
The parameter $x$ is itself small and fixed in terms of $s_2$, $s_3$, and $\epsilon$ through $\mathcal{N}_3(\boldsymbol{\alpha}) = 0$. For generic values of $s_2$ and $s_3$, expanding the normalisation condition $\mathcal{N}_3(\boldsymbol{\alpha})=0$ for small $\epsilon$ leads to the following three solutions
\begin{equation}\label{eq:solutionxm3}
x= 36 \, z \, \epsilon^{1/3}  (s_2+10 s_3)^{1/3}~,  \quad\quad z^3 = -1~.
\end{equation}
We thus recover the non-analytic behaviour observed in (\ref{eq:expansionStrS}). 

For the finely tuned combination $\mathcal{H}_3^{(1)}= s_2+10s_3=0$ (\ref{eq:defhypersurfaces}), we find the leading order behaviour
\begin{equation}
x = \pm 36   \sqrt{ 3s_2 \epsilon }~,
\end{equation}
recovering the non-analytic behaviour observed in (\ref{eq:finalBC}). To obtain the subleading $\epsilon^{3/2}$ dependence in (\ref{genericm3}), we must add a subleading piece $\tilde{s}\epsilon^\beta$, $\beta >1$ to one of the couplings. WLOG we take $\alpha_3$. Expanding the normalisation condition we infer that only $\beta={3/2}$ is consistent with the perturbative expansion. Our ansatz becomes (\ref{eq:finetuningBC})
\begin{equation}\label{eq:ansatzpert3/2}
\alpha_2^\epsilon= \alpha_{2,c}^{(3)}+s_2\epsilon~, \quad \alpha_3^\beta=\alpha_{3,c}^{(3)} -\frac{1}{10}s_2\epsilon+\tilde{s} \epsilon^{3/2}~, \quad u=12 +\tilde{x}\,\epsilon^{1/2}~.
\end{equation}
Adding further subleading terms will not change the leading non-analytic behaviour. Setting $s_2=1$ and $\tilde{s}=\pm 1/5$ we recover (\ref{eq:alpha3expansion}). Expanding $\mathcal{N}_3^{(1)}(\alpha_2^\epsilon,\alpha_3^\beta)$ for small $\epsilon$, we find (\ref{eq:finalBC})
\begin{equation}\label{eq:solutionxstarm3}
\mathcal{N}_3^{(1)}(\alpha_2^\epsilon,\alpha_3^\beta)=\left(- \frac{1}{1728}\tilde{x}^3+ \frac{9}{4}s_2 \tilde{x} - 270\tilde{s}\right)\epsilon^{3/2}+ \mathcal{O}(\epsilon^2)~.
\end{equation}
Evaluating the action in a perturbative expansion along (\ref{eq:ansatzpert3/2}), we obtain (\ref{eq:Skm})
\begin{equation}\label{Sna3}
\mathcal{F}^{(0)}_{3,\mathrm{n.a.}}(\alpha_2^\epsilon,\alpha_3^\beta)=  \frac{19}{40}-\frac{1}{2}\log 3+\frac{9}{10}s_2\epsilon-\frac{1}{2}H_3 \left(- \frac{1}{1728}\tilde{x}^3+ \frac{9}{4}s_2 \tilde{x} - 270\tilde{s}\right)\epsilon^{\boldsymbol{3/2}}+\frac{9}{2}\tilde{s}\,\epsilon^{\boldsymbol{3/2}}+ \mathcal{O}(\epsilon^2)~,
\end{equation}
recovering (\ref{eq:S3exact3/2}) for $s_2=1$ and $\tilde{s}=\pm 1/5$. From the above expression we infer that for vanishing $\tilde{s}$ the coefficient multiplying the leading non-analyticity vanishes, since its solution in $\tilde{x}=\tilde{x}_\star$ agrees with the solution of the leading order term in (\ref{eq:solutionxstarm3}).
The non-analytic behaviour in (\ref{Sna3}) is robust against other deformations of the ansatz (\ref{eq:ansatzpert3/2}).
\newline\newline
\textbf{Generic path for general $m$.}
In addition to the ($m-2$) critical exponents in (\ref{eq:Skm}) the matrix integrals (\ref{eq:partfunction}) exhibit one more multicritical exponent. 

To obtain (\ref{eq:m2_5/2}) for $m=2$ we observed the reaction of the action when allowing $\alpha_2$ to slightly deviate from its critical value $\alpha_{2,c}^{(2)}$ (\ref{eq:defalphac}). We can generalise this for the polynomials $V_m(M,\boldsymbol{\alpha})$ for $m\geq 3$. Here we observe the reaction of the action when allowing one arbitrary coupling to deviate away from the multicritical point. In other words we consider 
 \begin{equation}\label{eq:ansatz_Gamma_str}
\alpha_2= \alpha_{2,c}^{(m)},~ \ldots , ~\alpha_{m-1}= \alpha_{m-1,c}^{(m)}~, ~\alpha_m=\alpha_{m,c}^{(m)}+ \epsilon~,\quad 0<\epsilon \ll 1~,
\end{equation}
where WLOG $\alpha_m$ deviates away from its critical value $\alpha_{m,c}^{(m)}$.  The leading reaction of the normalisation condition away from its critical value $4m$ is of the form $\epsilon^{1/m}$ easily extracted from (\ref{eq:boundaryc}).\footnote{Comparing (\ref{eq:ansatz_CE}) to our ansatz (\ref{eq:ansatz_Gamma_str}) we set $s_n=0$, $n\leq m-1$ and $s_m=1$.} To avoid spurious critical exponents in $\mathcal{F}_{m,\mathrm{n.a.}}^{(0)}(\boldsymbol{\alpha})$, we need to expand the solution to the normalisation condition to order $\mathcal{O}(\epsilon)$. 
In addition to (\ref{eq:ansatz_Gamma_str}) we take the following ansatz\footnote{We would obtain the right critical exponent already if we stopped at linear order in $\epsilon$. However the coefficient is affected by terms up to order $\mathcal{O}(\epsilon^2)$}  for $u$
\begin{equation}\label{eq:ansatzStringS}
 u =4m +\tilde{A}\epsilon^{1/m}+\sum_{\ell=2}^{2m} \tilde{C}_\ell \,\epsilon^{\ell/m}+ \mathcal{O}\left(\epsilon^{{(2m+1)}/{m}}\right)~.
\end{equation}
The coefficients $\{\tilde{A}, \tilde{C}_\ell\}\in \mathbb{C}$ we obtain by comparing coefficients of equal powers of $\epsilon$ in the normalisation condition.
The coefficients $\tilde{A}$ are given by
\begin{equation}
\tilde{A}^{(n)}=e^{i\pi n/m}\left(\frac{(4m)^{2m}}{2m B(m,1/2)}\right)^{1/m}~, \quad n = 1,\ldots, m~.
\end{equation}
Upon choosing one $\tilde{A}^{(n)}$ the $\tilde{C}_{\ell}$ for $2\leq \ell\leq 2m-1$ are fixed uniquely. WLOG we choose $\tilde{A}^{(m)}$ and find
\begin{equation}
u= 4m\sum_{\ell=0}^{2m}(-1)^{\ell} m^{\ell} \binom{2m-1}{m-1}^{\ell/m}\epsilon^{\ell/m}+ \mathcal{O}\left(\epsilon^{(2m+1)/m}\right)~.
\end{equation}
The leading non-analyticity is thus given by 
\begin{equation}\label{eq:Gstr_general}
\mathcal{F}^{(0)}_{m,\mathrm{n.a.}}(\alpha_{2,c}^{(m)},\ldots, \alpha_{m-1,c}^{(m)},\alpha_m^\epsilon)  =S_c[\rho^{(m)}_{\mathrm{ext}}(\lambda,\boldsymbol{\alpha}_c)]-\frac{3}{4}+\beta_1\,\epsilon+ \beta_2\,\epsilon^2+ \beta_3\,\epsilon^{\bold{2+\frac{1}{m}}}+\mathcal{O}\left(\epsilon^{2+\frac{2}{m}}\right)~,
\end{equation}
with $\beta_3$ given by 
\begin{equation}\label{eq:beta3}
\beta_3= \frac{m^{2m+3}}{2(2m+1)}\binom{2m-1}{m-1}^{1+1/m}\mathcal{C}^{(2)}_{m}~.
\end{equation}
$\mathcal{C}^{(2)}_m$ are the Catalan numbers (\ref{eq:defCatalan}).
On the other side $\beta_1$ and $\beta_2$ are $\epsilon$-independent expressions in $m$.
From (\ref{eq:Gstr_general}) we infer the leading non-analyticity $\boldsymbol{2+1/m}$ which is intimately related to the critical exponent (\ref{eq:u_gamma_sm}) observed for large $k$ in the perturbative expansion. 
\newline\newline
{\textbf{Example $m=3$.}}  For the multicritical matrix integral with $m=3$ we make the ansatz
\begin{equation}\label{eq:ansatz_Gammastr_m3}
\alpha_2= \alpha_{2,c}^{(3)}~, \quad \alpha_3= \alpha_{3,c}^{(3)}+ \epsilon~, \quad u= 12+\tilde{A}\epsilon^{1/3}+\sum_{\ell=2}^{6} \tilde{C}_\ell \,\epsilon^{\ell/3}+ \mathcal{O}\left(\epsilon^{{7}/{3}}\right)~.
\end{equation}
Expanding the normalisation condition $\mathcal{N}_3(\alpha_{2,c},\alpha_3^\epsilon)=0$ (\ref{eq:BC}) we obtain to leading order 
\begin{equation}
\tilde{A}^{(n)}= 36\,e^{i\pi n/3}\times 10^{1/3}~, \quad n=1,2,3~.
\end{equation}
Choosing $\tilde{A}^{(3)}$ and repeating (\ref{eq:ansatz_Gammastr_m3}) after including subleading corrections in the normalisation condition we obtain the ansatz
\begin{equation}\label{eq:fullAnsatz_m3_Gammastr}
u= 12\sum_{\ell=0}^{6}(-1)^{\ell} 3^{\ell}(10)^{\ell/m}\epsilon^{\ell/3}+ \mathcal{O}\left(\epsilon^{7/3}\right)~.
\end{equation}
Expanding the action around (\ref{eq:fullAnsatz_m3_Gammastr}) we obtain 
\begin{align}
\mathcal{F}^{(0)}_{3,\mathrm{n.a.}}(\alpha_{2,c}^{(3)},\alpha_3^\epsilon) =\frac{19}{40}-\frac{1}{2}\log 3+\frac{9}{2}\epsilon-6075\epsilon^2+\frac{492075}{7}\times (10)^{1/3}\epsilon^{{\boldsymbol{7/3}}}+ \mathcal{O}\left(\epsilon^{8/3}\right)~.
\end{align}
We observe the leading critical exponent $\boldsymbol{7/3}$.
\newline\newline
\textbf{General remarks.} It is worth noting that only $\epsilon$ and $\epsilon^2$ appear at orders lower than the leading non-analytic behaviour of $S^{(r)}_m[\rho^{(m)}_{\mathrm{ext}}(\lambda,\boldsymbol{\alpha}^\epsilon)]$ (\ref{eq:Skm1}).  This is a consequence of our particular choice of deformation (\ref{eq:defhypersurfaces}). For $m$ prime the full set of critical exponents are elements of $\mathbb{Q}^+/\mathbb{Z}$. For non-prime $m$  some of the critical exponents are integers.  Whether or not we should refer to these integers as critical exponents is a subtle matter.\footnote{It can often happen that when a critical exponent is na\"ively integer valued there is in fact a logarithmic dependence on the coupling. 
Logarithmic behaviour is also present for the critical exponent of a two-matrix model \cite{Boulatov:1986sb}  whose continuum description has been argued to be the free fermion coupled to two-dimensional gravity. It is relatively straightforward to prove that our integer valued critical exponents are indeed integers exhibiting no logarithmic dependence on the coupling.} A point of concern for integer critical exponents is that they may be sensitive to analytic redefinitions of the couplings. 
For $m=4$, $k=2$ and WLOG taking $p=2$ we have 
\begin{equation}
u= 16+ \tilde{x} \epsilon^{1/2}~, \quad \alpha_{2}= -\frac{1}{8}+ \frac{560}{3}s_4 \epsilon+ \delta \epsilon^2~, \quad \alpha_3= \frac{1}{160}-28s_4 \epsilon~, \quad \alpha_{4}= - \frac{1}{8960}+ s_4\epsilon
\end{equation}
and  (\ref{eq:Skm}) reduces to 
\begin{equation}\label{eq:S24}
S^{(2)}_4[\rho^{(4)}_{\mathrm{ext}}(\lambda,\boldsymbol{\alpha}^\epsilon)]-\mathcal{S}_4^{(2)}[\boldsymbol{s},\epsilon,\epsilon^2]
=\frac{8}{5}\,\delta\,\epsilon^{{\boldsymbol{2}}}+ \mathcal{O}\left(\epsilon^{5/2}\right) ~,
\end{equation}
where 
\begin{equation}
\mathcal{S}_4^{(2)}[\boldsymbol{s},\epsilon,\epsilon^2]= S_c[\rho^{(4)}_{\mathrm{ext}}(\lambda,\boldsymbol{\alpha}_c)]+160 s_4\epsilon - 143360 s_4^2\epsilon^2~.
\end{equation}
If $s_4$ and $\delta$ obey
\begin{equation}
\frac{8}{5}\delta- 143360 s_4^2 = 0~,
\end{equation}
we can cancel the integer critical exponent $\boldsymbol{2}$. 

Furthermore we note that the matrix integral (\ref{eq:defVm}) admits $(m-1)$ distinct critical exponents only upon considering deformations away from the multicritical point. The only other critical exponent is that of a \textbf{square root} non-analyticity in the normalisation condition leading to $\mathcal{F}_{m,\mathrm{n.a.}}^{(0)}(\boldsymbol{\alpha})\sim  \epsilon^{\boldsymbol{5/2}}$ which occurs on surfaces connecting the $m^{\mathrm{th}}$ polynomial $V_m(M,\boldsymbol{\alpha})$ to a binomial matrix integral $\tilde{V}_n(M,\alpha)$, $n\leq m$ (\ref{eq:tVk}).

\begin{center} $\star \star \star$\end{center}
{\color{black}\textbf{Summary.}} In summary, the set of $(m-1)$ critical exponents for the $m^{\mathrm{th}}$ multicritical matrix integral (\ref{eq:defVm}) are given by ${\color{magenta}\boldsymbol{m/(m-r')}}$, $r'= 1,\ldots ,m-2$ (\ref{eq:Skm}) and ${\color{magenta}\boldsymbol{2+1/m}}$ (\ref{eq:Gstr_general}).

\section{Critical exponents in the continuum picture}\label{continuum_sec}

In section \ref{nonanalyticsec} we uncovered a set of non-analyticities arising from the deformation of multicritical matrix integrals (\ref{eq:defVm}) slightly away from the multicritical point (\ref{eq:defalphac}). We showed that the $m^{\mathrm{th}}$ multicritical matrix integral has $(m-1)$ distinct non-analyticities (\ref{eq:Skm}) and (\ref{eq:Gstr_general}). In this section we will uncover the same non-analyticities within the continuum picture of $\mathcal{M}_{2m-1,2}$ coupled to two-dimensional quantum gravity.  

\subsection{A minimal model refresher}
Minimal models are two-dimensional CFTs characterised by two coprime integers $(p,p')$ with $p,p'\geq 2$ and WLOG we assume $p>p'$.\footnote{In addition to the original work \cite{Polyakov:1984yq}, an excellent resource discussing their detailed properties is given by \cite{AlvarezGaume:1989vk, yellowbook}.} We will denote the $(p,p')$ minimal model by $\mathcal{M}_{p,p'}$. The central charge of $\mathcal{M}_{p,p'}$ is given by
\begin{equation}\label{eq:cminimalmodel}
c^{(p,p')} \equiv 1 - \frac{6(p-p')^2}{p p'}~.
\end{equation}
Each $\mathcal{M}_{p,p'}$ has a finite number of conformal primaries $\mathcal{O}_{r,s}$ whose (holomorphic) conformal dimension is given by
\begin{equation}\label{eq:CD_primaries}
\Delta_{r,s}= \frac{(rp'-sp)^2- (p-p')^2}{4pp'}~, \quad r=1,\ldots, p-1~, \quad s=1,\ldots , p'-1~.
\end{equation}
These obey
\begin{equation}\label{eq:relationCD}
\Delta_{r,s}=\Delta_{p-r,p'-s}=\Delta_{r+p,s+p'}~,
\end{equation}
such that the number of distinct conformal primaries is given by $n_{p,p'} \equiv (p-1)(p'-1)/2$. The identity operator $\mathbb{I}$ of vanishing conformal dimension is always present and given by $\mathcal{O}_{1,1} = \mathcal{O}_{p-1,p'-1}$. The anti-holomorphic conformal dimensions are denoted by $\bar{\Delta}_{r,s}$ and take the same values as (\ref{eq:CD_primaries}).

The simplest minimal model is $\mathcal{M}_{3,2}$, which is a two-dimensional CFT with central charge $c^{(3,2)}=0$ and the unique operator $\mathbb{I}$, of vanishing conformal dimension. 
\newline\newline
{\textbf{(Non)-unitary minimal models.}} It will prove useful to distinguish between unitary and non-unitary minimal models. In addition to a positive definite inner product, unitary minimal models have positive central charge and non-negative conformal dimensions. As shown in \cite{Friedan}, the unitary models are given by the series $\mathcal{M}_{m+1,m}$ with $m> 2$. Unitary minimal models have ${n_{m+1,m} = m(m-1)/2}$ primaries. 

Non-unitary minimal models have negative central charge. Although the highest weight states have positive norm, their Virasoro descendants have negative norm. The simplest example of a non-unitary minimal model is $\mathcal{M}_{5,2}$, the Yang-Lee model with $c^{(5,2)}= -22/5$. $\mathcal{M}_{5,2}$ has two conformal primaries of holomorphic conformal dimension $\Delta_{1,1} = 0$ and $\Delta_{2,1} = -1/5$. 

In what follows we will focus on $\mathcal{M}_{2m-1,2}$ with $m\ge 2$. The general expression for their central charge is 
\begin{equation}\label{eq:cmnu}
 c^{(2m-1,2)}= 1-\frac{3(3-2m)^2}{2m-1}~.
\end{equation}
The number of conformal primaries is $n_{2m-1,2} = (m-1)$, and their holomorphic dimensions are given by
\begin{equation}\label{eq:confdim}
\Delta_{r,1}= \frac{(2m-1-2r)^2- (2m-3)^2}{8(2m-1)}~, \quad r=1,\ldots ,m-1~.
\end{equation}
The conformal dimensions are increasingly negative for increasing $r$, and the lowest weight primary $\mathcal{O}_{\text{min}} \equiv \mathcal{O}_{m-1,1}$ has holomorphic conformal dimension
\begin{equation}\label{eq:Deltamin}
\Delta_{\mathrm{min}}= \frac{(m-1)(m-2)}{2(1-2m)}~.
\end{equation}
From (\ref{eq:cmnu}) and (\ref{eq:Deltamin}) we infer the large $m$ expansions
\begin{equation}\label{eq:c_MM}
c^{(2m-1,2)} = -6m+16 + \ldots~,\quad \Delta_{\mathrm{min}} = -\frac{m}{4} + \frac{5}{8} + \ldots~.
\end{equation}
Although $c^{(2m-1,2)}$ grows at large $m$, it has been argued \cite{Itzykson:1986pj,Itzykson:1986pk} that a better measure of the number of degrees of freedom is captured by
\begin{equation}
c_{\text{eff}}^{(2m-1,2)} \equiv c^{(2m-1,2)} - 24 \Delta_{\mathrm{min}} = 1 + \frac{3}{1-2m}~.
\end{equation}
We note that $c_{\text{eff}}^{(2m-1,2)}<1$ goes to one in the large $m$ limit.

\subsection{Critical exponents}

To compute critical exponents associated to a given conformal field theory, we consider the partition function of the theory deformed by a small amount of a particular conformal primary $\mathcal{O}_\Delta$. We first discuss critical exponents for CFTs on a fixed background, and then proceed to a fluctuating background.
\newline\newline
{\textbf{2d CFT on a fixed flat background.} From the perspective of a path-integral, we would like to compute
\begin{equation}\label{eq:pathintegral}
Z[\lambda_\Delta, \ell] = \int [\mathcal{D}{\Phi} ] e^{-S_{\text{CFT}}[{\Phi}] - \lambda_\Delta \int \dd^2x \, \mathcal{O}_\Delta}~,\quad \lambda_{\Delta} \in \mathbb{R}~,
\end{equation}
where $\ell$ denotes the size of the flat square on which the CFT resides. 
The conformal primary $\mathcal{O}_{\Delta}$ has dimension $(\Delta,\bar{\Delta})$ and for simplicity we take $\Delta=\bar{\Delta}$. The dimensionful scales of the problem are the volume of space $\ell^2$, the ultraviolet length scale $\ell_{\text{uv}}\ll \ell$, and the coupling $\lambda_\Delta$ whose holomorphic scaling dimension is $\Delta_\lambda = \Delta-1$. Following the line of argumentation from the scaling hypothesis $Z[\lambda_\Delta, \ell] =Z[q^{-\Delta_{\lambda}}\lambda_\Delta, q\ell]$, $q\in \mathbb{R}^+$ \cite{Kadanoff:1966wm}
 we would like the UV independent part of $\log Z[\lambda_\Delta, \ell]$ to be extensive in the volume. Given that $Z[\lambda_{\Delta}, \ell]$ is dimensionless, we must have
\begin{equation}\label{eq:cexponent_fixedb}
\log Z[\lambda_\Delta, \ell] = \mathcal{N}\ell^2 (\lambda_\Delta)^{\boldsymbol{\nu}_\Delta}~,\quad \boldsymbol{\nu}_\Delta \equiv 1/(1-\Delta)~.
\end{equation}
$\mathcal{N}$ is a normalisation constant independent of $\lambda_{\Delta}$. Notice that the critical exponent associated to the identity operator is simply $\boldsymbol{\nu}_0  = 1$. 
Furthermore, for $\Delta > 1$, corresponding to an irrelevant $\mathcal{O}_\Delta$, the critical exponent would be negative. 
\newline\newline
{\textbf{2d CFT on a fluctuating background.} We now consider a two-dimensional CFT with central charge $c_{\text{m}}<1 $ coupled to two-dimensional gravity.  Integrating over all metrics renders the extensivity condition of the scaling hypothesis somewhat subtle. In the Weyl gauge the two-dimensional metric is chosen to be $g_{ij} = e^{2b\varphi} \tilde{g}_{ij}$. The problem then maps to studying the matter CFT with central charge $c_{\mathrm{m}}$, trivially coupled to a Liouville CFT with central charge $c_L =26-c_{\mathrm{m}}$, and the $\mathfrak{b}\mathfrak{c}$-ghost system with central charge $c_g = -26$ \cite{Polyakov:1981rd}. The Liouville action is given by \cite{Polyakov:1981rd}
\begin{equation}
S_L[\varphi,\Lambda] = \frac{1}{4\pi} \int \dd^2 x \sqrt{\tilde{g}} \left( \tilde{g}^{ij} \partial_i \varphi \partial_j \varphi + Q R[\tilde{g}_{ij}] \varphi + 4\pi \Lambda \,e^{2b \varphi} \right)~,
\end{equation}
where $\tilde{g}_{ij}$ is taken to be the round metric on $S^2$ such that $R[\tilde{g}_{ij}] = 2$ and $\Lambda \geq 0$ is the cosmological constant. Moreover, $Q = b + 1/b$ with \cite{Seiberg:1990eb, Ginsparg:1993is, DiFrancesco:1993cyw, DioBeatrix}
\begin{equation}\label{eq:bQ}
b = \frac{\sqrt{25-c_{\text{m}}} - \sqrt{1-c_{\text{m}}}}{2\sqrt{6}}~, \quad Q= \sqrt{\frac{25-c_{\mathrm{m}}}{6}}~.
\end{equation}
The residual gauge invariance in the Weyl gauge enforces that all operators of the combined theory are spinless primaries with conformal dimension $\Delta = 1$. In the trivial ghost sector this is achieved by dressing the matter primaries of weight $\Delta_{\mathrm{m}}$ by a Liouville operator of weight $\Delta_L = 1-\Delta_{\mathrm{m}}$. \newline\newline
{\textbf{Unitary 2d CFT on a fluctuating background.} We now specify to a unitary two-dimensional CFT with $c_{\mathrm{m}} \in (0,1)$.
The simplest critical exponent corresponds to the matter identity whose coupling is $\Lambda$. The partition function of interest is\footnote{On a genus $h$ surface there is also a topological term proportional to $e^{\vartheta\chi}$, where $\vartheta$ is the coupling governing the genus expansion, see e.g. \cite{DioBeatrix}. As we are working on a fixed topology throughout this section we will drop this term.}
\begin{equation}\label{Lidentity}
\mathcal{Z}[\Lambda] = \int [\mathcal{D}\varphi] e^{-S_L[\varphi,\Lambda]} = \Lambda^{\boldsymbol{Q/b}}~.
\end{equation}
We indicate the partition function on a fluctuating background by $\mathcal{Z}$.
A simple derivation of the above follows from performing a shift in $\varphi$ \cite{David:1988hj,Distler:1988jt}. Due to the Liouville dressing, the critical exponent of the identity is no longer simply given by $\boldsymbol{\nu}_0  = 1$ (\ref{eq:cexponent_fixedb}), but rather \cite{Knizhnik:1988ak}
\begin{equation}
\boldsymbol{\nu}_{\text{grav}} \equiv Q/b =  \frac{1}{12}\sqrt{\left( 1 - c_{\text{m}}\right) \left(25 - c_{\text{m}} \right)} + \frac{25-c_{\text{m}}}{12}~.
\end{equation}
The critical exponent for $\boldsymbol{\nu}_{\text{grav}}$  informs us how to modify the scaling behaviour of length upon coupling to gravity. On a fixed background  the total scaling dimension of a length scale is minus one, whereas now we must take it to be $\boldsymbol{\nu}_{\text{grav}}/2$. $\boldsymbol{\nu}_{\text{grav}}$ is also known as the string scusceptibility. 

Now, rather than the identity we consider turning on a matter conformal primary $\mathcal{O}_\Delta$. The partition function of interest, in the Weyl gauge, becomes
\begin{equation}\label{eq:part_lambda_Delta}
\mathcal{Z}[\lambda_\Delta] = \int [\mathcal{D}\varphi][ \mathcal{D} \Phi] e^{-S_L[\varphi,\Lambda=0] -S_{\text{CFT}}[\Phi] - \lambda_\Delta \int \dd^2x \sqrt{\tilde{g}}  \,e^{2\sigma_\Delta \varphi} \mathcal{O}_\Delta}~,
\end{equation}
where we set $\Lambda=0$ since we are interested in turning on $\mathcal{O}_\Delta$ alone.
We further have 
\begin{equation}\label{eq:sigma_delta}
\sigma_\Delta \equiv \frac{\sqrt{25-c_{\text{m}}}-\sqrt{24 \Delta +1-c_{\text{m}}}}{2 \sqrt{6}}~,
\end{equation}
which ensures that the matter operator is dressed appropriately. We note that $b= \sigma_{\Delta=0}$. Upon shifting $\varphi \to \varphi - ( \log \lambda_\Delta)/2\sigma_\Delta$, \cite{David:1988hj,Distler:1988jt} and noting that the path-integration measure over $\varphi$ is invariant under such shifts, it is straightforward to deduce
\begin{equation}\label{genZ}
\mathcal{Z}[\lambda_\Delta] = \mathcal{N} \left( \lambda_{\Delta} \right)^{\boldsymbol{Q/\sigma_\Delta}}~,
\end{equation}  
where $\mathcal{N}$ is a $\lambda_\Delta$ independent normalisation. For more details we refer to \cite{DioBeatrix}.
\newline\newline
{\textbf{$\mathcal{M}_{2m-1,2}$ on a fluctuating background.} For non-unitary models with $c_{\mathrm{m}}\leq 0$, further care must be taken due to the presence of operators with negative conformal dimension. Our main interest is in $\mathcal{M}_{2m-1,2}$, whose most relevant operator $\mathcal{O}_{\text{min}}$ has negative conformal dimension $\Delta_{\mathrm{min}}$ (\ref{eq:Deltamin}). The lowest weight operator replaces the identity in that all other operators are `irrelevant' with respect to $\mathcal{O}_{\text{min}}$. Thus, in the non-unitary case we might be inclined \cite{Staudacher:1989fy,Brezin:1989db} to replace (\ref{Lidentity}) with
\begin{equation}\label{eq:cE_continuum}
\mathcal{Z}[\Lambda_{\text{min}}] = \int [\mathcal{D}\varphi][ \mathcal{D} \Phi] e^{-S_L[\varphi,\Lambda=0] -S_{\text{CFT}}[\Phi] - \Lambda_{\text{min}} \int \dd^2x  \sqrt{\tilde{g}}\, e^{2\sigma_{\text{min}} \varphi} \mathcal{O}_{\text{min}}}~,
\end{equation}
where $\sigma_{\mathrm{min}}\equiv \sigma_{\Delta_{\mathrm{min}}}$ (\ref{eq:sigma_delta}). Using similar techniques to those discussed previously one finds
\begin{equation}\label{eq:CFT_Gammastr}
\mathcal{Z}[\Lambda_{\text{min}}] = (\Lambda_{\text{min}})^{\boldsymbol{Q/\sigma_{{\mathrm{min}}}}}~.
\end{equation}
In effect, one is replacing $\sigma_{\Delta=0}$ with $\sigma_{\Delta_{\mathrm{min}}}$ (\ref{Lidentity}). Note that
\begin{equation}\label{eq:Qsigma}
{Q}/{\sigma_{{\mathrm{min}}}} = \boldsymbol{2 + {1}/{m}}~.
\end{equation}
Turning on other operators $\mathcal{O}_{r,1}$ while setting $\Lambda_{\text{min}}=0$ leads to
\begin{equation}\label{genZ2}
\mathcal{Z}[\lambda_{\Delta_{r,1}}] = \mathcal{N} \left( \lambda_{\Delta_{r,1}} \right)^{\boldsymbol{Q/\sigma_{\Delta_{r,1}}}}~,\quad Q/\sigma_{\Delta_{r,1}} = \bold{{(1+2m)}/{(r+1)}}~,
\end{equation}  
 where $r =1, 2, \ldots , m-2$. Finally we note the useful relation 
\begin{equation}\label{continuumcrit}
{\sigma_{\text{min}}}/{\sigma_{\Delta_{r,1}}} =  \bold{{m}/{(1 + r)}}~, \quad\quad r = 1,\ldots,m-2~.
\end{equation}
\newline
{\textbf{Summary.}} In summary, we obtain ($m-1$) critical exponents: Turning on the operator of lowest conformal dimension we obtain ${\color{magenta}\textbf{2+1/m}}$ (\ref{eq:Qsigma}). Turning on any of the other $(m-2)$ operators we obtain the critical exponents ${\color{magenta}\textbf{m/(1+r)}}$, $r=1,\ldots, m-2$ (\ref{continuumcrit}).
\newline\newline
{\textbf{A fixed ``area'' perspective.}} In order to compare to the perturbative discussion of section \ref{multiplePD} it proves instructive to consider the gravitational path integrals with a constraint fixing the total area of space to a fixed value $\upsilon$. This can be achieved by inserting a $\delta$-function inside of the gravitational path-integral (\ref{eq:pathintegral}). For two-dimensional conformal field theories with $c_{\text{m}}<1$, we have
\begin{equation}\label{fixedaream2}
\mathcal{Z}_{\text{area}}[\upsilon] = \mathcal{N} \, \upsilon^{\boldsymbol{-1-Q/b}} \times e^{-\upsilon \Lambda}~, \quad\quad \int \dd^2x \sqrt{\tilde{g}} \,e^{2b\varphi} = \upsilon~.
\end{equation}
where $\mathcal{N}$ is independent of $\upsilon$ and $\Lambda$. Integrating $\mathcal{Z}_{\text{area}}[\upsilon]$ against $\upsilon$, we recover $\mathcal{Z}[\Lambda]$ (\ref{Lidentity}). For $c_{\text{m}} = 0$, we note that $1+Q/b = \textbf{7/2}$, the value observed in (\ref{eq:F0m2}). For $c_{\text{m}}= c^{(2m-1,2)}$ in (\ref{eq:cmnu}), we find instead $1+Q/b = \textbf{3/2+m}$. 
Let us now consider those non-unitary minimal models whose lowest weight operator $\mathcal{O}_{\text{min}}$ is different from the identity. We can also consider fixing 
\begin{equation}\label{minfix}
\int \dd^2 x \sqrt{\tilde{g}} \, \mathcal{O}_{\text{min}} \, e^{2 \sigma_{\text{min}}\varphi}  = \upsilon~.
\end{equation}
This leads to the following partition function
\begin{equation}\label{mingrowth}
\mathcal{Z}_{\text{min}}[\upsilon] = \mathcal{N} \, \upsilon^{\boldsymbol{-1-Q/\sigma_{\text{min}}}} e^{-\upsilon \Lambda_{\text{min}}}~.
\end{equation}
For the Lee-Yang model $\mathcal{M}_{5,2}$ with $c^{(5,2)} = -22/5$, we have $1+Q/\sigma_{\text{min}} = \textbf{10/3}$. For general $\mathcal{M}_{2m-1,2}$ we have $1+Q/\sigma_{\text{min}} = (\bold{3+1/m})$.

\subsection{Comparison to matrix integrals} 

At this stage it behooves us to compare our results to those of the multicritical matrix integrals. We take inspiration from 't Hooft's diagrammatic picture \cite{tHooft:1974pnl}, whereby the perturbative diagrams of the matrix integrals correspond to discretised Riemann surfaces. Care must be taken in identifying the appropriate quantities between the matrix integrals and the continuum picture. 
\newline\newline
{\textbf{$\mathcal{M}_{3,2}$ on a fluctuating background.} Let us begin by discussing the simplest case, namely $m=2$. In this case the matrix diagrammatics (\ref{eq:F0m2}) indicates that the dependence on the number of vertices $k$ goes as $\sim k^{-\bold{7/2}} (\alpha_2/\alpha^{(2)}_{2,c})^{-k}$ at large $k$. One is motivated to identify the number of vertices, $k$, with the area of the surface in the continuum picture. Both are extensive quantities sensitive to the total number of points on the surface. In doing so, one finds a match between the behaviour of the fixed area partition function (\ref{fixedaream2}) and the matrix diagrammatics. This suggests that the identification of $k$ in the matrix diagrammatics and $\upsilon$ in the continuum is indeed sensible.\footnote{Although we do not discuss it here, this identification continues to be sensible for those matrix integrals argued to describe the unitary minimal models coupled to gravity. As an explicit example, the $\mathcal{M}_{4,3}$ model on a fluctuating background was studied in \cite{Boulatov:1986sb}, leading to the fixed area behaviour $\sim \upsilon^{-\bold{10/3}}$ which agrees with the prediction from the corresponding two-matrix model.} Going from the diagrammatics to the critical exponent is simply a matter of integrating (summing) over $\upsilon$ ($k$), and identifying $\Lambda \propto (\alpha_2-\alpha^{(2)}_{2,c})$.
\newline\newline
{\textbf{$\mathcal{M}_{2m-1,2}$ on a fluctuating background.}  We would like to compare the asymptotics at large vertex number from the multicritical matrix diagrammatics to the continuum picture. Part of the issue is that there are multiple couplings, and consequently multiple paths in coupling space to reach the multicritical point. Along the path (\ref{eq:gamma_starm}) which simultaneously tunes several couplings, the growth of vertices goes as $\sim k^{-(\bold{3+1/m})}t^k$ (\ref{eq:naF_diagrammatics}). Recalling (\ref{mingrowth}) and noting that for general $m$, $1+Q/\sigma_{\text{min}} = (\bold{3+1/m})$, we find evidence that such a tuning corresponds to fixing the extensive quantity (\ref{minfix}), rather than the area (\ref{fixedaream2}).  The remaining task is to identify $\Lambda_{\text{min}}$, and the additional $(m-2)$ couplings $\lambda_{\Delta_{r,1}}$ from the perspective of the multicritical matrix integral. This is precisely the problem of non-analyticities solved in section \ref{nonanalyticsec}. The non-analyticities found in the $m^{\text{th}}$ multicritical matrix integral correspond to the values $Q/\sigma_{\mathrm{min}}$ (\ref{eq:Qsigma}) and $\sigma_{\text{min}}/\sigma_{\Delta_{r,1}}$, $r=1,\ldots ,m-2$ (\ref{continuumcrit}) arising from $\mathcal{M}_{2m-1,2}$ on a fluctuating background. We thus identify $\Lambda_{\text{min}} = \epsilon$ in (\ref{eq:Gstr_general}), $\lambda_{\Delta_{r,1}} = \epsilon^{\sigma_{\text{min}}/Q}$ and $r' = m-r-1$  in (\ref{eq:Skm}).
We observe that $\sigma_{\text{min}}/Q$ is independent of $r$. 
Further to this, our hypersurface equation (\ref{eq:defhypersurfaces}) provides the detailed relation between the matrix deformation and the corresponding matter primary. 

\section{Remarks on a Hilbert space}\label{sec:Hilbert}

In this section we remark on the Hilbert space of $\mathcal{M}_{2m-1,2}$ coupled to two-dimensional gravity, and its manifestation from the matrix integral perspective. 

\subsection{$S^2$ considerations}

On a fixed background, $\mathcal{M}_{2m-1,2}$ has a finite number of  primaries equal to $n_{2m-1,2}= (m-1)$, each accompanied by an infinite tower of descendants. 
On a fluctuating background these operators must satisfy constraints arising from the diffeomorphism invariance. Additionally we need to consider the contribution from the Liouville and $\mathfrak{b}\mathfrak{c}$-ghost sector. 

Concretely we must identify the set of BRST invariant operators. This was examined in early work of Lian-Zuckerman (LZ) \cite{Lian:1991gk} and subsequent work \cite{Imbimbo:1991ia, Bouwknegt:1991mv, Kutasov:1991qx}. 
Under the assumption that the Liouville sector can be treated as a linear dilaton theory, it was noted that the BRST cohomology comprises of an {\it{infinite}} collection of operators. In particular LZ operators have a non-trivial ghost number and generally contain matter and Liouville descendants. Though also infinite, this infinity is far smaller than the infinite operator content of the matter theory on a fixed background arising from the Virasoro descendants. The origin of these operators is intimately connected to the presence of null operators in the Liouville sector \cite{Teschner:1995yf}\footnote{Indeed, one can always find primary operators in Liouville theory with central charge $c_L = 26 -c^{(2m-1,2)}$ whose conformal dimension lies on one of the values in the Kac table, and hence admit null states in their Verma module. The null operators are themselves primary.} and the matter sector.  In conformal gauge $\dd s^2= e^{2b \varphi(z,\bar{z})}\dd z\dd \bar{z}$ the LZ operators are given by
\begin{equation}
\mathcal{R}_{r,\pm}^{\mathrm{LZ}}(t)\equiv \mathcal{O}_{r,\pm}^{\mathrm{LZ}}(\mathfrak{b},\mathfrak{c},\varphi,\Phi; t)\otimes \mathcal{\bar{O}}_{r,\pm}^{\mathrm{LZ}}(\tilde{\mathfrak{b}},\tilde{\mathfrak{c}},\varphi, \Phi; t)\otimes \mathcal{O}_{r,1}\otimes e^{2\sigma_{\mathrm{LZ}}\varphi}~,
\end{equation}
where $t\in \mathbb{Z}$ and $\pm$ denote the particular LZ operator. The holomorphic conformal dimensions of these operators are given by 
\begin{equation}
 \Delta^{{\mathrm{LZ}}}_{r,\pm}(t)+\Delta_{r,1}+ \sigma_{\mathrm{LZ}}(Q-\sigma_{\mathrm{LZ}})=0~,
\end{equation}
where $Q=b+b^{-1}$ and LZ operators are graded by the ghost number. The LZ weights for $\mathcal{M}_{2m-1,2}$ are given by
\begin{equation}
\Delta^{{\mathrm{LZ}}}_{r,\pm}(t)= A_{r,\pm}(t)- \Delta_{r,1}-1~,
\end{equation}
where $A_{r,\pm}(t)$ are given by \cite{RochaCaridi}
\begin{equation}\label{eq:AtBt}
A_{r,\pm}(t)\equiv \frac{(4(2m-1)t+2r\pm(2m-1))^2-(2m-3)^2}{8(2m-1)}~.
\end{equation}
The anti-holomorphic conformal dimension has to be equal to the holomorphic conformal dimension. The argument $t$ is related to the ghost number, whereas the subscript $\pm$ indicates whether the ghost number is even $(+)$ or odd $(-)$. 
\newline\newline
{\textbf{Example $\mathcal{M}_{3,2}$.}} For $\mathcal{M}_{3,2}$ the LZ operators with ghost number $(\mathfrak{n}_b,\mathfrak{n}_c)= (0,0)$ and $(\mathfrak{n}_b,\mathfrak{n}_c)= (0,2)$ respecetively associated to  the matter primary $\mathcal{O}_{{1,1}}$ are \cite{Imbimbo:1991ia, Kutasov:1991qx} 
\begin{equation}\label{eq:LZ_M32}
\mathcal{R}_{1,+}^{\mathrm{LZ}}(0)= \mathbb{I}~,\quad\quad
\mathcal{R}_{1,+}^{\mathrm{LZ}}(0)=\mathfrak{c}(z)\partial^2\mathfrak{c}(z)\mathfrak{\tilde{c}}(\bar{z})\partial^2\mathfrak{\tilde{c}}(\bar{z})\otimes e^{2Q\varphi}~,
\end{equation}
with $\sigma_{\mathrm{LZ}} = 0$ and $\sigma_{\mathrm{LZ}} = Q$ respectively. The operators (\ref{eq:LZ_M32}) have non-trivial ghost number as compared to the vertex-operators considered in section \ref{continuum_sec} whose BRST invariant form takes $\mathcal{O}_{1,-}^{\mathrm{LZ}}(\mathfrak{b},\mathfrak{c},\varphi, \Phi;0)= \mathfrak{c}$ and $\mathcal{\bar{O}}_{1,-}^{\mathrm{LZ}}(\tilde{\mathfrak{b}},\tilde{\mathfrak{c}},\varphi, \Phi;0)=\tilde{\mathfrak{c}}$ and $\sigma_{\mathrm{LZ}}=b$.
On the other side the LZ operators with lowest ghost number $(\mathfrak{n}_b,\mathfrak{n}_c)= (1,1)$ associated to the matter primary $\mathcal{O}_{1,1}$ for $\mathcal{M}_{3,2}$ is
\allowdisplaybreaks
\begin{equation}\label{eq:R1p}
\mathcal{R}_{1,+}^{\mathrm{LZ}}(-1) = \Big(\mathfrak{b}(z)\mathfrak{c}(z)- b^{-1}\partial\varphi(z,\bar{z})\Big)\left(\mathfrak{\tilde{b}}(\bar{z})\mathfrak{\tilde{c}}(\bar{z})- b^{-1}\bar{\partial}\varphi(z,\bar{z})\right)\otimes e^{-b\varphi(z,\bar{z})}~,
\end{equation}
where combining (\ref{eq:bQ}) with $c^{(3,2)}$ (\ref{eq:c_MM}) we have $b= \sqrt{2/3}$. Besides the operator (\ref{eq:R1p}) there exists another operator $\mathcal{R}_{1,+}^{\mathrm{LZ}}(-1)$ with $\sigma_{LZ}= 2/b$.
To show the BRST invariance of these operators, it is useful to recall the (holomorphic) BRST current \cite{PolchinskiBook}
\begin{equation}
J_{\mathrm{BRST}}= \mathfrak{c}\,T^\varphi+\frac{1}{2} :\mathfrak{c}\, T^g:+ \frac{3}{2}\,\partial^2\,\mathfrak{c}~,
\end{equation}
where $T^\varphi$ and $T^g$ are the Liouville and ghost stress tensor respectively
\begin{equation}
T^\varphi= -(\partial\varphi)^2+ Q\partial^2\varphi~, \quad T^g=:(\partial\mathfrak{b})\mathfrak{c}:  - 2\partial\left( : \mathfrak{b}\mathfrak{c} :\right)~.
\end{equation}
In particular we find \cite{Imbimbo:1991ia}
\begin{equation}
\delta \mathcal{R}_{1,+}^{\mathrm{LZ}}(-1) 
= \mathfrak{c}(0)\tilde{\mathfrak{c}}(0)\otimes \left(\frac{3}{2}\,L_{-1}^2+ L_{-2}\right)\left(\frac{3}{2}\,\tilde{L}_{-1}^2+ \tilde{L}_{-2}\right) e^{-b\varphi(0,0)}~,
\end{equation}
where $L_{n}, \tilde{L}_n$ are the Virasoro generators, satisfying 
\begin{equation}
L_n\equiv \oint_{\mathcal{C}}\frac{\dd z}{2\pi i z}\,z^{n+2}T^\varphi(z)~,\quad \tilde{L}_n\equiv \oint_{\mathcal{C}}\frac{\dd \bar{z}}{2\pi i \bar{z}}\,\bar{z}^{n+2}\tilde{T}^\varphi(\bar{z})~.
\end{equation}
In other words the BRST variation leads to a null operator. 

We note that we consider the stress tensor of the Liouville action arising when assuming it is a linear dilaton theory. This is justified when calculating the critical exponents (\ref{eq:cE_continuum}) for which we set the cosmological constant to zero. The BRST transformation of a LZ operator can produce null operators in the matter or Liouville sector  \cite{Imbimbo:1991ia}, which must subsequently be set to zero.

One may ask whether the LZ operators contribute additional critical exponents for the theory on $S^2$. By the Riemann-Roch theorem, non-vanishing $\mathfrak{b}\mathfrak{c}$-correlation functions on a compact Riemann surface with Euler characteristic $\chi$ require \cite{PolchinskiBook}
\begin{equation}\label{eq:RR_theorem}
n_\mathfrak{c}- n_\mathfrak{b}= \frac{3}{2}\chi~.
\end{equation}
For $S^2$ we have $\chi=2$. 
Given that LZ operators have a non-trivial ghost number, generically different from $n_\mathfrak{c}- n_\mathfrak{b}=3$, we expect no new critical exponents from the LZ operators on an $S^2$ topology.
 
Assuming that some form of the operator-state correspondence holds for $S^2$ we are thus led to conclude that the associated Hilbert space is finite-dimensional. This might be related to observations on de Sitter space \cite{Erik_dS,Banks_dS}.\newline\newline
Contrarily to $S^2$ the torus $T^2$ has Euler characteristic $\chi=0$. As a consequence of the Riemann-Roch theorem (\ref{eq:RR_theorem}) we thus infer that the LZ operators contribute to the torus partition function. 

\subsection{$T^2$ considerations}

On the cylinder, the Hilbert space $\mathcal{H}_{T^2}$ lives on spatial $S^1$ constant time slices. States $|\Psi\rangle \in \mathcal{H}_{T^2}$ in the trivial ghost sector living on these spatial slices are subject to the Virasoro constraints
\begin{equation}\label{eq:Virasoro_constraints}
\left(L_0^{\mathrm{tot}}+\tilde{L}_0^{\mathrm{tot}}-2\right)|\Psi\rangle=0~, \quad \quad \left(L_0^{\mathrm{tot}}-\tilde{L}_0^{\mathrm{tot}}\right)|\Psi\rangle=0~,
\end{equation}
where $L_n^{\mathrm{tot}}$ and $\tilde{L}_n^{\mathrm{tot}}$ are the Virasoro generators for the matter and Liouville sector. The first equation in (\ref{eq:Virasoro_constraints}) is what replaces the Hamiltonian constraint in canonical quantum gravity \cite{DeWitt:1967yk}, while the second replaces the spatial diffeomorphism constraint. The above equations are the state analog of the constraint that vertex operators with trivial ghost contribution must have $\Delta\equiv\Delta_L+\Delta_{r,1}=1$ and $\Delta=\bar{\Delta}$. Other states in $\mathcal{H}_{T^2}$, as first pointed out by Lian-Zuckerman \cite{Lian:1991gk}, may also include non-trivial ghost excitations. 

One way to characterise $\mathcal{H}_{T^2}$ is through the torus partition function \cite{Kutasov:1990sv, Bershadsky:1990xb}. For fixed modular parameter $\tau= \tau_1+i\tau_2$ the states in the BRST cohomology contribute 
\begin{equation}
Z_{\mathrm{fixed}}[\tau_2]=(q\bar{q})^{\frac{1}{24}\left(26-1-c^{(2m-1,2)}\right)}\sum_{r=1}^{m-1}\sum_{t\in \mathbb{Z}}\left((q\bar{q})^{\Delta^{{\mathrm{LZ}}}_{r,+}(t)+\Delta_{r,1}}+(q\bar{q})^{\Delta^{{\mathrm{LZ}}}_{r,-}(t)+\Delta_{r,1}}\right)~,
\end{equation}
where $q=e^{2\pi i \tau_2}$ and we used (\ref{eq:AtBt}).
The overall shift encodes the Casimir energy from the ghost, Liouville and matter sector. The $\tau_1$-independence of $Z_{\mathrm{fixed}}[\tau_2]$ is due to the diffeomorphism constraint (\ref{eq:Virasoro_constraints}).

What remains to be done is integrate over the modular parameter $\tau_2$ and the zero modes of the Liouville sector
\begin{equation}
\mathcal{T}[\Lambda_{\mathrm{min}}]= \log\Lambda_{\mathrm{min}}\,\int_\mathcal{F}\frac{\dd^2 \tau}{\tau_2^{3/2}}\,Z_{\mathrm{fixed}}[\tau_2]~.
\end{equation}
$\mathcal{F}$ is the fundamental domain of the modular group. The power of $\tau_2$ is fixed by modular invariance and stems from the various zero modes in the $\mathfrak{b}\mathfrak{c}$-ghost and Liouville sector. The logarithm in $\Lambda$ stems from the volume of the Liouville zero mode, and essentially encodes the fact that the Liouville interaction imposes a cutoff in the Liouville field space. Evaluating $\mathcal{T}[\Lambda_{\mathrm{min}}]$ leads to \cite{Bershadsky:1990xb}
\begin{equation}\label{eq:Log_divergence_continuum}
\mathcal{T}[\Lambda_{\mathrm{min}}] =\frac{2m-2}{48m}\log\Lambda_{\mathrm{min}}~.
\end{equation}
Comparing to the first non-planar contribution of the matrix integral as presented in appendix \ref{app:np} we see that under the identification $\Lambda_{\mathrm{min}}=\epsilon$ the results agree (\ref{eq:F1_multicrit}). In this way the LZ states appear in the leading non-planar contribution of the $m^{\mathrm{th}}$ multicritical matrix integral.

\section{Discussion and open questions}\label{sec:discussion}
We summarise some open questions and speculative remarks.\newline\newline
\textbf{Large $m$ and Euclidean dS$_2$.} In \cite{Zamolodchikov:1982vx} it has been observed that upon coupline $\mathcal{M}_{2m-1,2}$ to gravity whilst fixing the area $\upsilon$ and turning on only the identity operator of $\mathcal{M}_{2m-1,2}$ exhibits a saddle point solution in the large $m$ limit. This saddle point solution is the round metric on $S^2$, which is Euclidean dS$_2$. 
Motivated by this, this work provides the basis to understand this observation from the matrix integral point of view. Recalling (\ref{eq:Skm}) for $r'=m-2$ we recover Zamolodchikov's continuum critical exponent from a matrix integral perspective. 
In the large $m$ limit and upon tuning the couplings to the multicritical point (\ref{eq:defalphac}) the polynomial $V_m(\lambda,\boldsymbol{\alpha})$ (\ref{eq:defVm}) reduces to\footnote{We would like to acknowledge Jorge Russo for useful discussions.} \cite{Ambjorn:2016lkl}
\begin{equation}
\lim_{m\rightarrow \infty}V_m(\lambda,\boldsymbol{\alpha}_c)= \frac{1}{2}\lambda^2\,_2F_2\left(1,1; \frac{3}{2},2; -\frac{\lambda^2}{4}\right)~.
\end{equation}
Moreover the width of the eigenvalue distribution  (\ref{eq:eigdistrV}) scales with $m$ and so becomes unbounded in the large $m$ limit. 
 We further remark that the most fine-tuned path where we switch on all the hypersurfaces and $\boldsymbol{s} \in \mathcal{H}_m^{(1)}\cup \ldots \cup \mathcal{H}_m^{(m-2)}$ (\ref{eq:defhypersurfaces}) corresponds to the identity operator in the continuum theory. 

As postulated in \cite{Gibbons:1976ue}, the logarithm of the Euclidean gravitational path-integral on a compact manifold for theories admitting a sphere saddle gives a semiclassical expansion of the entropy of the corresponding de Sitter solution. The presence of a semiclassical limit allows one to interpret the details of its expansion in terms of the classical saddle. 
The fixed area partition function on a genus $h=0$ surface (\ref{fixedaream2}) for large negative $c_{\mathrm{m}}$ can be written as 
\begin{equation}
\log \mathcal{Z}_{\text{area}}[\upsilon]= 2\vartheta+ \left(\frac{c_{\mathrm{m}}}{6}- \frac{25}{6} - \frac{6}{c_{\mathrm{m}}}+ \ldots \right)\log\frac{\upsilon}{\upsilon_0}~,
\end{equation}
where $\upsilon_0$ is a reference area. As noted in \cite{DioBeatrix}, to leading order the logarithmic term resembles an entanglement entropy for a two-dimensional CFT with central charge $c_{\mathrm{m}}$ \cite{Casini:2011kv, Holzhey:1994we,Calabrese:2004eu}, with the subleading corrections corresponding to contributions stemming from the CFT being coupled to dynamical gravity.  Moreover from the matrix perspective $2\vartheta = \log N^2$ suggesting that the size $N$ of the matrix also has an entropic interpretation. 
\newline\newline
\textbf{Non-unitarity \& torus Hilbert space.} As a consequence of the Riemann-Roch theorem (\ref{eq:RR_theorem}) the LZ operators contribute on $T^2$. In particular this implies that we have to deal with descendants of the Virasoro primaries of $\mathcal{M}_{2m-1,2}$. These are negative norm operators. The consequences of the non-unitarity from both the point of view of the MMI and the point of view of the continuum theory remain to be explored. 
\newline\newline 
{\textbf{Diagrammatics \& critical exponents.}} To evaluate the double sum (\ref{eq:F3_pert2}), capturing the diagrammatic expansion of the $m=3$ model, we introduced the path $\gamma_\star^{(3)}$ (\ref{eq:gamma_star}) in coupling space. This allowed us to explicitly determine the radius of convergence and for $m\geq 3$ using $\gamma_\star^{(m)}$ (\ref{eq:gamma_starm}) we observed the critical exponent $\mathcal{F}_{m,\mathrm{n.a.}}^{(0)}(\boldsymbol{\alpha})\big|_{\gamma_\star^{(m)}}\sim \epsilon^{\boldsymbol{2+1/m}}$ (\ref{eq:naF_diagrammatics}). However introducing the single parameter $t\in [0,1]$ connecting the origin in coupling space to the multicritical point prevents from observing the other $(m-2)$ critical exponents $\boldsymbol{m/(m-r')}$, $r'=1,\ldots ,m-2$, from a diagrammatic perspective. It would be interesting to uncover these.
\newline\newline
{\textbf{Disk topology.}} In \cite{Seiberg:1990eb, Fateev:2000ik, Teschner:1995yf, Zamolodchikov:2001ah} Liouville theory was studied on the disk topology. Upon taking the semiclassical limit $b\rightarrow 0$ the Liouville action admits a saddle point solution for $\varphi$. From the perspective of the physical metric $g_{ij}= e^{2b\varphi}\tilde{g}_{ij}$ this can be interpreted as the hyperbolic metric on the Poincar{\'e} disk (the Euclidean AdS$_2$ black hole)
\begin{equation}
\dd s^2= \frac{1}{\pi b^2\Lambda}\frac{1}{(1-\rho^2)^2}\left(\dd \rho^2+ \rho^2\dd \theta^2\right)~, \quad \rho \in [0,1)~.
\end{equation}
Since coupling $\mathcal{M}_{2m-1,2}$ to gravity implies $b= \sqrt{2/(2m-1)}$ (\ref{eq:bQ}) this semiclassical limit corresponds to a large $m$ limit. It would be interesting to explore relations between this saddle and recent discussions on JT gravity and matrix integrals \cite{Saad:2019lba}. 

For Dirichlet boundary conditions $\varphi$ diverges at the boundary of the disk and the relation to matrix integrals was studied in \cite{Alexandrov:2003nn}. It would be interesting to generalise this to the multicritical case. 
For Neumann boundary conditions one needs to further add the boundary term 
\begin{equation}
S_{\mathrm{bdy}}= \int_{S^1}\dd u \sqrt{h}\left(\frac{Q}{2\pi} K\varphi+ \Lambda_B e^{b\varphi}\right)~,
\end{equation}
to the bulk action, where $h$ is the induced metric on $S^1$ and $K$ is the extrinsic curvature at the boundary. 
The comparison to the matrix integral uses either the resolvent $\RR(z)$ or the loop operator $W_\ell$  \cite{Polyakov:1980ca}. The boundary cosmological constant $\Lambda_B$ is connected to $z$. For multicritical matrix integrals $\RR(z)$ is given by (\ref{eq:largezResolvent}), $W_\ell$ is given by (\ref{eq:loopopexp}). 
\newline\newline
{\textbf{Hartle-Hawking \& topology.}} As a final remark, it is interesting to note that the partition function $\mathcal{Z}[\Lambda]$ on $S^2$ only dominates (in absolute value) over the partition function $\mathcal{T}[\Lambda]$ on $T^2$ for sufficiently large $\Lambda$, while for small enough $\Lambda>0$, the $T^2$ partition function dominates. This is also true for higher genus partition functions. It would be interesting to understand if this has any consequences for the Hartle-Hawking picture \cite{Chen:2020tes, Hartle:1983ai}. Further to this, being matrix integrals rather than matrix path integrals there is no a priori indication for the existence of Hilbert space from the matrix integral perspective. It would be interesting to uncover a Lorentzian picture directly from the matrix integral \cite{DioGuille}.

\subsection*{Acknowledgements}

It is a great pleasure to acknowledge Tarek Anous, Teresa Bautista Solans, Frederik Denef, Diego Hofman, Austin Joyce, Yuan Miao, Antonio Rotundo, Jorge Russo, Jeremy van der Heijden, Erik Verlinde and G{\'e}rard Watts for illuminating discussions. We would also like to acknowledge the creators of the TikzLings package. D.A. is funded by the Royal Society under the grant The Atoms of a deSitter Universe. The work of B.M. is part of the research programme of the Foundation for Fundamental Research on Matter (FOM), which is financially supported by the Netherlands Organisation for Science Research (NWO). 

{\appendix

\section{Non-planar contribution}\label{app:np}
To compare to the log-divergence in (\ref{eq:Log_divergence_continuum}) we need to go beyond the planar approximation of the large $N$ limit of $\mathcal{F}_m(\boldsymbol{\alpha})$ (\ref{eq:Fm0_general}). Whereas the planar contribution is obtained from a large $N$ saddle point approximation, to find non-planar contributions one needs to make use of other techniques. We will use the method of orthogonal polynomials \cite{Bessis:1980ss}.  We will only provide minimalistic details, for a more detailed explanation of this method we refer for example to \cite{DioBeatrix}.
Two polynomials are said to be orthogonal with respect to a weight function $w(x)$ if they satisfy
\begin{equation}
\mathrm{ortho}_a:\quad \int \dd x \, w(x) \, p_n(x)p_m(x)= h_n\delta_{m,n} \label{eq:ortho}
\end{equation}
In addition to (\ref{eq:ortho}), orthogonal polynomials satisfy the {\it{three-term recurrence relation}}
\begin{align}\nonumber
\mathrm{ortho}_b: \quad  &x\, p_n(x)= A_n\, p_n(x) + S_n\, p_{n+1}(x) + R_n\, p_{n-1}(x)\quad\quad \text{for} \quad\quad n>0~,\\ \label{eq:3termsrec}
& x\, p_0(x)= A_0\, p_0(x)+ S_0\, p_1(x)~,
\end{align}
where $A_n$, $S_n$, and $R_n$ are some real constants. 
Focusing on monic polynomials
\begin{equation}\label{eq:monicp}
P_n(\lambda)\equiv \lambda^{n} + \sum_{j=0}^{n-1} a_j\lambda^j~, \quad\quad n=0,\ldots, N-1~,
\end{equation}
we obtain \cite{Bessis:1980ss}
\begin{align}\label{eq:FEinRn}
\frac{1}{N^2}\mathcal{F}_{m}(\boldsymbol{\alpha}) = -\frac{1}{N}\log \frac{h_0(\boldsymbol{\alpha})}{h_0(\boldsymbol{0})} - \frac{1}{N}\sum_{n=1}^{N-1}\left(1- \frac{n}{N}\right)\log \frac{R_n(\boldsymbol{\alpha})}{R_n(\boldsymbol{0})}~,
\end{align}
where we highlight the coupling dependency of $h_n$ (\ref{eq:ortho}) and $R_n$ (\ref{eq:3termsrec}) explicitly. \newline\newline
{\textbf{Example $m=3$.}} Using (\ref{eq:ortho}) and (\ref{eq:3termsrec})  for $\omega(x)= e^{-N V_3(x,\boldsymbol{\alpha})}$ we obtain \cite{DioBeatrix}
\begin{align} \label{eq:equationqI}
\frac{n}{N} &= {R_n(\boldsymbol{\alpha})}\Big[1  + \alpha_2\left(R_{n+1}(\boldsymbol{\alpha}) + R_n(\boldsymbol{\alpha}) + R_{n-1}(\boldsymbol{\alpha})\right)\cr
&+\alpha_3\big(2R_{n+1}(\boldsymbol{\alpha})+2R_n(\boldsymbol{\alpha})R_{n-1}(\boldsymbol{\alpha})+R_{n+1}(\boldsymbol{\alpha})R_{n-1}(\boldsymbol{\alpha})+R_{n-1}(\boldsymbol{\alpha})R_{n-2}(\boldsymbol{\alpha})\cr
&+R_n^2(\boldsymbol{\alpha})+R_{n+1}^2(\boldsymbol{\alpha})+R_{n-1}^2(\boldsymbol{\alpha})+R_{n+1}(\boldsymbol{\alpha})R_{n+2}(\boldsymbol{\alpha})\big)\Big]\;.
\end{align}
Let us now define the variables $\varepsilon\equiv {1}/{N}$ and $x\equiv n\varepsilon$. In the large $N$ limit, $x$ is well approximated by a continuous parameter. In view of this, it is convenient to set $r(x,\boldsymbol{\alpha})\equiv R_n(\boldsymbol{\alpha})$. We note that $r(x,\boldsymbol{\alpha})$ is also a function of $N$, but we suppress this dependence for notational simplicity. 
We can rewrite (\ref{eq:equationqI}) as 
\begin{multline}\label{eq:repsilon}
x= r(x,\boldsymbol{\alpha}) +\alpha_2 r(x,\boldsymbol{\alpha})\left[r(x+\varepsilon,\boldsymbol{\alpha})+r(x,\boldsymbol{\alpha}) +r(x-\varepsilon,\boldsymbol{\alpha})\right]+\alpha_3\Big[2r(x,\boldsymbol{\alpha})r(x+\epsilon,\boldsymbol{\alpha})\cr
+ 2r(x,\boldsymbol{\alpha})r(x-\epsilon,\boldsymbol{\alpha})+r(x+\epsilon,\boldsymbol{\alpha})r(x-\epsilon,\boldsymbol{\alpha})+r(x-\epsilon,\boldsymbol{\alpha})r(x-2\epsilon,\boldsymbol{\alpha})+r(x,\boldsymbol{\alpha})^2\cr
+r(x-\epsilon,\boldsymbol{\alpha})^2+r(x+\epsilon,\boldsymbol{\alpha})^2+r(x+\epsilon,\boldsymbol{\alpha})r(x+2\epsilon,\boldsymbol{\alpha})\Big]~.
\end{multline}
It follows from (\ref{eq:repsilon}) that $r(x,\boldsymbol{\alpha})$ is symmetric under $\varepsilon\leftrightarrow -\varepsilon$ and we can expand it in even powers of $\varepsilon$
\begin{equation}\label{eq:expre}
r(x,\boldsymbol{\alpha})= r_{0}(x,\boldsymbol{\alpha})+\varepsilon^2 \, r_{2}(x,\boldsymbol{\alpha}) +\varepsilon^4 \, r_{4}(x,\boldsymbol{\alpha})+\cdots~.
\end{equation}
To obtain the first non-planar contribution we only need $r_{0}(x,\boldsymbol{\alpha})$ and $r_{2}(x,\boldsymbol{\alpha})$ which we easily infer from (\ref{eq:repsilon}) by comparing powers of $\epsilon$.
\newline\newline
An equation similar to (\ref{eq:repsilon}) can  be obtained for $m\geq 3$ upon choosing $\omega(x)= e^{-N V_m(x,\boldsymbol{\alpha})}$, $m\geq 4$ (\ref{eq:ortho}).
Our final ingredient will be the Euler-Maclaurin formula
\begin{equation}\label{eq:eulerMac}
\frac{1}{N}\sum_{n=1}^N f\left(\frac{n}{N}\right)=  \int_0^1\dd x f(x)+\frac{1}{2N}f(x)\big|_0^1 +\sum_{n=1}^{p-1} \frac{B_{2n}}{(2n)!}\frac{1}{N^{2n}}f(x)^{(2n-1)}\big|_0^1+\mathcal{R}_N~.
\end{equation}
In the above, $f(x)$ is a $2p$ times continuously differentiable function, $\mathcal{R}_N$ is a remainder term scaling as $\mathcal{O}(1/N^{2p+1})$, and the $B_{2n}$ denote the Bernoulli numbers. Applying the Euler-Maclaurin formula to
\begin{equation}\label{eq:EulerMclaurinf}
f\left(x\right)= \left(1- x\right)\log \frac{r(x,\boldsymbol{\alpha})}{x}~,
\end{equation}
and expanding (\ref{eq:FEinRn}) in inverse powers of $N$, we find
\begin{align}\label{eq:Fexp} 
\frac{1}{N^2}\mathcal{F}_m(\boldsymbol{\alpha}) = &- \int_0^1\dd x(1-x)\log\frac{r(x,\boldsymbol{\alpha})}{x} -\frac{1}{N}\log \frac{h_0(\boldsymbol{\alpha})}{h_0(0)}+\frac{1}{2N}\lim_{x\rightarrow 0}\log\frac{r(x,\boldsymbol{\alpha})}{x}\cr
&-\frac{1}{12 N^2} \left((1-x)\log \frac{r(x,\boldsymbol{\alpha})}{x}\right)^{(1)}\Bigg|_0^1\end{align}
up to order $\mathcal{O}(1/N^4)$ corrections.
Expanding all three terms in (\ref{eq:expre}) and evaluating $h_0(\boldsymbol{\alpha}) $ for small $\boldsymbol{\alpha}$  we find  up to powers of order $\mathcal{O}(1/N^2)$ 
\begin{align}\label{eq:F0F1}\nonumber
\frac{1}{N^2}\mathcal{F}_m(\boldsymbol{\alpha})=&-\int_0^1 \dd x\,(1-x)\,\log\frac{r_{0}(x,\boldsymbol{\alpha})}{x} \\ 
&-\frac{1}{N^2}\left[\int_0^1 \dd x\,(1-x)\,\frac{r_{2}(x,\boldsymbol{\alpha})}{r_{0}(x,\boldsymbol{\alpha})}+\frac{1}{12}\left[(1-x)\log \frac{r_{0}(x,\boldsymbol{\alpha})}{x}\right]^{(1)}\Bigg|_0^1 -\frac{3}{4}\alpha_2 \right]~.
\end{align}
Note that we encounter an ambiguity in choosing $r_{0}(x,\boldsymbol{\alpha})$ since it is the solution of an $m^{\mathrm{th}}$ order polynomial. We pick the solution yielding the on-shell value (\ref{eq:Scriticality}) when evaluating the $\mathcal{O}(N^0)$ integral along $\gamma_\star^{(m)}$. 
Evaluating the second line in the above expression is in general difficult however to get the coefficient of the log-divergence we only care about the first integral. To further simplify our analysis we zoom into the multicritical point $\boldsymbol{\alpha}_c$ (\ref{eq:defalphac}). The non-analytic behaviour of (\ref{eq:F0F1}) occurring for $\boldsymbol{\alpha}= \boldsymbol{\alpha}_c$ close to the upper boundary, equals the non-analyticity observed upon considering small deformations away from the multicritical point only after evaluating the integral. We obtain for $m=3$ and $m=4$
\begin{equation}\label{eq:F1_multicrit}
\mathcal{F}_{3, \mathrm{n.a.}}^{(1)}(\boldsymbol{\alpha_c}) = \frac{1}{18}\log \epsilon~, \quad \mathcal{F}_{ 4,\mathrm{n.a.}}^{(1)}(\boldsymbol{\alpha_c}) = \frac{1}{16}\log \epsilon~, \quad \epsilon\ll 1~,
\end{equation}
where the subscript indicates the leading non-analyticity. For general $m\geq 3$ one finds \cite{Gross:1989vs}
\begin{equation}
\mathcal{F}_{m, \mathrm{n.a.}}^{(1)}(\boldsymbol{\alpha_c}) = \frac{2m-2}{24m}\log \epsilon~,\quad \epsilon\ll 1~,
\end{equation}
and the coefficient of the log agrees with the coefficient of the log of the torus partition function (\ref{eq:Log_divergence_continuum}) of the continuum theory. 
For $m=2$ and the generalisations thereof discussed in section \ref{m2review} on the other side we obtain 
\begin{equation}
\mathcal{\tilde{F}}_{m, \mathrm{n.a.}}^{(1)}(\boldsymbol{\alpha_c}) = \frac{1}{24}\log\epsilon~, \quad \epsilon\ll 1~.
\end{equation}
As a final remark we note that $\mathcal{{F}}_m^{(1)}(4\alpha_2,\ldots, 2m\alpha_m)$ counts leading non-planar diagrams. More explicitly it counts diagrams whose vertices are emanating four or $2m$ edges and which can fit on a surface of genus one. As an example, a perturbative analysis of (\ref{eq:F0F1}) using $r_0(x,\boldsymbol{\alpha})$ and $r_2(x,\boldsymbol{\alpha})$ for $m=3$ easily reveals 
\begin{equation}
\mathcal{{F}}_3^{(1)}(4\alpha_2,6\alpha_3) = \alpha_2+10\alpha_3-30\alpha_2^2- 2400\alpha_3^2-600\alpha_2\alpha_3+ \ldots ~.
\end{equation}

\section{Solutions of $\mathcal{N}_4(\boldsymbol{\alpha})=0$}\label{app:m4}

In this section, we discuss the solutions of the normalisation condition (\ref{eq:NCm4}) for the $m=4$ case:
\begin{equation}\label{eq:solutions_m4}
u_{1,\pm}^{(4)}=- \frac{2\alpha_3}{7\alpha_4}-S\pm \frac{1}{2}\sqrt{-4S^2-2p+\frac{q}{S}}~, \quad
u_{2,\pm}^{(4)}= -\frac{2\alpha_3}{7\alpha_4}+S\pm \frac{1}{2}\sqrt{-4S^2-2p-\frac{q}{S}}~,
\end{equation}
where 
\begin{equation}\label{pandqvals}
p\equiv \frac{24(14\alpha_2\alpha_4-5\alpha_3^2)}{245\alpha_4^2}~, \quad q\equiv \frac{64(5\alpha_3^3-21\alpha_2\alpha_3\alpha_4+49\alpha_4^2)}{1715\alpha_4^3}~.
\end{equation}
In the above, we have defined
\begin{eqnarray}
S &\equiv& \frac{1}{2}\sqrt{-\frac{2}{3}p-\frac{256}{105\alpha_4}\left(Q+\frac{\Delta_0}{Q}\right)}~,  \\
Q &\equiv& \sqrt[3]{\frac{\Delta_1+ \sqrt{\Delta_1^2-4\Delta_0^3}}{2}}~, \label{Qvals} \\
\Delta_0 &\equiv& \frac{3}{256}(3\alpha_2^2-10(\alpha_3+14\alpha_4))~, \\
\Delta_1&\equiv& -\frac{27}{4096}\left(2(\alpha_2^3-5\alpha_2\alpha_3-50\alpha_3^2)+35\alpha_4(1+8\alpha_2)\right)~,\\
D_4 &\equiv& -\frac{1}{27}\left(\Delta_1^2-4\Delta_0^3\right)~. \label{d4val}
\end{eqnarray}
Much of our interest lies in a solution of $\mathcal{N}_4(\boldsymbol{\alpha}) = 0$ that is regular in a small neighbourhood $\aleph_{\bold{0}}$ around the origin of coupling space $\boldsymbol{\alpha}=0$. To analyse the problem, we can consider approaching $\boldsymbol{\alpha}=0$ uniformly in all directions, and exploring the behaviour of the various solutions throughout $\aleph_{\bold{0}}$. An exhaustive analysis reveals that one must keep track of the various signs of $\boldsymbol{\alpha}$ and the special combination $(\alpha_3+14 \alpha_4)$. The term $(\alpha_3+14 \alpha_4)$ is already revealed in the form of $\Delta_0$ and can be seen to carry through into the more involved building blocks such as $Q$ and $S$. For instance, near $\aleph_{\bold{0}}$ we have
\begin{equation}
Q = \begin{cases} \frac{5^{2/3}}{8 \times 7^{1/3}} \left( \frac{\alpha_3+14 \alpha_4}{\alpha_4}\right)^{1/3} + \mathcal{O}(\alpha^{4/3})~, \quad & \alpha_4 \ge 0~,\\
\frac{3 \times 35^{1/3}}{16} \left(-\alpha_4\right)^{1/3}  + \mathcal{O}(\alpha^{4/3})~, \quad & \alpha_4 < 0~.
\end{cases}
\end{equation}
We find the following combination of solutions to be smooth near $\aleph_{\bold{0}}$
\begin{multline}\label{eq:B4}
B_4(\boldsymbol{\alpha}) = u^{(4)}_{1,+}\left(\Theta_{+++} + \Theta_{-++} + \Theta^+_{--+}  + \Theta^+_{+-+} \right) +  \\   u^{(4)}_{2,-}\left(\Theta_{---} +  \Theta_{++-} + \Theta_{-+-} + \Theta_{+--}  + \Theta_{--+}^- + \Theta_{+-+}^- \right)~,
\end{multline}
where we have introduced the notation
\begin{eqnarray}
\Theta_{\rho_2 \rho_3 \rho_4} &\equiv& \Theta(\rho_2 \alpha_2) \Theta(\rho_3 \alpha_3) \Theta(\rho_4 \alpha_4)~, \\
\Theta^{\rho}_{\rho_2 \rho_3 \rho_4} &\equiv& \Theta(\rho_2 \alpha_2) \Theta(\rho_3 \alpha_3) \Theta(\rho_4 \alpha_4)\Theta(\rho(\alpha_3+14\alpha_4))~.
\end{eqnarray}

\subsubsection*{Further properties of $\mathcal{N}_4(\boldsymbol{\alpha}) = 0$}

The solutions (\ref{eq:solutions_m4}) also reveal additional information. For instance, expanding the discriminant $D_4$ (\ref{d4val}) at small $\alpha_2$ and $\alpha_3$, we identify $\alpha_4 = -27/8960$ as the special value $\tilde{\alpha}_{4,c}$ in (\ref{eq:deftilde_ac}). Similarly, expanding the discriminant $D_4$ at small $\alpha_3$ and $\alpha_4$, we identify $\alpha_2 = -1/12$ as the special value $\alpha^{(2)}_{2,c}$ in (\ref{eq:defalphac}). Finally expanding for small $\alpha_2$ and $\alpha_4$ reveals $\alpha_3= -2/135$ as the special value $\tilde{\alpha}_{3,c}$ (\ref{eq:deftilde_ac}). Near $(\alpha_2,\alpha_3,\alpha_4) = (0,0,-27/8960)$,   $(\alpha_2,\alpha_3,\alpha_4)= (0,-2/135,0)$ as well as $(\alpha_2,\alpha_3,\alpha_4) = (-1/12,0,0)$, $\Delta_1$ remains non-vanishing such that the non-analytic behaviour of the solutions $u^{(4)}$ is that of a \textbf{square root}. Expanding the discriminant $D_4$ near $\alpha_2=-1/8$, reveals $\alpha_3 = 1/160$ as a special value, which we recognise as $\alpha^{(4)}_{3,c}$, one of the multicritical couplings (\ref{eq:defalphac}). At $(\alpha_2,\alpha_3) = (-1/8,1/160)$ we further have that $\Delta_1 = 0$, while $D_4$ goes as $(1 + 8960 \alpha_4)^3$ revealing the third multicritical value $\alpha_4 = -1/8960$. Also, at $(\alpha_2,\alpha_3) = (-1/8,1/160)$ we observe that $Q$ in (\ref{Qvals}) goes as $(1 + 8960 \alpha_4)^{1/2}$, and $p$ in (\ref{pandqvals}) goes as $(1 + 8960 \alpha_4)$. Expanding away from the multicritical point reveals distinct non-analytic behaviour in the solutions of $\mathcal{N}_4(\boldsymbol{\alpha}) = 0$. For instance, fixing $(\alpha_2,\alpha_3) = (-1/8,1/160)$ and deviating slightly away from $\alpha_4 =  -1/8960$, we uncover a \textbf{fourth root} non-analyticity.


\section{Non-analyticities: normalisation condition}\label{app:na}

In this appendix we prove the following.\newline\newline
{\textbf{Claim.}}
For 
\begin{equation}\label{eq:movingaway_app}
\boldsymbol{\alpha}^\epsilon \equiv \boldsymbol{\alpha} _c+ \boldsymbol{s}\,\epsilon~, \quad u = 4m+ \tilde{x}\, \epsilon^{\frac{1}{m-r'}}~,\quad \tilde{x}\in\mathbb{R}~,
\end{equation}
with $\boldsymbol{s}$ living on  $\mathcal{H}_{m}^{(1)}\cup \mathcal{H}_m^{(2)}\ldots \cup \mathcal{H}_m^{(r')}=0$, $r'= 1,\ldots m-2$, (\ref{eq:defhypersurfaces}) the normalisation condition (\ref{eq:BC}) reduces to (\ref{eq:bc})
\begin{align}\label{eq:bc_app}
\mathcal{N}_m^{(r')}(\boldsymbol{\alpha}^\epsilon)
   =&\left[(-1)^m\frac{\tilde{x}^m}{(4m)^m}- \frac{\tilde{x}^{r'}}{4^{r'}}m^2\,\mathcal{H}_{m}^{(r'+1)}\right]\,\epsilon^{\frac{m}{m-r'}}+ \mathcal{O}\left(\epsilon^{\frac{m+1}{m-r'}}\right)~.
\end{align}
{\textbf{Proof.}} Plugging (\ref{eq:movingaway}) with $\boldsymbol{s}$ constraint to live on the hypersurface $\mathcal{H}_{m}^{(1)}\cup \mathcal{H}_m^{(2)}\ldots \cup \mathcal{H}_m^{(r')}=0$ into the normalisation condition (\ref{eq:BC}) and expanding for small $\epsilon$ we obtain 
\begin{equation}\label{eq:bc_app_proof}
 \mathcal{N}_m^{(r')}(\boldsymbol{\alpha}^\epsilon)=(-1)^m\frac{\tilde{x}^m}{(4m)^m} \epsilon^\frac{m}{m-r'} - \sum_{n=2}^m \left(\frac{4^n}{2n B(n,1/2)}\sum_{\ell=0}^n \binom{n}{\ell}\left(\frac{\tilde{x}}{4m}\right)^\ell \, \epsilon^{\frac{\ell}{m-r'}+1}\right)\,m^n\, s_n~.
 \end{equation}
We apply a recursive argument by showing that the sum multiplying the term $\tilde{x}^{k-1}$, $k\geq 3$, vanishes for $\boldsymbol{s}$ constraint to
 \begin{equation}
\mathcal{H}_{m}^{(k-2)} \cup \mathcal{H}_{m}^{(k-1)}\cup \mathcal{H}_{m}^{(k)}=0~.
\end{equation}
For $\ell= k-1$ we have (\ref{eq:bc_app})
\allowdisplaybreaks
\begin{align}\nonumber\label{eq:recursivezero}
&\sum_{n=k-1}^m \frac{4^n}{2n B(n,1/2)}\binom{n}{k-1}\,m^n s_n=\frac{(4m)^{k-1}}{2(k-1)B(k-1,1/2)}s_{k-1}+ \sum_{n=k}^m \frac{4^n}{2n B(n,1/2)}\binom{n}{k-1}\,m^n s_n\\ \nonumber
&\overset{\mathcal{H}_{m}^{(k-2)}}{=} \sum_{n=k}^m \frac{4^n}{2n B(n,1/2)}\left(\binom{n}{k-1}- \binom{n-2}{k-3}\right)\,m^n s_n= \frac{(4m)^k}{kB(k,1/2)}s_k+\\ \nonumber
&+\sum_{n=k+1}^m \frac{4^n}{2n B(n,1/2)}\left(\binom{n}{k-1}- \binom{n-2}{k-3}\right)\,m^n s_n\\ \nonumber
&\overset{\mathcal{H}_{m}^{(k-1)}}{=} \sum_{n=k+1}^m \frac{4^n}{2n B(n,1/2)}\left(\binom{n}{k-1}- \binom{n-2}{k-3}- 2\binom{n-2}{k-2}\right)\,m^n s_n\\ 
&=\sum_{n=k+1}^m \frac{4^n}{2n B(n,1/2)}\binom{n-2}{k-1}\,m^n s_n= \mathcal{H}_{m}^{(k)}~.
\end{align}
A superscript over an equality sign means that we are using the condition $\boldsymbol{s}\in \mathcal{H}_m^{(j)}$. 
For $\ell=1$ the sum already vanishes on $\mathcal{H}_m^{(1)}\cup \mathcal{H}_m^{(2)}=0$. For $\ell=0$ it vanishes on $\mathcal{H}_m^{(1)}=0$.
If the directions $\boldsymbol{s}$ now lives on 
\begin{equation}
\bigcup_{j=1}^{r'}\mathcal{H}_m^{(j)}=0~,
\end{equation}
in (\ref{eq:bc_app_proof}) all terms of order $\mathcal{O}(\tilde{x}^{k})$, $k< r' $ vanish and we obtain (\ref{eq:bc_app}).

\section{Non-analyticities: action}\label{app:action}
In this appendix we prove the following.\newline\newline
{\textbf{Claim.}} Along 
\begin{equation}\label{eq:finetuningBC_app}
\alpha^\epsilon_{n\neq p}= \alpha_{c,n\neq p}^{(m)}+  s_{n\neq p}\,\epsilon~, \quad \alpha_p^\beta = \alpha_{c,p}^{(m)}+ s_p\, \epsilon+\tilde{s}\,\epsilon^{\frac{m}{m-r'}}~,\quad u = 4m+ \tilde{x}\,\epsilon^{\frac{1}{m-r'}}~.
\end{equation}
and using (\ref{eq:finalBC}) we obtain
\begin{multline}\label{eq:Skm1_app}
S^{(r')}_m[\rho^{(m)}_{\mathrm{ext}}(\lambda,\boldsymbol{\alpha}^\epsilon)] = \mathcal{S}_m^{(r')}[\boldsymbol{s},\epsilon,\epsilon^2]
 -\frac{1}{2}H_m\left[(-1)^m\frac{\tilde{x}^m}{(4m)^m}- \frac{\tilde{x}^{r'}}{4^{r'}}m^2\,\mathcal{H}_{m}^{(r'+1)}-\frac{(4m)^p}{2p B(p,1/2)}\tilde{s}\right]\,\epsilon^{\frac{m}{m-r'}}  \\ +\frac{(4m)^p}{2p^2 B(p,1/2)}\frac{m!p!}{(m+p)!}\,\tilde{s}\,\epsilon^{\frac{m}{m-r'}}+ \mathcal{O}\left(\epsilon^{\frac{m+1}{m-r'}}\right)~,
\end{multline}
where
\begin{multline}
\mathcal{S}_m^{(r')}[\boldsymbol{s},\epsilon,\epsilon^2]\equiv S_c[\rho^{(m)}_{\mathrm{ext}}(\lambda,\boldsymbol{\alpha}_c)]\cr
+\epsilon\left(
\sum_{n=2}^m\frac{(4m)^n}{4n^2 B(n,1/2)}s_n
+\sum_{\substack{n=1\\k\geq 2}}^m \frac{(-1)^{n+1}}{4n(n+k)}\binom{m}{n}\frac{(4m)^k}{B(k,1/2)}s_k+ \sum_{\substack{k=1\\n\geq 2}}^m(-1)^{k+1}\frac{k}{4n^2(n+k)}\binom{m}{k}\frac{(4m)^n}{B(n,1/2)}s_n \right)\cr
+\epsilon^2\sum_{n,k=2}^m \frac{s_{n}}{4n^2 B(n,1/2)}\frac{s_k}{2(n+k)}\frac{(4m)^{n+k}}{B(k,1/2)}~
\end{multline}
evaluated on (\ref{eq:hypersurfacek}). The action $S_c[\rho^{(m)}_{\mathrm{ext}}(\lambda,\boldsymbol{\alpha}_c)]$ at criticality was defined in (\ref{eq:Scriticality}). \newline\newline
{\textbf{Proof.}}
To proof (\ref{eq:Skm1_app}) we split the on-shell action (\ref{eq:Srhoext}) into pieces and write out the series expansion of the logarithm. This leads to 
\begin{multline}\label{eq:actionAppendix}
S^{(k)}_m[\rho_{\mathrm{ext}}(\lambda)]\cr
=\sum_{n=1}^m \frac{(2n)!}{4n}\,\alpha_{n}\,\omega_{n}^{(m)}+\sum_{n=1}^m\frac{\alpha_{n,c} u^{n}}{4n^2 B(n,1/2)}+\epsilon\sum_{n=2}^m\frac{s_n u^{n}}{4n^2 B(n,1/2)}-\sum_{\ell=1}^\infty\frac{(-1)^{\ell+1}}{2\ell}\frac{\tilde{x}^\ell}{(4m)^\ell}\,\epsilon^{\frac{\ell}{m-k}}~,
\end{multline}
where (\ref{eq:omegank})
\begin{equation}\label{eq:omegank_app}
\omega_{n}^{(m)}(\boldsymbol{\alpha})\equiv \frac{1}{(n!)^24^n}\sum_{k=1}^m \frac{\alpha_k u^{n+k}}{2(n+k)B(k,1/2)}~.
\end{equation}
and $B(n,1/2)$ is the beta function 
\begin{equation}
B(n,1/2)= \frac{4^n (n!)^2}{n (2n)!}~.
\end{equation}
\newline
{\textbf{$1^{\mathrm{st}}$ term.}}
For the first term in the action (\ref{eq:actionAppendix}) we find 
\begin{align}\label{eq:firstterm}
\sum_{n=1}^m \frac{(2n)!}{4n}\,\alpha_{n}\,\omega_{n}^{(m)}&=\frac{1}{2}(H_{2m}-H_m)+\frac{(-1)^{m+1}}{2}\frac{\tilde{x}^m}{(4m)^m}H_m\,\epsilon^{\frac{m}{m-r'}}+\frac{1}{2}H_m \frac{m^2}{4^{r'}}\mathcal{H}_{m}^{(r'+1)}\cr
&+ \sum_{n,k=1}^m\sum_{\ell>m}^{n+k}\frac{(-1)^{n+k}k}{2n(n+k)}\binom{m}{n}\binom{m}{k}\binom{n+k}{\ell}\frac{\tilde{x}^\ell}{(4m)^\ell}\,\epsilon^{\frac{\ell}{m-r'}}~\cr
&+\sum_{n,k=2}^m \frac{s_{n}}{4n^2 B(n,1/2)}\frac{s_k}{2(n+k)}\frac{(4m)^{n+k}}{B(k,1/2)}\,\epsilon^2+ \mathcal{O}\left(\epsilon^{\frac{2m-r'-1}{m-r'}}\right)~.
\end{align}
where we used i$1)$, i$2)$, i$3)$ (\ref{eq:equalities}) and c$2)$ (\ref{eq:conjectures}) and $H_m$ is the $m^{\mathrm{th}}$ harmonic number.
Additionally (\ref{eq:firstterm}) contains a coupling and $\tilde{x}$ dependent part. Using (\ref{eq:omegank_app}) we obtain along (\ref{eq:finetuningBC_app})
\begin{align}\nonumber
&\sum_{n,k=1}^m\sum_{\ell=0}^{n+k} \frac{\alpha_{n}}{4n^2 B(n,1/2)}\frac{\alpha_k}{2(n+k)}\frac{(4m)^{n+k}}{B(k,1/2)}\binom{n+k}{\ell}\frac{\tilde{x}^\ell}{(4m)^\ell}\,\epsilon^{\frac{\ell}{m-r'}}
=\sum_{n,k=1}^m \frac{\alpha_{n}}{4n^2 B(n,1/2)}\frac{\alpha_k}{2(n+k)}\frac{(4m)^{n+k}}{B(k,1/2)}\\
&+\sum_{n,k=1}^m\sum_{\ell=1}^{n+k} \frac{\alpha_{n}}{4n^2 B(n,1/2)}\frac{\alpha_k}{2(n+k)}\frac{(4m)^{n+k}}{B(k,1/2)}\binom{n+k}{\ell}\frac{\tilde{x}^\ell}{(4m)^\ell}\,\epsilon^{\frac{\ell}{m-r'}}~,
\end{align}
where the case of the $\alpha$'s in both sums equal to their critical value we already treated in obtaining (\ref{eq:firstterm}). For the other case using (\ref{eq:defalphac}) we have 
\begin{align} \label{eq:decomp_tealterm} \nonumber
&\sum_{n,k=1}^m\sum_{\ell=1}^{n+k} \frac{1}{4n(n+k)}\left[(-1)^{n+1}\binom{m}{n}\frac{(4m)^k}{B(k,1/2)}s_k+ (-1)^{k+1}\frac{k}{n}\binom{m}{k}\frac{(4m)^n}{B(n,1/2)}s_n\right]\binom{n+k}{\ell}\frac{\tilde{x}^\ell}{(4m)^\ell}\,\epsilon^{\frac{\ell}{m-r'}+1}\\ \nonumber
+&\sum_{n,k=1}^m \frac{1}{4n(n+k)}\left[(-1)^{n+1}\binom{m}{n}\frac{(4m)^k}{B(k,1/2)}s_k+ (-1)^{k+1}\frac{k}{n}\binom{m}{k}\frac{(4m)^n}{B(n,1/2)}s_n\right]\epsilon\\ \nonumber
+&\sum_{n,k=2}^m \frac{s_{n}}{4n^2 B(n,1/2)}\frac{s_k}{2(n+k)}\frac{(4m)^{n+k}}{B(k,1/2)}\,\epsilon^2\\ 
+&\sum_{n,k=2}^m\sum_{\ell=1}^{n+k}\binom{n+k}{\ell} \frac{s_{n}}{4n^2 B(n,1/2)}\frac{s_k}{2(n+k)}\frac{(4m)^{n+k}}{B(k,1/2)}\frac{\tilde{x}^\ell}{(4m)^\ell}\,\epsilon^{\frac{\ell}{m-r'}+2}~.
\end{align}
Only the sums in the first line of  (\ref{eq:decomp_tealterm}) and the last sum could contribute to the leading non-analyticity in (\ref{eq:Skm1_app}). We treat the sums independently. For the second sum in the second line we use that (\ref{eq:conjectures}) vanishes for $\ell <m$ and the first non-vanishing term arises for $\ell=m$ proportional to $\epsilon^{(2m-r')/(m-r')}$. Since our leading non-analyticity grows us $\mathcal{O}(\epsilon^{m/(m-r')})$, $r'=1,\ldots ,m-2$ the former is subleading with respect to the non-analyticity we are after.  
For the first sum of the second line we show that for  $\boldsymbol{s}\in \mathcal{H}_{m}^{(1)}\cup \mathcal{H}_{m}^{(2)} \ldots \cup \mathcal{H}_{m}^{(\ell)}=0$ we have 
\begin{equation}\label{eq:firsttermc}
\sum_{n=1}^m\sum_{k=2}^m \frac{(-1)^{n+1}}{4n(n+k)}\binom{m}{n}\frac{(4m)^k}{B(k,1/2)}s_k\binom{n+k}{\ell}\frac{\tilde{x}^{\ell}}{(4m)^{\ell}}=\frac{1}{2}H_m \frac{m^2}{4^{\ell}}\mathcal{H}_{m}^{(\ell+1)}~.
\end{equation}
We show (\ref{eq:firsttermc}) for $\ell=1$ and $\ell=2$, for general $\ell>2$ the logic stays the same. For $\ell= r'$ it leads to the claimed result (\ref{eq:Skm1_app}). For $\ell=1$ we have
\begin{align}
&\frac{1}{4m}\sum_{n=1}^m\sum_{k=2}^m \frac{1}{4n}\left[(-1)^{n+1}\binom{m}{n}\frac{(4m)^k}{B(k,1/2)}s_k\right]
=\frac{1}{4m}\sum_{k>2}\sum_{n=1}^m \frac{1}{4n}(-1)^{n+1}\binom{m}{n}\frac{(4m)^k}{B(k,1/2)}s_k\cr
+&\frac{1}{4m}\frac{(4m)^2}{B(2,1/2)}s_2\sum_{n=1}^m\frac{(-1)^{n+1}}{4n}\binom{m}{n}
\overset{\mathcal{H}_{m}^{(1)}}{=}\frac{1}{2}H_m \frac{m^2}{4}\mathcal{H}_{m}^{(2)}~.
\end{align}
For $\ell=2$ we have 
\begin{align}
&\frac{1}{(4m)^2}\sum_{k=2}^m\sum_{n=1}^m \frac{(-1)^{n+1}}{4n(n+k)}\binom{m}{n}\frac{(4m)^k}{B(k,1/2)}\binom{n+k}{2}s_k\cr
\overset{\mathcal{H}_{m}^{(1)}}{=}&\frac{1}{(4m)^2}\sum_{\substack{n=1 \\k\geq 3}}^m\frac{(-1)^{n+1}}{4n}\binom{m}{n}\frac{(4m)^k}{B(k,1/2)}\left[\frac{1}{n+k}\binom{n+k}{2}- \frac{2}{k(n+2)}\binom{n+2}{2}\right]s_k\cr
\overset{\mathcal{H}_{m}^{(2)}}{=}&\frac{1}{(4m)^2}\sum_{\substack{n=1 \\k\geq 3}}^m\frac{(-1)^{n+1}}{4n}\binom{m}{n}\frac{(4m)^k}{B(k,1/2)}\bigg[\frac{1}{n+k}\binom{n+k}{2}- \frac{2}{k(n+2)}\binom{n+2}{2}\cr
&- \frac{6(k-2)}{2k}\left(\frac{1}{n+3}\binom{n+3}{2}- \frac{2}{3(n+2)}\binom{n+2}{2}\right)\bigg]s_k\cr
=&\frac{1}{(4m)^2}\sum_{\substack{n=1 \\k\geq 3}}^m\frac{(-1)^{n+1}}{4n}\binom{m}{n}\frac{(4m)^k}{B(k,1/2)}\frac{1}{k}\binom{k-2}{2}s_k\cr
= &\frac{1}{(4m)^2}\sum_{n=1}^m\frac{(-1)^{n+1}}{2n}\binom{m}{n}\sum_{k\geq 4}\frac{(4m)^k}{2kB(k,1/2)}\binom{k-2}{2}s_k=\frac{1}{2}H_m \frac{m^2}{4^2}\mathcal{H}_{m}^{(3)}~.
\end{align}
Finally the last sum in (\ref{eq:decomp_tealterm}). For any fixed $\ell$ it vanishes on  $\mathcal{H}_{m}^{(1)}\cup \ldots \cup \mathcal{H}_{m}^{(\ell+1)}=0$. We start by showing this for $\ell=1$:
\begin{align}
&\sum_{n,k=2}^m \frac{s_{n}}{4n^2 B(n,1/2)}\frac{s_k}{2(n+k)}\frac{(4m)^{n+k}}{B(k,1/2)}\binom{n+k}{1}= \frac{(4m)^2}{B(2,1/2)}s_2 \sum_{n=2}^m\frac{(4m)^ns_{n}}{4n^2 B(n,1/2)}\cr
+&\sum_{k>2}\sum_{n=2}^m \frac{(4m)^ns_{n}}{4n^2 B(n,1/2)}\frac{(4m)^{k}}{2B(k,1/2)}s_g=\sum_{k>2}\sum_{n=2}^m \frac{(4m)^ns_{n}}{4n^2 B(n,1/2)}\frac{(4m)^k}{B(k,1/2)}\left(\frac{1}{2}- \frac{1}{k}\right)\cr
=&\sum_{n=2}^m \frac{(4m)^ns_{n}}{4n^2 B(n,1/2)}\sum_{k=3}^m\frac{(4m)^k}{2kB(k,1/2)}\binom{k-2}{1}s_k~\propto \mathcal{H}_m^{(2)}~.
\end{align}
Now for $\boldsymbol{s} \in \mathcal{H}_m^{(1)} \cup \mathcal{H}_m^{(2)}$ we obtain non-analyticity $\mathcal{O}\left(\epsilon^{m/(m-2)}\right)$, i.e. $k=2$. However from (\ref{eq:decomp_tealterm}) we infer  a leading contribution of order $\mathcal{O}\left(\epsilon^{{(2m-3)}/{(m-2)}}\right)$.
For $m>3$ which is the only case in which $k=2$ is allowed the exponent is therefore bigger than $m/(m-2)$ and so we do not get a contribution violating $\mathcal{O}\left(\epsilon^{m/(m-2)}\right)$ as a leading non-analyticity. For $\ell>1$ the logic is the same. 
\newline\newline
\textit{Identities for harmonic numbers $H_m$.}
\begin{align}\label{eq:equalities}
\mathrm{i}1)~&\sum_{n,k=1}^m\frac{(-1)^{n+k}k}{2n(n+k)}\binom{m}{n}\binom{m}{k}\binom{n+k}{m}= \frac{(-1)^{m+1}}{2}H_m~,\quad
\mathrm{i}2)~\sum_{n=1}^m \frac{(-1)^{n+1}}{2n}\binom{m}{n}= \frac{1}{2}H_m~,\cr
\mathrm{i}3)~&\sum_{n,k=1}^m \frac{(-1)^{n+k}k}{2n(n+k)}\binom{m}{n}\binom{m}{k}= \frac{1}{2}(H_{2m}-H_m)~,
\end{align}
\textit{Conjectures.}
\begin{align}\label{eq:conjectures}
\mathrm{c}1)\quad&\sum_{n,k=1}^m\sum_{\ell=1}^{m-1}\frac{(-1)^{n+k}k}{2n(n+k)}\binom{m}{n}\binom{m}{k}\binom{n+k}{\ell}\frac{1}{(4m)^\ell}=0~,\cr
\mathrm{c}2)\quad&\sum_{k=1}^m\sum_{n=2}^m \frac{(-1)^{k+1}}{4n(n+k)}\frac{k}{n}\binom{m}{k}\frac{(4m)^n}{B(n,1/2)}s_n\binom{n+k}{\ell}\frac{1}{(4m)^\ell}\bigg|_{\ell= 1,\ldots m-1}=0~.
\end{align}
{\textbf{$2^{\mathrm{nd}}$ term.}} For the second term in (\ref{eq:actionAppendix}) we start by rewriting the normalisation condition (\ref{eq:BC})
\begin{equation}
1- \sum_{n=1}^m\frac{\alpha_{n,c} u^{n}}{2n B(n,1/2)}= (-1)^m (4m)^{-m}(u-4m)^m~.
\end{equation}
Dividing both sides by $2u$ and integrating with respect to $u$ and $\eta>0$ we have
\begin{equation}
\int_\eta^{u'} \dd u\left(\frac{1}{2u}- \sum_{n=1}^m\frac{\alpha_{n,c} u^{n-1}}{4n B(n,1/2)}\right)=(-1)^m (4m)^{-m} \int_\eta^{u'} \dd u\,\frac{(u-4m)^m}{2u}~.
\end{equation}
We then get 
\begin{align}
\frac{1}{2}\log(u'/\eta) - \sum_{n=1}^m\frac{\alpha_{n,c} u'^{n}}{4n^2 B(n,1/2)}&=\frac{1}{2} \int_\epsilon^{u'} \dd u\frac{1}{u}\left(1-\frac{u}{4m}\right)^m=\frac{1}{2} \int_\epsilon^{u'} \dd u\,\sum_{\ell=0}^m (-1)^\ell \binom{m}{\ell}\frac{u^{\ell-1}}{(4m)^\ell}\cr
&= \sum_{\ell=1}^m \frac{(-1)^\ell}{2\ell}\binom{m}{\ell}\frac{u'^{\ell}}{(4m)^\ell}+ \frac{1}{2}\log(u'/\eta)~.
\end{align}
From this we conclude
\begin{align}
&\sum_{n=1}^m\frac{\alpha_{n,c} u^{n}}{4n^2 B(n,1/2)}= \sum_{\ell=1}^m \frac{(-1)^{\ell+1}}{2\ell}\binom{m}{\ell}\frac{u^{\ell}}{(4m)^\ell}
= \sum_{n=1}^m\frac{(-1)^{n+1}}{2n}\frac{\tilde{x}^n}{(4m)^n}\,\epsilon^{\frac{n}{m-k}}+ \frac{1}{2}H_m~.
\end{align}
The $\tilde{x}$ dependent term therefore exactly cancels the $4^{\mathrm{th}}$ term in (\ref{eq:actionAppendix}).\newline\newline
{\textbf{$3^{\mathrm{rd}}$ term.}} For the third term in the on-shell action (\ref{eq:actionAppendix}) we find for $\boldsymbol{s}\in \mathcal{H}_m^{(1)}\cup \ldots  \mathcal{H}_m^{(r')}$, following the same reasoning as in appendix \ref{app:na}
\begin{align}
\epsilon\sum_{n=2}^m\frac{s_n u^{n}}{4n^2 B(n,1/2)}=\epsilon\sum_{n=2}^m\frac{(4m)^n}{4n^2 B(n,1/2)}s_n+ \frac{\tilde{x}^{r'+1}}{(4m)^{r'+1}}\epsilon^{\frac{m+1}{m-r'}}\frac{m^{r'+2}}{2(k+1)}\mathcal{H}_{m}^{(r'+1)}~.
\end{align}

\subsection*{Fine-tuning}
We get the fine-tuning by realising that in the on-shell action (\ref{eq:Srhoext}) only the first and second term are affected to order $\mathcal{O}(\epsilon^{m/(m-r')})$ when fine-tuning $\alpha_p^{\beta}$ (\ref{eq:finetuningBC_app}}). Clearly the only contribution of the second term scaling as $\tilde{s}\,\epsilon^{m/(m-r')}$ is
\begin{equation}
 \sum_{n=1}^m\frac{\alpha_n u^{n}}{4n^2 B(n,1/2)}= \tilde{s}\frac{(4m)^p}{4p^2 B(p,1/2)}\,\epsilon^{\frac{m}{m-r'}}+ \mathcal{O}\left(\epsilon^{\frac{m}{m-r'}+1}\right)~.
\end{equation}
From the first term of the on-shell action (\ref{eq:Srhoext}) on the other side we get the contribution 
\begin{align}
&\left(\frac{(2p)!}{4p}\frac{\tilde{s}}{(p!)^24^p}\sum_{n=1}^m\frac{\alpha_{n,c}}{2(n+p)}\frac{(4m)^{n+p}}{B(n,1/2)}+\tilde{s}\sum_{n=1}^m \frac{(2n)!}{4n(n!)^2}\frac{\alpha_{n,c}}{4^n}\frac{(4m)^{n+p}}{2(n+p)B(p,1/2)}\right)\,\epsilon^{\frac{m}{m-r'}}~\cr
=&\left(\frac{(2p)!}{4p}\frac{m^p}{(p!)^2}\tilde{s}\sum_{n=1}^m(-1)^{n+1}\binom{m}{n}\frac{n}{(n+p)}+\frac{(4m)^p}{B(p,1/2)}\tilde{s}\sum_{n=1}^m(-1)^{n+1} \binom{m}{n}\frac{1}{4n(n+p)B(n,1/2)}\right)\,\epsilon^{\frac{m}{m-r'}}\cr
=&\left(\frac{(4m)^p}{4p^2 B(p,1/2)}\frac{m!p!}{(m+p)!}+ \frac{(4m)^p}{B(p,1/2)}\sum_{n=1}^m\binom{m}{n}\frac{ (-1)^{n+1}}{4n(n+p)}\right)\,\tilde{s}\,\epsilon^{\frac{m}{m-r'}}\cr
=&\left(\frac{(4m)^p}{2p^2 B(p,1/2)}\frac{m!p!}{(m+p)!}-\frac{(4m)^p}{4p^2 B(p,1/2)}\right)\,\tilde{s}\,\epsilon^{\frac{m}{m-r'}}+ H_m \frac{(4m)^p}{4p B(p,1/2)} \tilde{s}\,\epsilon^{\frac{m}{m-r'}}~.
\end{align}

}

\section{Hypergeometric functions}\label{app:bestiary}
We collect some useful properties about (regularised, generalised) hypergeometric functions. For $|z|<1$, $a_1, a_2, b_1 \in \mathbb{C}$ the hypergeometric function is defined by the power series
\begin{equation}\label{eq:def2F1}
_2F_1\left(a_1,a_2;b_1;z\right)\equiv \sum_{n=0}^\infty \frac{a_1^{(n)}a_2^{(n)}}{b_1^{(n)}}\frac{z^n}{n!}~, \quad x^{(n)}\equiv x(x+1)\cdots (x+n-1)= \frac{\Gamma(x+n)}{\Gamma(x)}~.
\end{equation}
In particular for $a_1=a_2=b_1=1$ we obtain the geometric series
\begin{equation}
_2F_1\left(1,1;1;x\right)= \sum_{n=0}^\infty  x^n~.
\end{equation}
For $a_1, \ldots a_p, b_1,\ldots b_q \in \mathbb{C}$ and $|z|<1$, (\ref{eq:def2F1}) generalises to the generalised hypergeometric function 
\begin{equation}
_p F_q \left[\begin{matrix} a_1 & a_2 & \cdots & a_p \\ b_1 & b_2 & \cdots & b_q \end{matrix}\,;\, z\right]\equiv \sum_{n=0}^\infty \frac{a_1^{(n)}\cdots a_p^{(n)}}{b_1^{(n)} \cdots b_q^{(n)}}\frac{z^n}{n!}~.
\end{equation}
Finally we define the regularised hypergeometric 
\begin{equation}\label{eq:def_regF}
_p \tilde{F}_q \equiv  \frac{1}{\Gamma(b_1)\cdots \Gamma(b_q)}\,_p F_q \left[\begin{matrix} a_1 & a_2 & \cdots & a_p \\ b_1 & b_2 & \cdots & b_q \end{matrix}\,;\, z\right]~.
\end{equation}
{\textbf{Normalisation condition $m=3$.}} 
To prove the conjecture  (\ref{eq:u3_a2a3}) we study
\begin{equation}\label{eq:u3_a2a3App}
u^{(3)}_\star(\alpha_2,\alpha_3)=4\sqrt{\pi} \sum_{k=0}^\infty\left(\frac{20\alpha_3}{3\alpha_2}\right)^{k}\frac{1}{\Gamma(2+k)}\,_3\tilde{F}_2\left(1,-k,1+k;\frac{1}{2}-\frac{k}{2}, 1-\frac{k}{2}; \frac{9\alpha_2^2}{40\alpha_3}\right)~,
\end{equation}
along the path $\gamma_\star^{(3)}$ (\ref{eq:gamma_star}). We start on the right hand side. 
\begin{equation}
_3\tilde{F}_2\left(1,1-k,k;1-\frac{k}{2}, \frac{3}{2}-\frac{k}{2}; \frac{3}{4}\right)= \lim_{(\rho_1,\rho_2)\rightarrow 1}\,_3 \tilde{F}_2\left(1,1-k,k;\rho_2-\frac{k}{2},\frac{3\rho_1}{2}-\frac{k}{2};\frac{3}{4}\right)~.
\end{equation}
Using the definition of the regularised hypergeometric function (\ref{eq:def_regF}) we obtain 
\begin{multline}\label{eq:wfc}
_3 \tilde{F}_2\left(1,1-k,k;\rho_2-\frac{k}{2},\frac{3\rho_1}{2}-\frac{k}{2};\frac{3}{4}\right)=\frac{1}{\Gamma\left(\rho_2-\frac{k}{2}\right)\Gamma\left(\frac{3\rho_1}{2}-\frac{k}{2}\right)}\cr
\times \left(1+ \sum_{\ell=1}^{k-1}(-1)^\ell\left( \frac{3}{4}\right)^\ell\frac{k^{(\ell)}(k-1)_\ell}{\prod_{n=0}^{\ell-1}\left(\frac{k}{2}-n-\frac{3\rho_1}{2}\right)\left(\frac{k}{2}-n-\rho_2\right)}\right)~,
\end{multline}
where $(k)_{\ell}$ is the falling factorial 
\begin{equation}
(k)_{\ell}\equiv \frac{\Gamma(k+1)}{\Gamma(k-\ell+1)}~,
\end{equation}
leading to 
\begin{equation}
u^{(3)}_\star(\alpha_2,\alpha_3)\big|_{\gamma_\star^{(3)}}= 12\sum_{k=1}^\infty\frac{(-t)^{k-1}}{\Gamma(k+1)}\sum_{\ell=0}^{k-1}\frac{(-3)^\ell}{3^{2k-1}}\frac{\Gamma(k+\ell)}{\Gamma(k-\ell)\Gamma(2+2\ell-k)}
\end{equation}
We are left to show for $k\geq 1$
\begin{align}\label{eq:bh_u3}
-12\frac{(-1)^k}{\Gamma(k+1)}\sum_{\ell=0}^{k-1}(-3)^\ell\frac{\Gamma(k+\ell)}{3^{2k-1}\Gamma(k-\ell)\Gamma(2+2\ell-k)}
= -12(-1)^k\binom{1/3}{k}~.
\end{align}
By writing out the fractional binomial coefficient we obtain
\begin{align}\label{eq:binomial_product}
 -12(-1)^k\binom{1/3}{k}&= -12 (-1)^k \frac{\Gamma(4/3)}{\Gamma(k+1)\Gamma(4/3-k)}= -12\frac{(-1)^k}{3^k\Gamma(k+1)}\prod_{n=0}^{k-1}(1-3n)~.
\end{align}
The last product we write in terms of stirling numbers $s_1(n,k)$ for the first kind. The Stirling numbers $s_1(n,k)$ enumerate $(-1)^{n-k}$ times the number of partitions of the symmetric group $S_n$ with exactly $k$ cycles. By definition, they are also the coefficients of the falling factorial 
\begin{equation}
(x)_n \equiv x(x-1)(x-2)\cdots (x-n+1) = \sum_{k=0}^n s_1(n,k) x^k~.
\end{equation}
Applying this to the product in (\ref{eq:binomial_product}) with $x= 1/3$ we find
\begin{align}
\prod_{n=0}^{k-1}(1-3n)&= (1-3)(1-3 \times 2)(1-3\times 3)\cdots (1-3 (k-1))= \sum_{n=0}^{k+1} 3^{k-n-1}\, s_1(k+1,n) \cr
&=\sum_{n=0}^{k+1} 3^n\,s_1(k+1,k+1-n) ~,
\end{align}
where in going to the last line we substituted $n\rightarrow k-n+1$. 
We now define 
\begin{align}
L_k\equiv \sum_{n=0}^{k}3^n s_1(k+1,k+1-n)~,\quad
R_k\equiv k! \sum_{n=0}^{k}3^n\frac{(-1)^n}{3^k}\binom{n+k}{n}\binom{n}{k-n}~.
\end{align}
We show that both expressions satisfy 
\begin{equation}\label{eq:recursion_LR}
L_{k}=(1-3k)L_{k-1}~, \quad R_{k}=(1-3k)R_{k-1}~, \quad k\geq 1~.
\end{equation}
We start with $L_k$. Using the recursion relation for the Stirling numbers 
\begin{equation}
s_1(k+1,n+1)= s_1(k,n)- ks_1(k,n+1)~,
\end{equation}
and $s_1(k,0)=0$ we find
\begin{align}
L_k&\equiv \sum_{n=0}^{k}3^n s_1(k+1,k+1-n)=  \sum_{n=0}^{k}3^n s_1(k,k-n)-k  \sum_{n=0}^{k}3^n s_1(k,k+1-n)\cr
&= \sum_{n=0}^{k-1}3^n s_1(k,k-n)+3^{k}s_1(k,0)- k\sum_{n=0}^{k}3^n s_1(k,k+1-n)= (1-3k)L_{k-1}~.
\end{align}
We now show the same recursion equation for $R_k$. We have
\begin{align}
R_k\equiv k! \sum_{n=0}^{k}3^n\frac{(-1)^n}{3^k}\binom{n+k}{n}\binom{n}{k-n}=\frac{(-3)^k}{(-1/3)!}\left(-\frac{1}{3}+k\right)!= \frac{(-3)^k}{(-1/3)!}\Gamma\left(k+\frac{2}{3}\right)
\end{align}
which implies
\begin{align}
R_{k-1}= -\frac{1}{3}\frac{(-3)^k}{(-1/3)!}\left(-\frac{4}{3}+k\right)!= -\frac{1}{3}\frac{(-3)^k}{(-1/3)!}\Gamma\left(-\frac{1}{3}+k\right)= \frac{1}{(1-3k)}R_k~.
\end{align}
Since $L_0=R_0$ and $L_1= R_1$ so (\ref{eq:recursion_LR}) completes the proof. 
\newline\newline
{\textbf{On-shell action $m=3$.}}
We now also rewrite the regularised hypergeometric appearing in (\ref{eq:F3_pert2}).
Following the same logic as for the normalisation condition we have 
\begin{align}
\,_3 \tilde{F}_2\left(1,-k,k;\frac{1}{2}-\frac{k}{2},1-\frac{k}{2};\frac{3}{4}\right)= \sum_{n=1}^k(-1)^n \frac{3^n}{2^k\sqrt{\pi}}\frac{k\Gamma(k+n)}{\Gamma(1+2n-k)\Gamma(1-n+k)}~,
\end{align} 
and consequently we can write $\mathcal{F}^{(3)}(\alpha_2,\alpha_3)$ along $\gamma_\star^{(3)}$ as 
\begin{align}
\mathcal{F}^{(3)}\left(\alpha_2,\alpha_3\right)\big|_{\gamma_\star^{(3)}}
&=-\sum_{k=1}^\infty \frac{(-1)^k}{3^{2k}\Gamma(k+3)}\sum_{\ell=0}^{k-1} \frac{(-3)^{\ell+1}\Gamma(k+\ell)}{\Gamma(k-\ell)\Gamma(3+2\ell-k)}(k+\ell)\,t^k~.
\end{align}
Comparing to the normalisation condition (\ref{eq:bh_u3}) we thus indeed obtain the scaling $\sim k^{-\boldsymbol{10/3}}$.

 \begingroup
  \section*{References}\label{references}
 \renewcommand{\section}[2]{}

\endgroup


\begin{thebibliography}{99}

\subsection*{Matrix integrals}

\bibitem{Kazakov:1989bc}
V.~Kazakov,
``The Appearance of Matter Fields from Quantum Fluctuations of 2D Gravity,''
Mod. Phys. Lett. A \textbf{4}, 2125 (1989)
doi:10.1142/S0217732389002392

\bibitem{Staudacher:1989fy}
M.~Staudacher,
``The Yang-lee Edge Singularity on a Dynamical Planar Random Surface,''
Nucl. Phys. B \textbf{336}, 349 (1990)
doi:10.1016/0550-3213(90)90432-D


\bibitem{tHooft:1974pnl} 
  G.~'t Hooft,
  ``A Two-Dimensional Model for Mesons,''
  Nucl.\ Phys.\ B {\bf 75}, 461 (1974).
  doi:10.1016/0550-3213(74)90088-1

\bibitem{Brezin:1977sv} 
  E.~Brezin, C.~Itzykson, G.~Parisi and J.~B.~Zuber,
  ``Planar Diagrams,''
  Commun.\ Math.\ Phys.\  {\bf 59}, 35 (1978).
  doi:10.1007/BF01614153

\bibitem{Eguchi:1981kk}
T.~Eguchi and H.~Kawai,
``Number of Random Surfaces on the Lattice and the Large $N$ Gauge Theory,''
Phys. Lett. B \textbf{110}, 143-147 (1982)
doi:10.1016/0370-2693(82)91023-1


\bibitem{Gross:1989vs}
D.~J.~Gross and A.~A.~Migdal,
``Nonperturbative Two-Dimensional Quantum Gravity,''
Phys. Rev. Lett. \textbf{64}, 127 (1990)
doi:10.1103/PhysRevLett.64.127


\bibitem{Brezin:1989db} 
  E.~Brezin, M.~R.~Douglas, V.~Kazakov and S.~H.~Shenker,
  ``The Ising Model Coupled to 2-$D$ Gravity: A Nonperturbative Analysis,''
  Phys.\ Lett.\ B {\bf 237}, 43 (1990).
  doi:10.1016/0370-2693(90)90458-I

\bibitem{Polyakov:1980ca}
A.~M.~Polyakov,
``Gauge Fields as Rings of Glue,''
Nucl.\ Phys.\ B \textbf{164}, 171-188 (1980)
doi:10.1016/0550-3213(80)90507-6;
Dyson-Schwinger equations approach to the large-$N$ limit: Model systems and string representation of Yang-Mills theory, Wadia, Spenta R.,
Phys. Rev. D, 1981
  doi: 10.1103/PhysRevD.24.970;
  A.~A.~Migdal,
  ``Loop Equations and 1/N Expansion,''
  Phys.\ Rept.\  {\bf 102}, 199 (1983).
  doi:10.1016/0370-1573(83)90076-5;
  J.~Ambj{\o}rn and Y.~M.~Makeenko,
  ``Properties of Loop Equations for the Hermitean Matrix Model and for Two-dimensional Quantum Gravity,''
  Mod.\ Phys.\ Lett.\ A {\bf 5}, 1753 (1990).
  doi:10.1142/S0217732390001992
  

\bibitem{Boulatov:1986sb}
D.~Boulatov and V.~Kazakov,
``The Ising Model on Random Planar Lattice: The Structure of Phase Transition and the Exact Critical Exponents,''
Phys. Lett. B \textbf{186}, 379 (1987)
doi:10.1016/0370-2693(87)90312-1


\bibitem{Alexandrov:2003nn}
S.~Y.~Alexandrov, V.~A.~Kazakov and D.~Kutasov,
``Nonperturbative effects in matrix models and D-branes,''
JHEP \textbf{09}, 057 (2003)
doi:10.1088/1126-6708/2003/09/057
[arXiv:hep-th/0306177 [hep-th]].



\bibitem{Ambjorn:2016lkl}
J.~Ambj{\o}rn, T.~Budd and Y.~Makeenko,
``Generalized multicritical one-matrix models,''
Nucl. Phys. B \textbf{913}, 357-380 (2016)
doi:10.1016/j.nuclphysb.2016.09.013
[arXiv:1604.04522 [hep-th]].
%
%


\bibitem{Bessis:1980ss} 
  D.~Bessis, C.~Itzykson and J.~B.~Zuber,
  ``Quantum field theory techniques in graphical enumeration,''
  Adv.\ Appl.\ Math.\  {\bf 1}, 109 (1980).
  doi:10.1016/0196-8858(80)90008-1


\bibitem{DiFrancesco:2004qj}
P.~Di Francesco,
``2D quantum gravity, matrix models and graph combinatorics,''
[arXiv:math-ph/0406013 [math-ph]], 

\bibitem{Caracciolo}
S.~Caracciolo and A.~Sportiello,
``Spanning Forests on Random Planar Lattices,''
J. Statist. Phys. \textbf{135}, 1063-1104 (2009)
doi:10.1007/s10955-009-9733-1
[arXiv:0903.4432 [hep-th]].
R.~Bondesan, S.~Caracciolo and A.~Sportiello,
``Critical Behaviour of Spanning Forests on Random Planar Graphs,''
J. Phys. A \textbf{50}, no.7, 074003 (2017)
doi:10.1088/1751-8121/aa546f
[arXiv:1608.02916 [cond-mat.stat-mech]].

\bibitem{DiFrancesco:1993cyw}
P.~Di Francesco, P.~H.~Ginsparg and J.~Zinn-Justin,
``2-D Gravity and random matrices,''
Phys. Rept. \textbf{254}, 1-133 (1995)
doi:10.1016/0370-1573(94)00084-G
[arXiv:hep-th/9306153 [hep-th]].


\bibitem{DioBeatrix}
D.~Anninos and B.~M\"uhlmann,
``Notes on Matrix Models,''
J. Stat. Mech. \textbf{2008}, 083109 (2020)
doi:10.1088/1742-5468/aba499
[arXiv:2004.01171 [hep-th]].

\bibitem{Ginsparg:1993is}
P.~H.~Ginsparg and G.~W.~Moore,
``Lectures on 2-D gravity and 2-D string theory,''
[arXiv:hep-th/9304011 [hep-th]].
  
\bibitem{Gross:1989vs} 
  D.~J.~Gross and A.~A.~Migdal,
  ``Nonperturbative Two-Dimensional Quantum Gravity,''
  Phys.\ Rev.\ Lett.\  {\bf 64}, 127 (1990).
  doi:10.1103/PhysRevLett.64.127

 \bibitem{enumerative_combinatorics}
 R.P. Stanley (2011), Vol. 1: Enumerative combinatorics. 
 Cambridge University Press.

\bibitem{Bell}
Bell E.T. (1927-1928); ``Partition Polynomials''. Annals of Mathematics 29 (1/4): 38-46.
doi: 10.2307/1967979 

\subsection*{Minimal models}


\bibitem{Itzykson:1986pj}
C.~Itzykson and J.~B.~Zuber,
``Two-Dimensional Conformal Invariant Theories on a Torus,''
Nucl. Phys. B \textbf{275}, 580-616 (1986)
doi:10.1016/0550-3213(86)90576-6

 
\bibitem{Polyakov:1984yq}
A.~M.~Polyakov, A.~A.~Belavin and A.~B.~Zamolodchikov,
``Infinite Conformal Symmetry of Critical Fluctuations in Two-Dimensions,''
J. Statist. Phys. \textbf{34}, 763 (1984)
doi:10.1007/BF01009438
  
\bibitem{Friedan}
D.~Friedan, Z.~a.~Qiu and S.~H.~Shenker,
``Conformal Invariance, Unitarity and Two-Dimensional Critical Exponents,''
Phys. Rev. Lett. \textbf{52}, 1575-1578 (1984)
doi:10.1103/PhysRevLett.52.1575
  
\bibitem{RochaCaridi}
A. Rocha-Caridi, 
in ``Vertex Operators in Mathematics and Physics''
Proceedings of a Conference November 10–17, 1983
Editors: Lepowsky, J., Mandelstam, S., Singer, I.M. (Eds.


\bibitem{Cardy:1985yy}
J.~L.~Cardy,
``Conformal Invariance and the Yang-lee Edge Singularity in Two-dimensions,''
Phys. Rev. Lett. \textbf{54}, 1354-1356 (1985)
doi:10.1103/PhysRevLett.54.1354


\bibitem{AlvarezGaume:1989vk}
L.~Alvarez-Gaume, G.~Sierra and C.~Gomez,
``Topics in conformal field theory,''
CERN-TH-5540-89.

\bibitem{yellowbook}
P. di Francesco, P. Mathieu, D. S{\'e}n{\'e}chal,
``Conformal field theory,''
Springer Science+ Business Media, LLC, 1997

\bibitem{Kadanoff:1966wm}
L.~P.~Kadanoff,
``Scaling laws for Ising models near T(c),''
Physics Physique Fizika \textbf{2}, 263-272 (1966)
doi:10.1103/PhysicsPhysiqueFizika.2.263


\bibitem{LeeYang}
Yang, C. N. , Lee, T. D.
``Statistical Theory of Equations of State and Phase Transitions. I. Theory of Condensation,''
Phys. Rev.,87, ssue = {3}, {404--409}pg, {1952},
  doi = {10.1103/PhysRev.87.404}
  
\bibitem{vonGehlen:1994rp}
G.~von Gehlen,
``NonHermitian tricriticality in the Blume-Capel model with imaginary field,''
[arXiv:hep-th/9402143 [hep-th]].

\bibitem{Blume}
M. Blume, 
``Theory of the First-Order Magnetic Phase Change in U${\mathrm{O}}_{2}$,''
Phys. Rev. 141, issue 2, pages: 517-524, 1966
  doi = {10.1103/PhysRev.141.517};
H.W. Capel, 
``On the possibility of first-order phase transitions in Ising systems of triplet ions with zero-field splitting",
Physica 32, number 5, pages: 966-988, 1966
doi = {10.1016/0031-8914(66)90027-9}


  \subsection*{ 2d quantum gravity and Liouville theory}

\bibitem{Seiberg:1990eb}
N.~Seiberg,
``Notes on quantum Liouville theory and quantum gravity,''
Prog. Theor. Phys. Suppl. \textbf{102}, 319-349 (1990)
doi:10.1143/PTPS.102.319

\bibitem{Fateev:2000ik}
V.~Fateev, A.~B.~Zamolodchikov and A.~B.~Zamolodchikov,
``Boundary Liouville field theory. 1. Boundary state and boundary two point function,''
[arXiv:hep-th/0001012 [hep-th]].


\bibitem{Teschner:1995yf}
J.~Teschner,
``On the Liouville three point function,''
Phys. Lett. B \textbf{363}, 65-70 (1995)
doi:10.1016/0370-2693(95)01200-A
[arXiv:hep-th/9507109 [hep-th]].

\bibitem{Zamolodchikov:2001ah}
A.~B.~Zamolodchikov and A.~B.~Zamolodchikov,
``Liouville field theory on a pseudosphere,''
[arXiv:hep-th/0101152 [hep-th]].


\bibitem{Zamolodchikov:1982vx} 
  A.~B.~Zamolodchikov,
  ``On The Entropy Of Random Surfaces,''
  Phys.\ Lett.\  {\bf 117B}, 87 (1982).
  doi:10.1016/0370-2693(82)90879-6


\bibitem{Knizhnik:1988ak} 
  V.~G.~Knizhnik, A.~M.~Polyakov and A.~B.~Zamolodchikov,
  ``Fractal Structure of 2D Quantum Gravity,''
  Mod.\ Phys.\ Lett.\ A {\bf 3}, 819 (1988).
  doi:10.1142/S0217732388000982

\bibitem{David:1988hj} 
  F.~David,
  ``Conformal Field Theories Coupled to 2D Gravity in the Conformal Gauge,''
  Mod.\ Phys.\ Lett.\ A {\bf 3}, 1651 (1988).
  doi:10.1142/S0217732388001975
  
\bibitem{Distler:1988jt} 
  J.~Distler and H.~Kawai,
  ``Conformal Field Theory and 2D Quantum Gravity,''
  Nucl.\ Phys.\ B {\bf 321}, 509 (1989).
  doi:10.1016/0550-3213(89)90354-4


\bibitem{Polyakov:1981rd}
A.~M.~Polyakov,
``Quantum Geometry of Bosonic Strings,''
Phys. Lett. B \textbf{103}, 207-210 (1981)
doi:10.1016/0370-2693(81)90743-7

%
%

\bibitem{Moore:1991ir}
G.~W.~Moore, N.~Seiberg and M.~Staudacher,
``From loops to states in 2-D quantum gravity,''
Nucl. Phys. B \textbf{362}, 665-709 (1991)
doi:10.1016/0550-3213(91)90548-C

\bibitem{Moore:1991ag}
G.~W.~Moore and N.~Seiberg,
``From loops to fields in 2-D quantum gravity,''
Int. J. Mod. Phys. A \textbf{7}, 2601-2634 (1992)
doi:10.1142/S0217751X92001174

\bibitem{Belavin:2014hsa}
A.~A.~Belavin and V.~A.~Belavin,
``Frobenius manifolds, Integrable Hierarchies and Minimal Liouville Gravity,''
JHEP \textbf{09}, 151 (2014)
doi:10.1007/JHEP09(2014)151
[arXiv:1406.6661 [hep-th]].

\bibitem{Belavin:2005af}
A.~Belavin and A.~B.~Zamolodchikov,
``Moduli integrals, ground ring and four-point function in minimal Liouville gravity,''

\bibitem{Belavin:2008kv}
A.~A.~Belavin and A.~B.~Zamolodchikov,
``On Correlation Numbers in 2D Minimal Gravity and Matrix Models,''
J. Phys. A \textbf{42}, 304004 (2009)
doi:10.1088/1751-8113/42/30/304004
[arXiv:0811.0450 [hep-th]].

\subsection*{JT gravity}

\bibitem{Jackiw:1984je}
R.~Jackiw,
``Lower Dimensional Gravity,''
Nucl. Phys. B \textbf{252}, 343-356 (1985)
doi:10.1016/0550-3213(85)90448-1;
C.~Teitelboim,
``Gravitation and Hamiltonian Structure in Two Space-Time Dimensions,''
Phys. Lett. B \textbf{126}, 41-45 (1983)
doi:10.1016/0370-2693(83)90012-6

\bibitem{Saad:2019lba}
P.~Saad, S.~H.~Shenker and D.~Stanford,
``JT gravity as a matrix integral,''
[arXiv:1903.11115 [hep-th]].

\bibitem{Johnson:2019eik}
C.~V.~Johnson,
``Nonperturbative Jackiw-Teitelboim gravity,''
Phys. Rev. D \textbf{101}, no.10, 106023 (2020)
doi:10.1103/PhysRevD.101.106023
[arXiv:1912.03637 [hep-th]].

\bibitem{Chen:2020tes}
Y.~Chen, V.~Gorbenko and J.~Maldacena,
``Bra-ket wormholes in gravitationally prepared states,''
[arXiv:2007.16091 [hep-th]].


\subsection*{Hilbert space}

\bibitem{Hartle:1983ai}
J.~B.~Hartle and S.~W.~Hawking,
``Wave Function of the Universe,''
Adv. Ser. Astrophys. Cosmol. \textbf{3}, 174-189 (1987)
doi:10.1103/PhysRevD.28.2960

\bibitem{Bershadsky:1990xb}
M.~Bershadsky and I.~R.~Klebanov,
``Genus one path integral in two-dimensional quantum gravity,''
Phys. Rev. Lett. \textbf{65}, 3088-3091 (1990)
doi:10.1103/PhysRevLett.65.3088;
M.~Bershadsky and I.~R.~Klebanov,
``Partition functions and physical states in two-dimensional quantum gravity and supergravity,''
Nucl. Phys. B \textbf{360}, 559-585 (1991)
doi:10.1016/0550-3213(91)90416-U


\bibitem{Kutasov:1990sv}
D.~Kutasov and N.~Seiberg,
``Number of degrees of freedom, density of states and tachyons in string theory and CFT,''
Nucl. Phys. B \textbf{358}, 600-618 (1991)
doi:10.1016/0550-3213(91)90426-X


\bibitem{Lian:1991gk}
B.~H.~Lian and G.~J.~Zuckerman,
``New selection rules and physical states in 2-D gravity: Conformal gauge,''
Phys. Lett. B \textbf{254}, 417-423 (1991)
doi:10.1016/0370-2693(91)91177-W


\bibitem{Imbimbo:1991ia}
C.~Imbimbo, S.~Mahapatra and S.~Mukhi,
``Construction of physical states of nontrivial ghost number in c \ensuremath{<} 1 string theory,''
Nucl. Phys. B \textbf{375}, 399-420 (1992)
doi:10.1016/0550-3213(92)90038-D

\bibitem{Bouwknegt:1991mv}
P.~Bouwknegt, J.~G.~McCarthy and K.~Pilch,
``Fock space resolutions of the Virasoro highest weight modules with c \ensuremath{<}= 1,''
Lett. Math. Phys. \textbf{23}, 193-204 (1991)
doi:10.1007/BF01885497
[arXiv:hep-th/9108023 [hep-th]].

\bibitem{Kutasov:1991qx}
D.~Kutasov, E.~J.~Martinec and N.~Seiberg,
``Ground rings and their modules in 2-D gravity with c \ensuremath{<}= 1 matter,''
Phys. Lett. B \textbf{276}, 437-444 (1992)
doi:10.1016/0370-2693(92)91664-U
[arXiv:hep-th/9111048 [hep-th]].

%


\bibitem{Itzykson:1986pk}
C.~Itzykson, H.~Saleur and J.~B.~Zuber,
``Conformal Invariance of Nonunitary Two-dimensional Models,''
Europhys. Lett. \textbf{2}, 91 (1986)
doi:10.1209/0295-5075/2/2/004

  \bibitem{PolchinskiBook}
 Polchinski, J. (1998). String Theory.  Vol. 1: An Introduction to the Bosonic String. (Cambridge Monographs on Mathematical Physics). 
 Cambridge: Cambridge University Press. doi:10.1017/CBO9780511816079
 

\bibitem{DeWitt:1967yk}
B.~S.~DeWitt,
``Quantum Theory of Gravity. 1. The Canonical Theory,''
Phys. Rev. \textbf{160}, 1113-1148 (1967)
doi:10.1103/PhysRev.160.1113;
B.~S.~DeWitt,
``Quantum Theory of Gravity. 2. The Manifestly Covariant Theory,''
Phys. Rev. \textbf{162}, 1195-1239 (1967)
doi:10.1103/PhysRev.162.1195

\subsection*{de Sitter space and finiteness}

\bibitem{DioGuille}
D.~Anninos and G.~A.~Silva,
``Solvable Quantum Grassmann Matrices,''
J. Stat. Mech. \textbf{1704}, no.4, 043102 (2017)
doi:10.1088/1742-5468/aa668f
[arXiv:1612.03795 [hep-th]].

 \bibitem{Banks_dS}
  T.~Banks, B.~Fiol and A.~Morisse,
  ``Towards a quantum theory of de Sitter space,''
  JHEP {\bf 0612} (2006) 004
  doi:10.1088/1126-6708/2006/12/004
  [hep-th/0609062];

\bibitem{Erik_dS} 
  M.~K.~Parikh and E.~P.~Verlinde,
  ``De Sitter holography with a finite number of states,''
  JHEP {\bf 0501} (2005) 054
  doi:10.1088/1126-6708/2005/01/054
  [hep-th/0410227];
  
\bibitem{Dong:2010pm}
X.~Dong, B.~Horn, E.~Silverstein and G.~Torroba,
Class. Quant. Grav. \textbf{27}, 245020 (2010)
doi:10.1088/0264-9381/27/24/245020
[arXiv:1005.5403 [hep-th]].
  
\bibitem{Gibbons:1977mu} 
  G.~W.~Gibbons and S.~W.~Hawking,
  ``Cosmological Event Horizons, Thermodynamics, and Particle Creation,''
  Phys.\ Rev.\ D {\bf 15}, 2738 (1977).
  doi:10.1103/PhysRevD.15.2738
  
\bibitem{Gibbons:1976ue}
G.~W.~Gibbons and S.~W.~Hawking,
``Action Integrals and Partition Functions in Quantum Gravity,''
Phys. Rev. D \textbf{15}, 2752-2756 (1977)
doi:10.1103/PhysRevD.15.2752

\bibitem{Holzhey:1994we} 
  C.~Holzhey, F.~Larsen and F.~Wilczek,
  ``Geometric and renormalized entropy in conformal field theory,''
  Nucl.\ Phys.\ B {\bf 424}, 443 (1994)
  doi:10.1016/0550-3213(94)90402-2
  [hep-th/9403108].
  
\bibitem{Calabrese:2004eu} 
  P.~Calabrese and J.~L.~Cardy,
  ``Entanglement entropy and quantum field theory,''
  J.\ Stat.\ Mech.\  {\bf 0406}, P06002 (2004)
  doi:10.1088/1742-5468/2004/06/P06002
  [hep-th/0405152].

\bibitem{Casini:2011kv} 
  H.~Casini, M.~Huerta and R.~C.~Myers,
  ``Towards a derivation of holographic entanglement entropy,''
  JHEP {\bf 1105}, 036 (2011)
  doi:10.1007/JHEP05(2011)036
  [arXiv:1102.0440 [hep-th]].

 
\end{thebibliography}
\end{document}